\documentclass[amssymb,prd,superscriptaddress,aps,nofootinbib,twocolumn,showpacs]{revtex4-1}
\usepackage{graphicx, epsfig, amssymb} %include figure files
\usepackage{amsmath, amsfonts}
\usepackage{bm} %include bold math: \bm{} creates bold letters in math mode
\usepackage[breaklinks]{hyperref}
\usepackage{color}
\usepackage{enumerate}
\usepackage{multirow}

\def\nn{\nonumber}

\def\be{\begin{equation}}
\def\ee{\end{equation}}
\def\beq{\begin{eqnarray}}
\def\eeq{\end{eqnarray}}
\begin{document}

\title{Tidal Love numbers of a slowly spinning neutron star}

%%%%
\author{Paolo Pani}\email{paolo.pani@roma1.infn.it}
\affiliation{Dipartimento di Fisica, ``Sapienza'' Universit\`a di Roma \& Sezione INFN Roma1, P.A. Moro 5, 00185, Roma, Italy.}
\affiliation{CENTRA, Departamento de F\'{\i}sica, Instituto Superior T\'ecnico, Universidade de Lisboa, Avenida~Rovisco Pais 1, 1049 Lisboa, Portugal.}

\author{Leonardo Gualtieri}\email{leonardo.gualtieri@roma1.infn.it}
\affiliation{Dipartimento di Fisica, ``Sapienza'' Universit\`a di Roma \& Sezione INFN Roma1, P.A. Moro 5, 00185, Roma, Italy.}

\author{Valeria Ferrari}\email{valeria.ferrari@roma1.infn.it}
\affiliation{Dipartimento di Fisica, ``Sapienza'' Universit\`a di Roma \& Sezione INFN Roma1, P.A. Moro 5, 00185, Roma, Italy.}
%%%

\begin{abstract}
By extending our recent framework to describe the tidal deformations of a spinning compact object, we compute for the first time the tidal Love numbers of a spinning neutron star to linear order in the angular momentum.
The spin of the object introduces couplings between electric and magnetic distortions and new classes of spin-induced (``rotational'') tidal Love numbers emerge. We focus on stationary tidal fields, which induce axisymmetric perturbations. 
We present the perturbation equations for both electric-led and magnetic-led rotational Love numbers for generic multipoles and explicitly solve them for various tabulated equations of state and for a tidal field with an electric (even parity) and magnetic (odd parity) component with $\ell=2,3,4$. 
For a binary system close to the merger, various components of the tidal field become relevant. In this case we find that an octupolar magnetic tidal field can significantly modify the mass quadrupole moment of a neutron star. Preliminary estimates, assuming a spin parameter $\chi\approx0.05$, show modifications $\gtrsim10\%$ relative to the static case, at an orbital distance of five stellar radii.
Furthermore, the rotational Love numbers as functions of the moment of inertia are much more sensitive to the equation of state than in the static case, where approximate universal relations at the percent level exist. For a neutron-star binary approaching
the merger, we estimate that the approximate universality of the induced
mass quadrupole moment deteriorates from $1\%$ in the static case to roughly $6\%$ when $\chi\approx0.05$.
Our results suggest that spin-tidal couplings can introduce important corrections to the gravitational waveforms of spinning neutron-star binaries approaching the merger.

%%%
\end{abstract}


\maketitle

%%%%%%%%%%%%%%%%%%%%%%%%% 
\section{Introduction}
%%%%%%%%%%%%%%%%%%%%%%%%%
In a previous paper~\cite{Pani:2015hfa} (henceforth Paper~I) we initiated the fascinating study of the tidal
deformations of a spinning compact object (see also Ref.~\cite{Landry:2015zfa} which appeared at the same time with
Paper~I). Our framework --~and that independently developed in Refs.~\cite{Landry:2015zfa,Poisson:2014gka}~-- is based
on a perturbative expansion in the angular moment which is valid for slowly spinning objects. When applied to the case
of spinning black holes, these studies have shown that the deformations of the multipole moments of a Kerr black holes
immersed in a tidal field are all zero\footnote{At least in the axisymmetric case to second order in the
  spin~\cite{Pani:2015hfa} and generically to first order in the spin~\cite{Landry:2015zfa}.}. In other words, the tidal
Love numbers --~i.e., the deformation of the multipole moments per unit tidal field~\cite{Murraybook,PoissonWill}~-- of
a spinning black hole are zero, as in the static case~\cite{Binnington:2009bb,Damour:2009vw,Guerlebeck:2015xpa}.

In this follow-up, we focus on the tidal Love numbers of a spinning neutron star (NS). The main motivation for such
analysis comes from the prospects of measuring the tidal Love numbers through gravitational-wave (GW) detections of
compact binaries~\cite{Flanagan:2007ix,Hinderer:2007mb} (see also
\cite{Baiotti:2010xh,Baiotti:2011am,Vines:2011ud,Pannarale:2011pk,Vines:2010ca,Lackey:2011vz,Lackey:2013axa,Favata:2013rwa,Yagi:2013baa,Maselli:2013mva,Maselli:2013rza}).

The tidal Love numbers encode the deformability of a self-gravitating object immersed in a tidal environment and depend
sensibly on the object internal structure~\cite{Murraybook,PoissonWill}. For static NSs, these numbers depend on the NS
mass and on the equation of state (EoS) of the matter composing the star. Therefore, measuring the NS deformability
through GW detections would help to constrain the behavior of matter at ultranuclear
density~\cite{Lattimer:2004pg,Hinderer:2009ca,Postnikov:2010yn,Damour:2012yf,Maselli:2013rza}.

Previous studies on the tidal deformability of NSs have considered only nonspinning objects.
The scope of the present paper is to quantify the importance of spin corrections to the NS tidal deformability by computing explicitly a new class of tidal Love numbers that was introduced in Paper~I and in Ref.~\cite{Landry:2015zfa}.

%%%%%%%%%%%%%%%%%%%%%%%%%
\subsubsection*{Executive summary}
%%%%%%%%%%%%%%%%%%%%%%%%%
For the reader's convenience, we summarize here the structure of the paper and our main results.
In Sec.~\ref{sec:tidal} we introduce the tidal Love numbers of a rotating body, extending the formalism developed in Paper~I.
Section~\ref{sec:estimate} is devoted to an estimate of spin effects for the tidal Love numbers, anticipating the main results obtained in the rest of the paper [cf. Fig.~\ref{fig:ratio}]. Our estimates show that --~for a NS-NS binary approaching the merger~-- the mass quadrupole moment of one star with spin\footnote{Here $\chi\equiv J/M^2$ is the dimensionless spin parameter, $M$ and $J$ being the mass and angular momentum of the star, respectively.} $\chi\approx0.05$ can deviate roughly by $13\%$ relative to the static case [cf. Eq.~\eqref{dM1}].

The equations governing the tidal perturbations are discussed in Sec.~\ref{sec:eqs} and in the Appendixes~\ref{app:Kojima} and \ref{app:sources}. In Sec.~\ref{sec:exterior} and in Appendix~\ref{app:exterior} we present the full exterior solution of a tidally deformed spinning object to first order in the spin for both an electric and a magnetic component of the tidal field with $\ell=2,3,4$. Such an exterior solution is specialized to the case of a spinning black hole in Appendix~\ref{app:BH}, where we also show that the corresponding tidal Love numbers are precisely zero in this case. This result complements the analyses presented in Ref.~\cite{Poisson:2014gka} and in Paper~I.

The final expressions for the rotational Love numbers of a spinning NS and the matching procedure to explicitly compute
them are presented in Sec.~\ref{sec:LoveNS}. Section~\ref{sec:results} is devoted to a detailed numerical analysis of
the rotational Love numbers in various cases and contains the main results of this work. In Fig.~\ref{fig:rot_VS_C} we
show some of the rotational Love numbers as functions of the compactness for various EoSs. In Fig.~\ref{fig:rot_universal} we discuss to which level these new Love numbers are
independent of the EoS, showing that the approximate universality~\cite{Yagi:2013bca,Yagi:2013awa,Yagi:2013sva} that
exists in the static\footnote{It is important to stress that the tidal Love numbers entering the nearly universal relations found in Refs.~\cite{Yagi:2013bca,Yagi:2013awa,Yagi:2013sva} are those of \emph{nonspinning} NSs, whereas the moment of inertia and the mass quadrupole moment refer to rotating stellar configurations. The spin corrections to the tidal Love numbers of a NS are computed in this work for the first time.} case deteriorates as the spin increases.

In the main text we focus on those rotational Love numbers which are associated with perturbations that do not break the reflection symmetry on the equatorial plane (and are therefore more relevant phenomenologically). The case of equatorial-symmetry breaking is briefly discussed in Appendix~\ref{app:noequatorial}. Along the way we clarify some questions related to the tidal Love numbers, for example in Appendix~\ref{app:comment} we reveal some inconsistency in the computation of the magnetic tidal Love numbers that appeared in the literature.
Through this work, we use geometrized $G=c=1$ units. Spacetime indices are
denoted by Greek letters, while space indices are denoted by Latin letters, i.e., $x^\mu=(x^0,x^a)$; $x^0$ is the time
coordinate associated with the unperturbed compact object.

%%%%%%%%%%%%%%%%%%%%%%%%%
\section{Tidal Love numbers of a spinning object}\label{sec:tidal}
%%%%%%%%%%%%%%%%%%%%%%%%%

%%%%%%%%%%%%%%%%%%%%%%%%%
\subsection{Definition}
%%%%%%%%%%%%%%%%%%%%%%%%%
Let us consider a (generically spinning) compact object immersed in a generic tidal environment. We adopt the same
decomposition of the tidal field as in Ref.~\cite{Binnington:2009bb}. Namely, we define the symmetric and trace-free
electric and magnetic tidal multipole moments as ${\cal E}_{a_1\dots a_\ell}\equiv [(\ell-2)!]^{-1}\langle
C_{0a_10a_2;a_3\dots a_\ell}\rangle$ and
${\cal B}_{a_1\dots a_\ell}\equiv [\frac{2}{3}(\ell+1)(\ell-2)!]^{-1}\langle\epsilon_{a_1 b c} C^{bc}_{a_2 0;a_3\dots a_\ell}\rangle$,
where $C_{abcd}$ is the Weyl tensor, a semicolon denotes a covariant derivative, $\epsilon_{abc}$ is the permutation
symbol, the angular brackets denote symmetrization of the indices $a_i$ and all traces are removed.
It can be shown that the electric (respectively, magnetic) moments ${\cal E}_{a_1\dots a_\ell}$ (respectively, ${\cal B}_{a_1\dots a_\ell}$) can be decomposed in a basis of even (respectively, odd) parity spherical harmonics with the symmetry axis aligned with the spin vector~\cite{Binnington:2009bb,Poisson:2014gka}. 
We denote by ${\cal E}^{(\ell)}_{m}$ and  ${\cal B}^{(\ell)}_{m}$ the amplitude of the electric and magnetic components of the external tidal field with harmonic indices $(\ell,m)$, where $m$ is the azimuthal number. The structure of the external tidal field is entirely encoded in the coefficients ${\cal E}^{(\ell)}_{m}$ and  ${\cal B}^{(\ell)}_{m}$ (cf. Ref.~\cite{Binnington:2009bb} for details).

As a result of the external perturbation, the mass and current multipole moments\footnote{We consider the Geroch-Hansen multipole moments~\cite{Geroch:1970cd,Hansen:1974zz}, which are equivalent~\cite{Gursel} to the multipole moments defined by Thorne~\cite{Thorne:1980ru} using asymptotically mass-centered Cartesian coordinates.} ($M_\ell$ and $S_\ell$, respectively) of the compact object will be deformed. In linear perturbation theory, these deformations are proportional to the applied tidal field. In the nonrotating case mass (current) multipoles have even (odd) parity, and therefore they 
only depend on electric (magnetic) components of the tidal field;
we can define the standard tidal Love numbers as
%%%
\begin{equation}
 \lambda_E^{(\ell)}\equiv \frac{\partial M_\ell}{\partial {\cal E}^{(\ell)}_ {m}}\,,\qquad 
 \lambda_M^{(\ell)}\equiv \frac{\partial S_\ell}{\partial {\cal B}^{(\ell)}_ {m}}\,, \label{Love0}
\end{equation}
%%%
and these quantities are independent of the tidal field.  The definitions above agree, modulo some normalization factor,
with the standard ones~\cite{Hinderer:2007mb,Binnington:2009bb,Damour:2009vw}, usually denoted as electric and magnetic
tidal Love numbers, respectively.  Because the object is spherically symmetric, the azimuthal number $m$ is
degenerate. Furthermore, parity and the angular momentum number $\ell$ are conserved: an electric (i.e. even parity) tidal field with harmonic index $\ell$ can only deform the mass multipole moment of
order $\ell$, whereas a magnetic (i.e. odd parity) tidal field with harmonic index $\ell$ can only deform the current
multipole moment of order $\ell$.

\begin{figure}[t]
\begin{center}
\includegraphics[width=6.5cm]{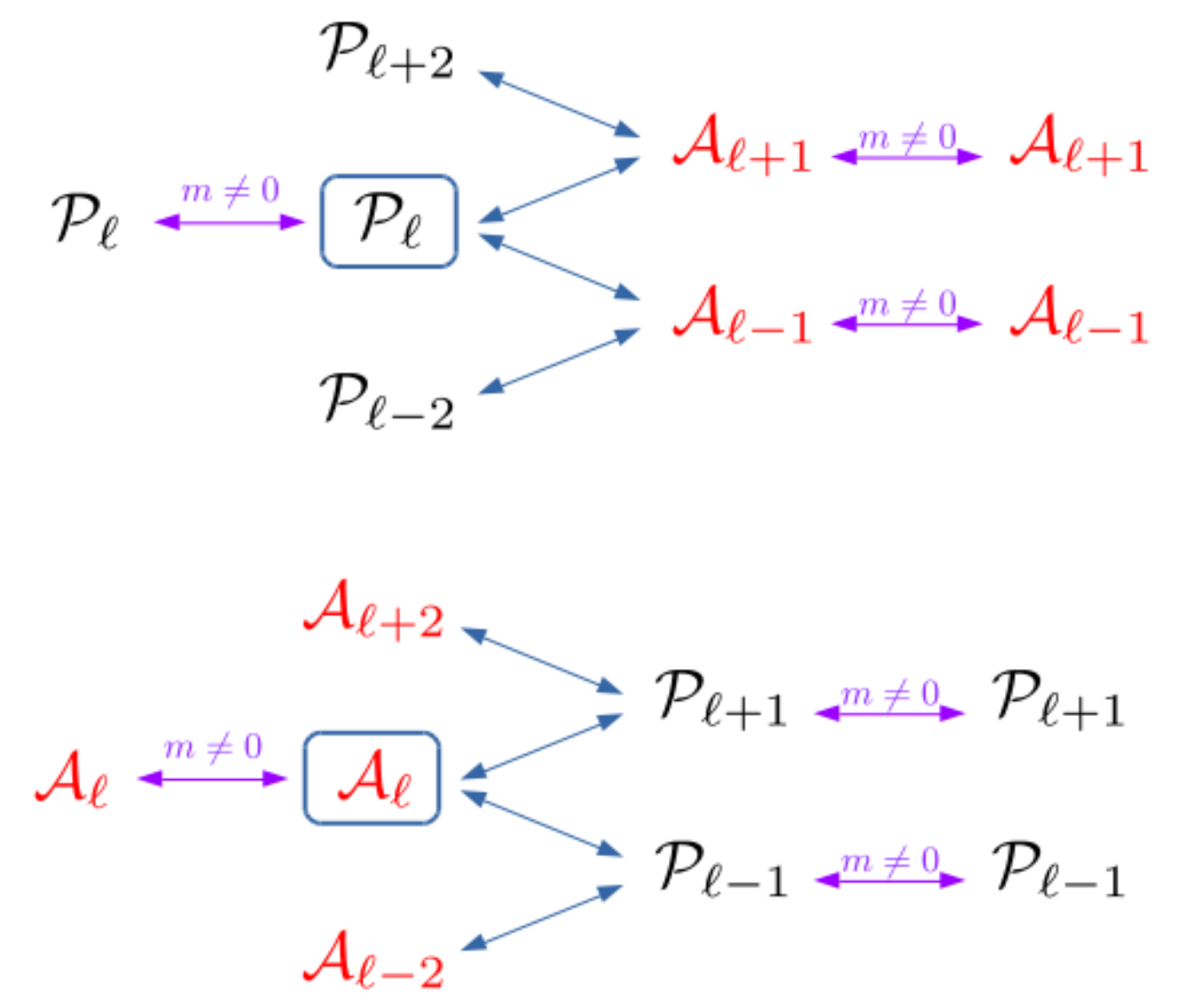}
\caption{(color online). Scheme of the spin-tidal coupling in the slow-rotation approximation to second order in the spin for the electric-led system (top diagram) and for the magnetic-led system (bottom diagram). ${\cal P}_\ell$ and ${\cal A}_\ell$ generically denote polar (i.e., even parity, or electric) and axial (i.e., odd parity, or magnetic) perturbations with harmonic index $\ell$. The quantity enclosed in a box denotes the component of the external tidal field. Perturbations with different parity and harmonic index are coupled to the external perturbations by arrows, each arrow (taken in either direction) denoting a coupling of linear order in the spin. For example, in the top diagram ${\cal P}_\ell$ sources ${\cal P}_{\ell+2}$ deformations to second order in the spin, since two arrows are needed to go from ${\cal P}_\ell$ to ${\cal P}_{\ell+2}$. Purple horizontal arrows denote Zeeman-like couplings which are nonzero only in the nonaxisymmetric case, $m\neq0$ (cf. Paper~I and the review~\cite{Pani:2013pma} for details.)}
\label{fig:scheme}
\end{center}
\end{figure}

Such degeneracy is broken when the central object is spinning. In such case there exist some selection rules which are discussed in detail in Paper~I and are schematically depicted in Fig.~\ref{fig:scheme}. Each arrow in this figure denotes a coupling of linear order in the spin. 
Therefore, to first order in the spin, electric (respectively, magnetic) tidal perturbations with multipolar index $\ell$ source magnetic (respectively, electric) deformations with multipolar index $\ell\pm1$ and vice versa. In particular, the mass quadrupole ($\ell=2$, electric) moment of the star acquires ${\cal O}(\chi)$ corrections proportional to the magnetic octupolar ($\ell=3$) tidal field. 

Furthermore, a nonaxisymmetric tidal field with index $\ell$ would deform the multipole moment of the same parity and with the same harmonic index $\ell$ to first order in the spin through a Zeeman-like splitting term proportional to the azimuthal number $m$ and denoted with a horizontal purple line in Fig.~\ref{fig:scheme}.

Finally, to second order in the spin, an electric (respectively, magnetic) tidal field with multipolar index $\ell$ sources electric (respectively, magnetic) deformations with multipolar indices $\ell$ and $\ell\pm2$. Thus, for instance, the mass quadrupole moment
acquires a ${\cal O}(\chi^2)$ correction proportional to the electric quadrupolar tidal field.

On the light of such selection rules, we can define a more generic set of tidal Love numbers as follows:
%%%
\begin{eqnarray}
&& \lambda_{E,-}^{(\ell\ell'm)}\equiv \frac{\partial M_\ell}{\partial {\cal B}^{(\ell')}_{m}}\,,\qquad 
 \lambda_{M,-}^{(\ell\ell'm)}\equiv \frac{\partial S_\ell}{\partial {\cal E}^{(\ell')}_{m}}\,, \label{Lovem}\\
 %%%%
 &&\lambda_{E,+}^{(\ell\ell'm)}\equiv \frac{\partial M_\ell}{\partial {\cal E}^{(\ell')}_{m}}\,,\qquad 
 \lambda_{M,+}^{(\ell\ell'm)}\equiv \frac{\partial S_\ell}{\partial {\cal B}^{(\ell')}_{m}}\,. \label{Lovep}
 %%%
\end{eqnarray}
%%%
The Love numbers with the ``$+$'' subscript measure how a mass (respectively, current) multipole moment is deformed by the presence of an electric (respectively, magnetic) tidal field, i.e. by a field of the same parity. On the other hand, the Love numbers with the ``$-$'' subscript measure how a mass (respectively, current) multipole moment is deformed by the presence of a magnetic (respectively, electric) tidal field, i.e. by a field of the opposite parity.
The definition just introduced generalizes the one given in Paper~I. Such generalization is necessary to accommodate the presence of a magnetic tidal field or of a tidal component with $\ell>2$; both possibilities were not considered in Paper~I. 

Because of the spin selection rules mentioned above (see also Paper~I and Ref.~\cite{Pani:2013pma}), the Love numbers in Eqs.~\eqref{Lovem} and \eqref{Lovep} enjoy various interesting properties (cf. Fig.~\ref{fig:scheme}): 
\begin{enumerate}
 \item At zeroth order in the spin, the Love numbers in Eq.~\eqref{Lovem} vanish, whereas the Love numbers in Eq.~\eqref{Lovep} are different from zero only when $\ell'=\ell$ and reduce to those defined in Eq.~\eqref{Love0}. As mentioned above, the azimuthal number $m$ is degenerate in this case.
 \item At first order in the spin, the Love numbers in Eq.~\eqref{Lovem} are different from zero only when $\ell'=\ell\pm1$, whereas the Love numbers in Eq.~\eqref{Lovep} are different from zero only when $\ell'=\ell$ and in the nonaxisymmetric case, $m\neq0$. These latter corrections correspond to the Zeeman splitting mentioned above.
 \item At second order in the spin, the Love numbers in Eq.~\eqref{Lovem} acquire new corrections only in the nonaxisymmetric case, whereas the Love numbers in Eq.~\eqref{Lovep} are different from zero both when $\ell'=\ell$ and when $\ell'=\ell\pm2$ for any value of $m$. The terms $\lambda_{E,+}^{(\ell\ell m)}$ and $\lambda_{M,+}^{(\ell\ell m)}$ include ${\cal O}(\chi^2)$ corrections to the static Love numbers in Eq.~\eqref{Love0}.
\end{enumerate}
%%%%%

In this work we focus on axisymmetric tidal perturbations to first order in the spin. In this case, the Love numbers in Eq.~\eqref{Lovep} reduce to the static case, Eq.~\eqref{Love0}, and one is left with the spin-induced (``rotational'') Love numbers in Eq.~\eqref{Lovem} with $m=0$. 
An explicit computation of the second-order and nonaxisymmetric corrections is left for future work. To simplify the notation, in the following we will introduce the shorthand notation
%%%
\begin{eqnarray}
&& \delta \lambda_{E}^{(\ell\ell')}\equiv \lambda_{E,-}^{(\ell\ell'0)}\,,\qquad \delta \lambda_{M}^{(\ell\ell')}\equiv \lambda_{M,-}^{(\ell\ell'0)}\,. \label{Love1}
 %%%
\end{eqnarray}
%%%
For example, $\delta \lambda_E^{(23)}$ denotes the correction to $\lambda_E^{(2)}$ arising from the coupling to the axisymmetric magnetic octupolar ($\ell=3$) component of the tidal field to first order in the spin. Among the various rotational Love numbers that we will compute, $\delta \lambda_E^{(23)}$ is the most relevant one from a phenomenological point of view because it is associated with a deformation of the quadrupole moment of the star.

%%%%%%%%%%%%%%%%%%%%%%%%%
\subsection{Estimating spin corrections to the tidal Love numbers}\label{sec:estimate}
%%%%%%%%%%%%%%%%%%%%%%%%%
%%%%%%%%
\begin{figure}[t]
\begin{center}
\includegraphics[width=8.2cm]{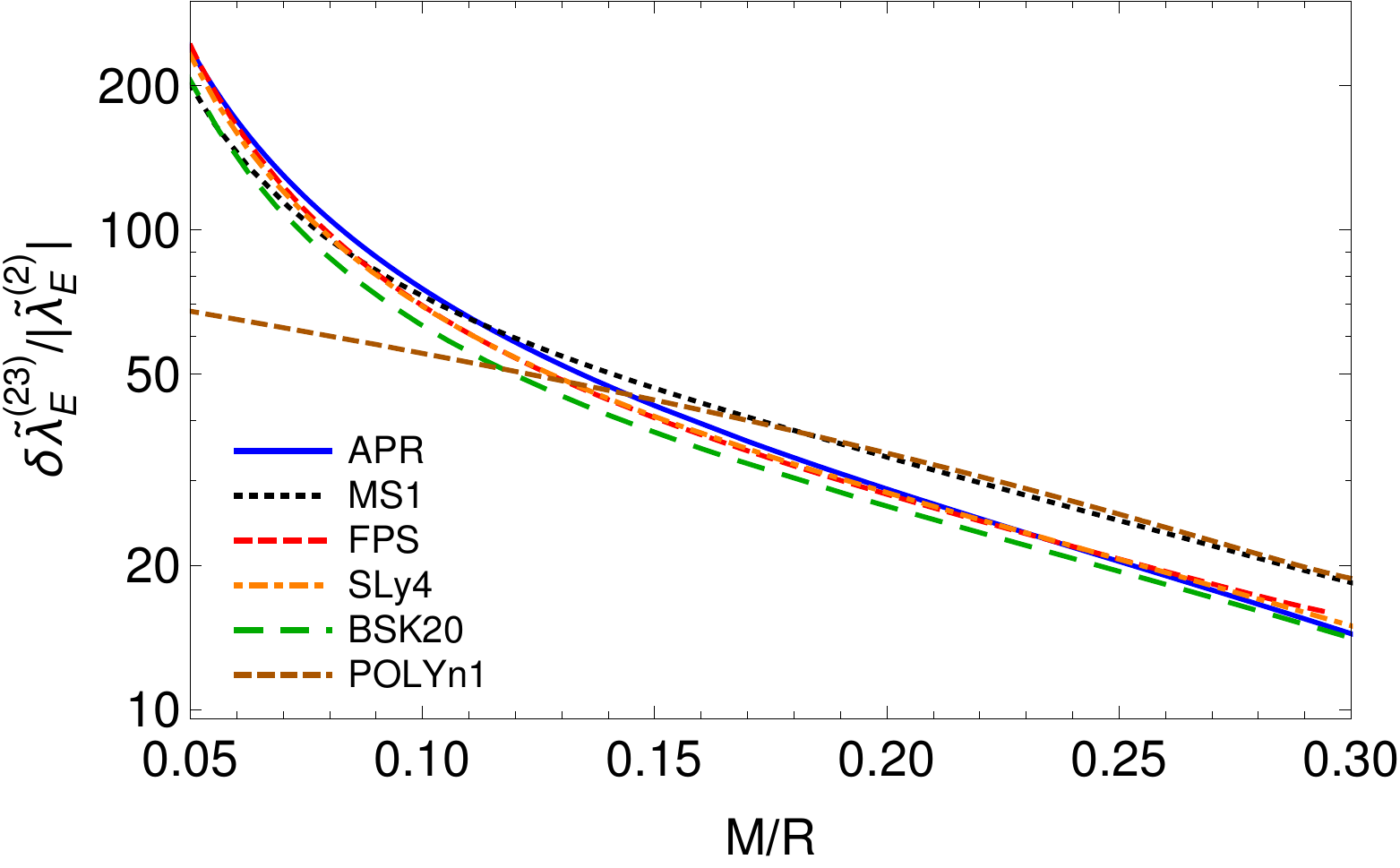}
\caption{(color online). Ratio between the (dimensionless) rotational tidal Love number $\delta\tilde\lambda_E^{(23)}$ and the standard electric quadrupolar tidal Love number $\tilde\lambda_E^{(2)}$ for the various EoSs considered in this work. This ratio is crucial for the estimate~\eqref{dM1}.}
\label{fig:ratio}
\end{center}
\end{figure}
%%%%%%%%%
Although the distribution of NS spin is uncertain, old NSs in the late stages of a binary inspiral are expected to rotate rather slowly. The fastest spinning NS observed so far is the most massive companion of the double pulsar system PSR J0737-3039A~\cite{2003Natur.426..531B}, with a spin period of $\approx 23\,{\rm ms}$, which corresponds to $\chi\sim0.02-0.05$, depending on the EoS~\cite{PhysRevD.86.084017,2013PhRvD..88b1501K}. Such rotation rate is not expected to decrease substantially as this system approaches the merger (see Ref.~\cite{Dietrich:2015pxa} for a recent discussion). Furthermore, the observation of numerous isolated millisecond pulsars suggests that higher spin rates (as high as $\chi\sim0.1$~\cite{Dietrich:2015pxa}) might alos be found in NS-NS binary systems. In light of this uncertainty, our estimates in this paper are based on the assumption $\chi\approx0.05$\footnote{Obviously, because we consider linear perturbations in the spin, the magnitude of the corrections discussed in the rest of the paper would be double in the unlikely case of a spinning NS in a binary system with $\chi\approx0.1$ near the merger.}.

It is therefore relevant to discuss in which regime the ${\cal O}(\chi)$ spin corrections computed in this paper are phenomenologically relevant. To answer this question, let us consider --~as a representative example~-- an equal-mass NS-NS binary with total mass $\sim2M$ at large orbital separation $r_0\gg M$. To the leading order, the $\ell=2$ components of the tidal field scale as (cf. e.g. Ref.~\cite{Taylor:2008xy})
%%%
\begin{eqnarray}
 {\cal E}_{m}^{(2)}&\sim& \frac{M}{r_0^3}\,,\qquad  {\cal B}_{m}^{(2)}\sim \frac{M}{r_0^3}\sqrt{\frac{M}{r_0}}\,,  \label{E2B2}
%  \quad \dot {\cal E}_{m}^{(2)}\sim \frac{M}{r_0^3}\left(\frac{M}{r_0}\right)^{3/2}\,,
\end{eqnarray}
%%%
(modulo a prefactor that depends on $m$) for the electric component and the magnetic component, respectively. Note that the latter component is suppressed by a factor $v=\sqrt{M/r_0}$ relative to the electric component.
On the other hand, the $\ell=3$ components involve a covariant derivative of the tidal tensor~\cite{PoissonWill} and 
are therefore suppressed by a factor $1/r_0$ relative to their $\ell=2$ counterparts, i.e.
%%%
\begin{eqnarray}
 {\cal E}_{m}^{(3)}&\sim& \frac{M}{r_0^4}\,,\qquad {\cal B}_{m}^{(3)}\sim\frac{M}{r_0^4}\sqrt{\frac{M}{r_0}}\,.\label{E3B3}
\end{eqnarray}
%%%
% 

As discussed above, spin corrections introduce new couplings between the multipole moments of a spinning NS and various components of the external tidal field. In Sec.~\ref{sec:moments}, we will show that the deformed quadrupole moment of a spinning NS in an axisymmetric tidal environment to linear order in the spin reads 
%%%
\begin{eqnarray}
\frac{M_2}{M^3}&\sim&\frac{2M^3}{r_0^3}\sqrt{\frac{\pi}{5}}\left[\tilde\lambda_E^{(2)}-9 \sqrt{\frac{10}{7}}\chi\left(\frac{M}{r_0}\right)^{3/2}\delta\tilde\lambda_E^{(23)}\right],\label{M2}
\end{eqnarray}
%%%
%%%
where tilded quantities are made dimensionless and spin independent by dividing the corresponding quantity by suitable powers of $M$ and $\chi$ [as explained below Eqs.~\eqref{dlambdaE23}--\eqref{dlambdaE43}]. 
As we will show, relations similar to Eq.~\eqref{M2} can be obtained also for other (mass and current) multipole moments and involve the novel family of tidal Love numbers defined in Eqs.~\eqref{Lovem}--\eqref{Lovep}. The main purpose of the present work is to compute such spin-induced corrections for a NS described by a realistic EoS to first order in the spin.

To quantify the effect of such spin-induced corrections, we can estimate the ${\cal O}(\chi)$ axisymmetric tidal correction to $M_2$ relative to the static case. Namely, from Eq.~\eqref{M2} we obtain
%%%
\begin{eqnarray}
 \frac{\delta M_2}{ M_2^{(0)}}&\sim&   13\% \left(\frac{\delta\tilde\lambda_E^{(23)}}{30\tilde\lambda_E^{(2)}}\right) \left(\frac{\chi}{0.05}\right)\left(\frac{C}{0.2}\right)^{3/2}\left(\frac{5R}{r_0}\right)^{3/2}\,,\nn\\ \label{dM1}
\end{eqnarray}
%%%
where $M_2^{(0)}$ is the induced quadrupole moment~\eqref{M2} for $\chi=0$, $\delta M_2$ is its ${\cal O}(\chi)$ correction, $C=M/R$ is the stellar compactness, and we have normalized the parameters to $\chi\sim 0.05$, $C\sim0.2$ and $r_0\sim 5R$. The latter represents an extrapolation because the expression above is only valid when the source of the tidal field is at large distance from the central object, $r_0\gg R$, but it should nevertheless capture the correct order of magnitude. 

The estimate~\eqref{dM1} is one of the main results of this paper. It is clear that the relative amplitude depends on
the ratio $\delta\tilde\lambda_E^{(23)}/\tilde\lambda_E^{(2)}$ which we show in Fig.~\ref{fig:ratio} as a function of
the NS compactness, anticipating the main computation presented in the rest of this work. Interestingly, for a typical
NS compactness, $M/R\approx 0.2$, $\delta\tilde\lambda_E^{(23)}\approx 30\tilde\lambda_E^{(2)}$ and, from
Eq.~\eqref{dM1}, we find that, for $\chi\approx0.05$, the ${\cal O}(\chi)$ correction to $M_2$ is about $13\%$ of
its static counterpart\footnote{Our estimates take into account only the axisymmetric components of the tidal field. Since the background is axisymmetric, perturbations with different azimuthal number $m$ are decoupled and nonaxisymmetric perturbations would simply introduce extra terms in Eq.~\eqref{M2} which are proportional to new Love numbers.}. Because the amplitude of higher-order
tidal multipoles increases near the merger, we expect such correction to become even larger. This simple estimate shows
that spin corrections to the tidal Love numbers are non-negligible and they might considerably affect the gravitational
waveforms near the merger.

Finally, another relevant question concerns second-order spin corrections. The relative amplitude of first- and second-order spin corrections to the tidal Love numbers depends on two competitive effects: on the one hand the corrections linear in the spin are larger than the quadratic corrections when $\chi\ll1$, as expected for old NSs in the late stages of a binary inspiral; on the other hand, the magnetic octupolar component of the tidal field is suppressed relative to the quadrupolar electric component, at least when the orbital velocity is small, $v\ll1$. 
It is therefore possible that ${\cal O}(\chi^2)$ spin corrections can also be relevant even in the $\chi\ll1$ regime, at least for intermediate orbital distances.
Clearly, in order to quantify the relative magnitude of second-order spin corrections (as well as of nonaxisymmetric ${\cal O}(\chi)$ corrections), it is necessary to compute other rotational Love numbers (for instance $\tilde\lambda_{E,+}^{(22m)}$). We leave this important task for future work, focusing here on first-order, axisymmetric corrections.

%%%%%%%%%%%%%%%%%%%%%%%%%
\section{Perturbation equations}\label{sec:eqs}
%%%%%%%%%%%%%%%%%%%%%%%%%
The study of linear perturbations of a slowly rotating star has been initiated in Refs.~\cite{ChandraFerrari91,Kojima:1992ie,1993ApJ...414..247K,1993PThPh..90..977K,Ferrari:2007rc} and was recently extended in the context of black-hole perturbation theory~\cite{Pani:2012vp,Pani:2012bp,Pani:2013ija} (see~\cite{Pani:2013pma} for a review). Here we follow extensively the formalism developed in Paper~I, to which we refer for further details. In particular, we make the same working assumptions discussed in Paper~I; namely, we consider a slowly rotating object, whose spin axis coincides with the axis of symmetry of the tidal field, which is assumed to be axisymmetric. The tidal field is weak and varies slowly in time, and in practice we consider only stationary tidal perturbations. As we will show, axisymmetry is in fact required for consistency of the field equations in the stationary case. 
As explained in Paper~I, our approximations are valid for a compact binary at a large orbital distance. Furthermore, NSs
in a binary system are expected to rotate slowly so that the slow-rotation approximation is well justified (see, e.g., \cite{2013PhRvD..88b1501K,Dietrich:2015pxa}). At variance with Paper~I, we limit our
analysis to first order in the spin, but consider both an electric and a magnetic tidal component with generic multipole
$\ell$.

%%%%%%%%%%%%%
\subsection{Spinning background}\label{sec:spback}
%%%%%%%%%%%%%
We consider a slowly rotating object whose line element, to first order in the spin, can be written as~\cite{Hartle:1967he,Hartle:1968si}
\begin{eqnarray}
 ds^2&&=-e^\nu dt^2+\frac{dr^2}{1-2{\cal M}/r}-2\sin^2\vartheta\omega r^2 dtd\varphi+r^2d\Omega^2\,,\nn\\ \label{metric}
\end{eqnarray}
where $d\Omega^2=d\vartheta^2+\sin^2\vartheta d\varphi^2$, and the radial functions $\nu$ and ${\cal M}$ are of zeroth order in rotation, whereas $\omega$ is of first order.
%%%%%%%%%%%%%%
The stress-energy tensor of the perfect fluid is
\begin{equation}
 T^{\mu\nu}=(P+\rho)u^\mu u^\nu+P g^{\mu\nu}\,,\label{Tmunu}
\end{equation}
%%%
where $P=P(r)$ and $\rho=\rho(r)$ are the background pressure and energy density, respectively. To first order in the spin, $u^\mu=e^{-\nu/2}(1,0,0,\Omega)$ where $\Omega$ is the fluid angular velocity and the normalization constant ensures $u^2=-1$.
%%%
Einstein's equations\footnote{In our geometrized units $\kappa=4\pi$ but we will leave it unspecified to keep track of the coupling to matter in the equations.}, $G_{\mu\nu}=2\kappa T_{\mu\nu}$, reduce to the classical Tolman-Oppenheimer-Volkoff equations for the background variables
%%%
\begin{eqnarray}
 \nu '&=& \frac{2 \left({\cal M}+\kappa  r^3 P\right)}{r (r-2 {\cal M})}\,,\qquad {\cal M}'= \kappa  r^2 \rho \,,\\
 P'&=&-\frac{(P+\rho ) \left({\cal M}+\kappa  r^3 P\right)}{r (r-2 {\cal M})}\,,\\
 {\bar \omega}''&=& \frac{\kappa  r (P+\rho) \left(r {\bar \omega}'+4 {\bar \omega}\right)}{r-2 {\cal M}}-\frac{4}{r} {\bar \omega}'\,,
\end{eqnarray}
%%%
where we defined $\bar\omega=\Omega-\omega$. Assuming a barotropic equation of state in the form $P=P(\rho)$, these equations can be integrated numerically with standard methods.
%%%

%%%%%%%%%%%%%
\subsection{Tidal perturbations}
%%%%%%%%%%%%%
Metric and fluid perturbations are decomposed in spherical harmonics according to their parity as discussed in
Appendix~\ref{app:Kojima}. Since we are considering stationary tidal fields, we assume no explicit time dependence in
the perturbations and no fluid motion\footnote{Recently, it has been noted that an irrotational (rather than a static)
  fluid provides a more realistic configuration for a stationary tidally distorted
  object in a binary system~\cite{Landry:2015cva}. Such condition only affects the odd-parity sector of the
  perturbations, but has dramatic consequences for the magnetic tidal Love numbers. Here we consider only static fluids,
  this assumption is consistent with the field equations to first order in the spin. An extension of the work of
  Ref.~\cite{Landry:2015cva} to account for spin corrections is left for the future.}, setting
$U^{(\ell)}=V^{(\ell)}=R^{(\ell)}=0$ in the fluid perturbations $\delta u^\mu$, cf. Eqs.~\eqref{ut}--\eqref{uphi}.

By using this decomposition, solving the gravitational equations perturbatively, and making use of the orthogonality properties of the spherical harmonics~\cite{Kojima:1992ie,Pani:2013pma}, we obtain a system of ordinary differential equations for the radial variables where various spin couplings between different multipolar indices $\ell$ and between perturbations with opposite parity are present.  

The equations separate into two groups according to their parity. In Paper~I, we defined the so-called polar-led and
axial-led systems, which describe the metric deformations induced by a purely electric and a purely magnetic tidal
field, respectively.  It can be shown that consistency of the field equations requires stationary perturbations to also be axisymmetric. In other words, the absence of explicit time dependence in the perturbations given in
Appendix~\ref{app:Kojima} also imposes $m=0$, where $m$ is the azimuthal number of the perturbations, which is conserved
by virtue of the axisymmetry of the background.  On the contrary, tidal perturbations with $m\neq0$ must have a weak
time dependence (cf. e.g. Ref.~\cite{Poisson:2014gka} and Paper I).  Therefore, to first order in the spin and in the
stationary case, the polar-led and axial-led systems given in Paper~I schematically reduce to

\begin{eqnarray}
 \left\{ \begin{array}{l}
{\cal P}_{{L}}=0\\
{\cal A}_{{L}+1}+\epsilon_a {\cal Q}_{{{L}+1}}\tilde{\cal P}_{{L}}=0\\
{\cal A}_{{L}-1}+\epsilon_a {\cal Q}_{{{L}}}\tilde{\cal P}_{{L}}=0  \end{array}\right.\,,\label{polar_led} \\
%%%%%%%%%
\left\{ \begin{array}{l}
          {\cal A}_{{L}}=0\\
{\cal P}_{{L}+1}+\epsilon_a {\cal Q}_{{{L}+1}}\tilde{\cal A}_{{L}}=0\\
{\cal P}_{{L}-1}+\epsilon_a {\cal Q}_{{{L}}}\tilde{\cal A}_{{L}}=0
         \end{array}\right.\label{axial_led}\,, 
\end{eqnarray}
%%%
where $\epsilon_a$ is a bookkeeping parameter for the order of the spin perturbations, ${\cal A}_{\ell}$, $\tilde {\cal A}_{\ell}$ (respectively, ${\cal P}_{\ell}$, $\tilde {\cal P}_{\ell}$) are linear combinations of the axial (respectively, polar) perturbations with multipolar index $\ell$ and ${\cal Q}_\ell=\ell/\sqrt{4\ell^2-1}$ are related to the Clebsch-Gordan coefficients in quantum mechanics\footnote{In the context of quantum mechanics, the spin selection rules are enforced by the addition rules for angular momenta~\cite{Pani:2013pma}.}.
%%%
The first set of equations in Eqs.~\eqref{polar_led} and \eqref{axial_led} is of zeroth order in the spin and corresponds to the well-studied static case~\cite{Hinderer:2007mb,Binnington:2009bb,Damour:2009vw,Damour:2009va}, whereas the remaining two sets of equations in Eqs.~\eqref{polar_led} and \eqref{axial_led} are spin-induced corrections.
As clear from the structure of the above equations, polar perturbations with multipolar index $\ell$ source axial perturbations with multipolar index $\ell+1$ and $\ell-1$ to first order in the spin and vice versa, cf. Paper~I for details.

Remarkably, it is possible to reduce the systems above to a simple set of differential equations for a generic multipole $\ell$. To simplify the notation, we will expand any variable to first order in the spin, e.g. $H_2^{(\ell)}$ and $\delta H_2^{(\ell)}$ will denote a function of zeroth order in the spin and its first-order correction, respectively.
%%%%%%%%%%%%%%%%%%%%%%%%%%%%%%%
\subsection{Electric-led tidal perturbations}
%%%%%%%%%%%%%%%%%%%%%%%%%%%%%%%
For any $\ell\geq2$, the polar-led system~\eqref{polar_led} reduces to the following coupled ordinary differential equations (ODEs)
%%%

%%%%%
\begin{equation}
  \left\{ \begin{array}{l}
 {\cal D}_{(\ell)}\left[{H_0^{(\ell)}}\right]=0\\
 {\cal D}^\ast_{(\ell+1)} \left[\delta h_0^{(\ell+1)}\right]={S^\ast_+}^{(\ell)}\\ 
 {\cal D}^\ast_{(\ell-1)} \left[\delta h_0^{(\ell-1)}\right]={S^\ast_-}^{(\ell)}\end{array}\right.\,,\label{polar_led2}
\end{equation}
%%%
where we defined the differential operators\footnote{Henceforth, we will use an asterisk to denote quantities related to magnetic perturbations.}
%%%
\begin{eqnarray}
 {\cal D}_{(\ell)}&\equiv& \frac{d^2}{dr^2}+C_1^{(\ell)} \frac{d}{dr}+C_0^{(\ell)}\,, \label{D}\\ 
 {\cal D}^\ast_{(\ell)}&\equiv& \frac{d^2}{dr^2}+{C_1^\ast}^{(\ell)} \frac{d}{dr}+{C_0^\ast}^{(\ell)}\,. \label{Ds}
\end{eqnarray}
%%%
The coefficients $C_i^{(\ell)}$ and ${C_i^\ast}^{(\ell)}$ ($i=1,2$) and the source terms ${S^\ast_\pm}^{(\ell)}$ are given in Appendix~\ref{app:sources}. As expected, the source terms depend on ${H_0^{(\ell)}}$, i.e. polar perturbations with multipolar index $\ell$ source axial perturbations with $\ell\pm1$. The remaining variables are algebraically related to $H_0^{(\ell)}$ and $\delta h_0^{(\ell\pm1)}$ by relations given in Appendix~\ref{app:sources}. In particular, $H_1^{(\ell)}=0={\delta h_1^{(\ell\pm1)}}$ so that the perturbed metric remains invariant under $t\to-t$ and $\varphi\to-\varphi$. The perturbations also depend on the speed of sound in the fluid, ${c_s}=\sqrt{\partial P/\partial\rho}$. 

Note that, in principle, the equations governing axial perturbations with $\ell=1$ are different from those with $\ell\geq2$. When $\ell=1$, one can use a residual gauge freedom to set $h_1^{(1)}=0=\delta h_1^{(1)}$~\cite{1970ApJ...159..847C,1989ApJ...345..925L}. In the stationary case, it turns out that $\delta h_1^{(\ell\pm1)}=0$ for any $\ell$ so that the equations above also include the special case $\ell=1$. As we discuss below, this property does not hold in the magnetic-led case.

%%%%%%%%%%%%%%%%%%%%%%%%%%%%%%%
\subsection{Magnetic-led tidal perturbations}
%%%%%%%%%%%%%%%%%%%%%%%%%%%%%%%
For any $\ell>2$, the axial-led system~\eqref{axial_led} reduces to the following coupled ODEs
\begin{equation}
\left\{ \begin{array}{l}
   {\cal D}^\ast_{(\ell)} \left[h_0^{(\ell)}\right]=0\\
 {\cal D}_{(\ell+1)}\left[{\delta H_0^{(\ell+1)}}\right]={S}^{(\ell)}_+\\ 
 {\cal D}_{(\ell-1)}\left[{\delta H_0^{(\ell-1)}}\right]={S}^{(\ell)}_-\end{array}\right.\,,\label{axial_led2}
\end{equation}
%%%
where again the source terms ${S}^{(\ell)}_\pm$ are given in Appendix~\ref{app:sources} and depend on $h_0^{(\ell)}$ and its derivatives. In this case, axial perturbations with multipolar index $\ell$ source polar perturbations with $\ell\pm1$.

%%%%%%%%%%%%%%%%%%%%%%%%%%%%%%%
% \subsection{Magnetic-led system for $\ell=2$}
%%%%%%%%%%%%%%%%%%%%%%%%%%%%%%%
Polar equations with $\ell=1$ are governed by a different set of equations and must therefore be discussed separately. When $\ell=1$, one can set $H_1^{(1)}=0=\delta H_1^{(1)}$ in the metric ansatz without loss of generality. For this reason, the differential operator ${\cal D}_{(1)}$ and the source term ${S}^{(1)}$ are different from those presented above and their explicit form is given in Appendix~\ref{app:sources}.

%%%%%%%%%%%%%%%%%%%%%%%%%
\subsection{Exterior solution}\label{sec:exterior}
%%%%%%%%%%%%%%%%%%%%%%%%%
Although the set of equations just presented is valid for any $\ell\ge2$, finding the vacuum solution without specifying the value of $\ell$ is rather challenging. Indeed, to zeroth order in the spin the polar and axial solutions of the first equation in~\eqref{polar_led2} and \eqref{axial_led2}, respectively, can be written in terms of special functions~\cite{Hinderer:2007mb,Binnington:2009bb,Damour:2009vw}
%%%
\begin{eqnarray}
 {H_0^{(\ell)}}&=& A\, P_\ell^2(r/M-1)+B  \, Q_\ell^2(r/M-1)\,,\\
 {h_0^{(\ell)}}&=& C\, \frac{r^2}{M^2}\, {}_2 F_1[1-\ell,\ell+1,4,r/(2M)]\nn\\
 &&+D \, G_{2,2}^{2,0}\left(\frac{r}{2M}\left|
\begin{array}{c}
 1-\ell,\ell+2 \\
 -1,2 \\
\end{array}\right.
\right) \,.
\end{eqnarray}
%%%
where $P_\ell^2$ and $Q_\ell^2$ are associated Legendre functions, ${}_2 F_1$ is the hypergeometric function, $G_{2,2}$ is the Meijer function, and $A$, $B$, $C$, $D$ are integration constants.
When plugged into the source terms, the corresponding first-order equations do not appear to have an analytical solution for generic $\ell$. However, the special functions above simplify enormously when $\ell$ is an integer (i.e. in the case of interest) and therefore the exterior perturbation equations can be solved \emph{analytically} for any specific integer value, $\ell=2,3,4,...$ 

The explicit vacuum solution to the $\ell=2$ electric-led sector reads (cf. Paper I)
\begin{widetext}
%%%
\begin{eqnarray}
 H_0^{(2)}&=& \alpha_2 y(y-2)- \gamma_2\left\{-3+\frac{1}{2-y}-\frac{1}{y}+3 y+\frac{3}{2} (y-2) y
   \log\left[1-2/y\right]\right\}  \,,\\
%%%%%%%%%%%%%%%%%%%%%  
 \delta h_0^{(3)}&=& -\frac{M \chi}{6720 y^2}  \left[-128 \sqrt{35} y (5y-4) \alpha _2+3 \left(8 \sqrt{35}
   \left(-16-44 y+90 y^2+270 y^3-945 y^4+405 y^5\right.\right.\right.\nn\\
   &&\left.\left.\left.+y \left(64-80 y+1080 y^3-1350 y^4+405
   y^5\right) \tanh^{-1}\left[\frac{1}{1-y}\right]\right) \gamma _2\right.\right..\nn\\
   &&\left.\left.
   +35 y \left(8+20
   y+60 y^2-210 y^3+90 y^4+15 y^3 \left(8-10 y+3 y^2\right)
   \log\left[1-2/y\right]\right) \gamma _{32}\right)\right] \,, \\
%%%%%%%%%%%%%%%%%%%%%  
 \delta h_0^{(1)}&=& -\frac{M \chi}{30 y^2}  \left[-20 \sqrt{15} y^3 \alpha _2+\sqrt{15} \left(4-30 y^2+30 y^3+3 y
   \left(4-10 y^2+5 y^3\right) \log\left[1-2/y\right]\right) \gamma
   _2-30 y \gamma _{12}\right]\,,
\end{eqnarray}
%%%
\end{widetext}
where $y=r/M$ and $\alpha_2$, $\gamma_2$, $\gamma_{23}$, $\gamma_{21}$ are integration constants\footnote{Our notation is the following. The constants $\alpha_i$ and $\gamma_i$ are proportional to ${\cal O}(\chi^0)$ perturbations, whereas $\gamma_{ij}$ are associated with ${\cal O}(\chi)$ perturbations with $\ell=i$ which are sourced by $\ell=j$ tidal perturbations with opposite parity. As before, we denote quantities related to the magnetic sector with an asterisk.}. The remaining nonvanishing perturbations can be obtained from the solution above using the algebraic relations presented in Appendix~\ref{app:sources}. 

Note that, in the solution above, we have already fixed the constants $\alpha_{23}$ and $\alpha_{21}$ to avoid spurious components of the tidal field which arise from the homogeneous system associated with Eq.~\eqref{polar_led2} (cf. Paper~I for details). With this choice, the external tidal field has a purely electric, quadrupolar component.

Likewise, the vacuum solution to the $\ell=2$ magnetic-led sector reads
%%%%
\begin{widetext}
\begin{eqnarray}
%%%%%%%%%%%%%%%%%%%%%  
  h_0^{(2)}&=& -\frac{M}{24 y} \left[-24 (y-2) y^3 \alpha _2^\ast+\left(-4-4 y-6 y^2+6 y^3+3 (y-2) y^3
   \log\left[1-2/y\right]\right) \gamma _2^\ast\right]\,,\\
 %%%%%%%%%%%%%%%%%%%%%  
 \delta H_0^{(3)}&=& \frac{\chi}{140 (y-2) y^4}  \left[-72 \sqrt{35} (y-2) y^4 \alpha _2^\ast+\sqrt{35} \left(2
   \left(4+2 y-6 y^2-23 y^3-61 y^4+455 y^5-420 y^6+105 y^7\right)\right.\right.\nn\\
   &&\left.\left.+3 y^4 \left(-6-137
   y+280 y^2-175 y^3+35 y^4\right) \log\left[1-2/y\right]\right) \gamma
   _2^\ast+70 y^3 \left(2 \left(-2-10 y+65 y^2-60 y^3+15 y^4\right)\right.\right.\nn\\
   &&\left.\left.+15 (y-2)^2
   (y-1) y^2 \log\left[1-2/y\right]\right) \gamma _{32}^\ast\right] \,,\\
 %%%%%%%%%%%%%%%%%%%%%  
 \delta H_0^{(1)}&=& -\frac{\chi}{120 (y-2)^2 y^3}  \left[-24 \sqrt{15} (y-2)^2 y^3 (10y-9) \alpha _2^\ast+\sqrt{15}
   \left(16-4 y-32 y^2-234 y^4+60 y^5\right.\right.\nn\\
   &&\left.\left.+3 (y-2)^2 y^3 (10y-9)\log\left[1-2/y\right]\right) \gamma _2^\ast-120 y^3 \gamma
   _{12}^\ast\right] \,,
\end{eqnarray}
\end{widetext}
%%%$
where again $\alpha_2^\ast$, $\gamma_2^\ast$, $\gamma_{23}^\ast$, $\gamma_{21}^\ast$ are integration constants. As in the electric-led case, the other constants $\alpha_{23}^\ast$ and $\alpha_{21}^\ast$ have been fixed by requiring that the tidal field has a purely magnetic, quadrupolar component.

The equations above represent the generic exterior solution of a slowly spinning object immersed in an axisymmetric, quadrupolar tidal field, whose electric and magnetic components are proportional to $\alpha_2$ and $\alpha_2^\ast$, respectively. This solution was computed in Ref.~\cite{Landry:2015zfa} using a different notation and in the generic, nonaxisymmetric case.
The explicit vacuum solutions for the cases with $\ell=3$ and $\ell=4$ are instead new and given in Appendix~\ref{app:exterior}.

To identify the electric and magnetic components of the tidal field we can expand the exterior solution at large distances. In the axisymmetric case, we can extract ${\cal E}_0^{(\ell)}$ and ${\cal B}_0^{(\ell)}$ from the asymptotic behavior of the $g_{tt}$ and $g_{t\varphi}$ components, namely~\cite{Binnington:2009bb}

%%%%
\begin{eqnarray}
 g_{tt}&\to& -\frac{2}{\ell(\ell-1)}{\cal E}_0^{(\ell)} Y^{\ell0}(\vartheta) r^\ell+\dots\,, \label{E0l}\\
 g_{t\varphi}&\to& \frac{2}{3\ell(\ell-1)}{\cal B}_0^{(\ell)} S_{\varphi}^\ell(\vartheta) r^{\ell+1}+\dots\,, \label{B0l}
\end{eqnarray}
%%%%
where $S_\varphi^\ell$ is one of the vector spherical harmonics defined in Eq.~\eqref{vecspheharm}. The normalization of the dimensionless constants $\alpha_\ell$ and $\alpha_\ell^\ast$ in the analytical solution just presented has been chosen such that
%%%
\begin{equation}
 {\cal E}_0^{(\ell)}\equiv -\frac{\ell(\ell-1)}{2}\frac{\alpha_\ell}{M^\ell} \,, \qquad  {\cal B}_0^{(\ell)}\equiv \frac{3\ell(\ell-1)}{2}\frac{\alpha_\ell^\ast}{M^\ell}  \,.\label{EB}
\end{equation}

%%%%%%%%%%%%%%%%%%%%%%%%%
\section{Rotational Love numbers of a slowly rotating NS}\label{sec:LoveNS}
%%%%%%%%%%%%%%%%%%%%%%%%%
With the explicit exterior solution at hand, we can now compute the spin-induced (rotational) Love numbers.  As explained in detail in Paper~I, the constants $\alpha_i$ and $\alpha_i^\ast$ are associated with the $\ell=i$ electric and magnetic components of the external tidal field [cf. Eq.~\eqref{EB}], whereas the constants $\gamma_i$ and $\gamma_{ij}$ (respectively, $\gamma_i^\ast$ and $\gamma_{ij}^\ast)$ are associated with the deformation of the mass (respectively, current) $\ell=i$ multipole moments\footnote{The separability between external tidal field and object's response suffers from some subtle ambiguity that we discussed in detail in Paper~I. Here we will follow the prescription of that paper, which seems robust enough at least to first order in the spin. In practice, because the Love numbers scale with high powers of the inverse compactness, $\sim(R/M)^q$~\cite{Hinderer:2007mb,Binnington:2009bb,Damour:2009vw}, where the power $q$ depends on $\ell$ and on the parity (for example $\tilde\lambda_E^{(2)}\sim (R/M)^5$, see Fig.~\ref{fig:lambda_nonrot} below) such ambiguity can at most affect the Love numbers by a small offset, $\sim (R_{\rm BH}/M_{\rm BH})^q\ll (R_{\rm NS}/M_{\rm NS})^q$, with little observational consequences.}.

To show this fact explicitly, we adopt the procedure explained in Paper~I to compute the Geroch-Hansen multipole moments~\cite{Geroch:1970cd,Hansen:1974zz} of the tidally distorted object. In brief, the procedure is based on considering the asymptotically flat solution of the metric perturbations, setting $\alpha_i=\alpha_i^\ast=0$ in the equations above. Using this solution, the multipole moments can be computed in a gauge-invariant way by evaluating the binding energy of a point particle in a circular geodesic of an axisymmetric, asymptotically-flat spacetime~\cite{Ryan:1995wh,Ryan:1997hg,Pappas:2012nt}. The multipole moments can then be identified by comparing the binding energy order by order in a (gauge-invariant) low-velocity expansion.

%%%%%%%%%%%%%%%%%%%%%%%%%
\subsection{Multipole moments of a tidally distorted NS} \label{sec:moments}
%%%%%%%%%%%%%%%%%%%%%%%%%

Further details on this procedure are given in Paper~I, here we simply present the final result for the first multipole moments\footnote{The tidal perturbations also source a deformation of $S_1$, which can be reabsorbed in the definition of the NS angular momentum.}:
\begin{eqnarray}
 \frac{M_2}{M^3}&=& -\frac{2 \gamma _2}{\sqrt{5 \pi }}+\chi  \left(\frac{75 \gamma _3^\ast}{16 \sqrt{7 \pi }}-\frac{2 \gamma _{23}^\ast}{\sqrt{5
   \pi }}\right)\,, \label{M2b} \\
 \frac{S_3}{M^4} &=& -\frac{3 \gamma _3^\ast}{4 \sqrt{7 \pi }}\nn\\
 &&-\frac{6\chi}{\sqrt{\pi }}  \left[\frac{\sqrt{5}}{3} \gamma _2+ \gamma _4+\frac{\sqrt{7}}{44} \left(\gamma _{32}+\gamma _{34}\right)\right] \,, \label{S3b}\\
 \frac{M_4}{M^5} &=& -\frac{4 \gamma _4}{7 \sqrt{\pi }}+\frac{\chi  \left(17 \sqrt{7} \gamma _3^\ast-672 \gamma _{43}^\ast\right)}{1176 \sqrt{\pi }}\,. \label{M4b}
\end{eqnarray}

With the multipole moments at hand, the Love numbers in Eqs.~\eqref{Love0} and \eqref{Lovem} can easily be computed. By using Eq.~\eqref{EB}, to zeroth order in the spin we obtain
%%%
%%%
\begin{eqnarray}
 \lambda_E^{(2)} &=&\frac{2 \gamma_2 M^5 }{\sqrt{5 \pi } \alpha_2},\quad \lambda_M^{(3)} = -\frac{\gamma _3^\ast M^7}{12 \sqrt{7 \pi }\alpha _3^\ast},\quad  \lambda_E^{(4)} = \frac{2 \gamma _4 M^9}{21 \sqrt{\pi } \alpha _4}, \nn \\ \label{Love0b}
\end{eqnarray}
%%%
%%%
whereas the new ${\cal O}(\chi)$ rotational Love numbers read
%%%
\begin{eqnarray}
 \delta\lambda_E^{(23)} &=& \chi M^6\frac{ \left(75 \sqrt{35} \gamma_3^\ast-224 \gamma_{23}^\ast\right)}{1008\sqrt{5 \pi } \alpha_3^\ast}\,,  \label{dlambdaE23}\\
 \delta\lambda_M^{(32)} &=& \chi M^6 \frac{\left(44 \sqrt{35} \gamma _2+21 \gamma _{32}\right)  }{28 \sqrt{7 \pi } \alpha _2}\,, \\
 \delta\lambda_M^{(34)} &=& \chi M^8 \frac{3 \left(44 \gamma _4+\sqrt{7} \gamma _{34}\right)  }{168 \sqrt{\pi } \alpha _4}\,, \\
 \delta\lambda_E^{(43)} &=&  \chi M^8 \frac{ \left(17 \sqrt{7} \gamma _3^\ast-672 \gamma _{43}^\ast\right)}{10584 \sqrt{\pi }\alpha _3^\ast}\,. \label{dlambdaE43}
\end{eqnarray}
%%%
The dimensionless quantities previously introduced are defined by dividing the expressions above by suitable powers of $\chi$ and $M^n$ (where, e.g., $n=6$ for $\delta\lambda_E^{(23)}$ and $\delta\lambda_M^{(32)}$, whereas $n=8$ for $\delta\lambda_M^{(34)}$ and $\delta\lambda_E^{(43)}$) in order to make them pure numbers. For example, $\delta\tilde\lambda_E^{(23)}\equiv \delta\lambda_E^{(23)}/(M^6\chi)$. 

In terms of these dimensionless quantities, the axisymmetric deformation of the mass quadrupole moment to linear order in the spin can be expressed as 
%%%%
\begin{equation}
 \frac{M_2}{M^3}=\tilde\lambda_E^{(2)} {\cal E}_0^{(2)}M^2+\chi \delta\tilde\lambda_E^{(23)} {\cal B}_0^{(3)}M^3\,. \label{M2final}
\end{equation}
%%%%
We stress that the above expression contains only the axisymmetric components of the tidal perturbations; the full expression to first order in the spin would also contain extra terms associated with $m\neq0$ perturbations. Clearly, to estimate Eq.~\eqref{M2final}, we need to evaluate the axisymmetric components of the quadrupolar electric tidal field and of the octupolar magnetic tidal field. For concreteness, let us consider an equal-mass NS-NS binary with total mass $\sim2M$ in a quasicircular motion at orbital distance $r_0$. The electric tidal tensor ${\cal E}_{ab}$ and the magnetic tidal tensor ${\cal B}_{abc}$ in such configuration were computed in Ref.~\cite{2009PhRvD..80l4039J}. By comparing the large-distance behavior of the perturbed black-hole metric derived in Ref.~\cite{2009PhRvD..80l4039J} (cf. Eqs.~(3.2)-(3.3) in that paper) to our leading-order expressions~\eqref{E0l}--\eqref{B0l}, we obtain, to Newtonian order\footnote{Note that, to leading Newtonian order, the perturbed black-hole metric computed in Ref.~\cite{2009PhRvD..80l4039J} coincides with that of a perturbed NS at large distance. For this reason the results of Ref.~\cite{2009PhRvD..80l4039J} are directly applicable to our case.},
%%%%
\begin{equation}
 {\cal E}_0^{(2)}=\frac{2M}{r_0^3}\sqrt{\frac{\pi}{5}} \,, \qquad {\cal B}_0^{(3)}= -\frac{18M}{r_0^4}\sqrt{\frac{2\pi}{7}}\left(\frac{M}{r_0}\right)^{1/2}\,, \label{E02B03}
\end{equation}
%%%%
which complement Eqs.~\eqref{E2B2} and~\eqref{E3B3} by including the correct prefactors for a circular equal-mass binary\footnote{We note that our expression for ${\cal E}_0^{(2)}$ coincides with that derived, e.g., in Ref.~\cite{Binnington:2009bb} after taking into account their different normalization of the spherical harmonics relative to ours (we use the standard definition of \cite{jackson1962classical}).}. Finally, by inserting Eq.~\eqref{E02B03} into Eq.~\eqref{M2final}, we obtain the expression anticipated in Eq.~\eqref{M2}.
As previously discussed, by evaluating Eq.~\eqref{M2} and using the numerical results presented in Sec.~\ref{sec:results} below, we estimate that the axisymmetric ${\cal O}(\chi)$ spin correction to the quadrupole moment $M_2$ for $\chi\approx0.05$ is about $13\%$ when $r_0\sim 5 R$, cf. Eq.~\eqref{dM1}.

So far we have focused on perturbations which do not break the reflection symmetry with respect to the equatorial plane at $\vartheta=\pi/2$. However, an axisymmetric electric (respectively, magnetic) tidal field with odd (respectively, even) values of $\ell$ generates perturbations which break the equatorial symmetry of the background~\eqref{metric}. Such components of the tidal field would induce multipole moments such as $M_3$, $S_2$, etc... Our formalism can directly accommodate these new Love numbers, although they are only relevant  for tidal sources that break the equatorial symmetry. 
%%%

A binary system of two nonspinning compact objects or of two spinning compact objects with (anti)aligned spins enjoys an equatorial symmetry relative to the orbital plane. However, in more general configurations --~e.g. if the angular momenta of the two objects are not (anti)aligned~-- the equatorial symmetry can be broken. This misalignment might happen after the NS formation due to supernova kicks, although subsequent tidal dissipation can contribute to realign the spins (cf. e.g. Ref.~\cite{Gerosa:2013laa} and references therein). 
For completeness, we discuss this problem in Appendix~\ref{app:noequatorial} where we also present some results for the rotational Love numbers with equatorial-symmetry breaking.

%%%%%%%%%%%%%%%%%%%%%%%%%
\subsection{Matching and numerical procedure}
%%%%%%%%%%%%%%%%%%%%%%%%%
The above Love numbers can be directly computed once the constants $\gamma$'s are known. The values of these constants are fixed by a continuous matching of the numerical interior solution describing the deformed compact star with the analytical exterior solution.
Since we include only linear corrections in the tidal field,
the ratios of the $\gamma$'s to the corresponding $\alpha$'s are pure numbers independent of the tidal field itself.

For example, to zeroth order in spin, for the correction to the quadrupole mass moment one obtains~\cite{Hinderer:2007mb}
% \begin{widetext}
 %%%
\begin{eqnarray}
 \tilde\lambda_E^{(2)}  &=& \frac{-4}{\sqrt{5 \pi }} (1-2 C)^2 \left[(2 C-1) \beta_{2 P}-2 C+2\right]\nn\\
 &\times&\left[3 (1-2 C)^2 \left((2 C-1) \beta _{2 P}-2 C+2\right)\log(1-2C) \right.\nn\\
 &&\left. +2 C \left( \left(4 C^4+6 C^3-22 C^2+15 C-3\right) \beta _{2 P} \right.\right.\nn\\
 &&\left.\left.+4 C^4-4 C^3+26 C^2-24 C+6\right)\right]\,, 
\end{eqnarray}
%%%
% \end{widetext}
where $C=M/R$, $\beta_P^{(2)}=R{H_0^{(2)}}'/H_0^{(2)}$ evaluated at the radius and we recall that tilded quantities are made dimensionless by dividing the corresponding quantity by suitable powers of $M$, e.g. $\tilde\lambda_E^{(2)}\equiv \lambda_E^{(2)}/M^5$. In the small compactness limit, one recovers the Newtonian result, $\tilde\lambda_E^{(2)}\sim C^{-5}$~\cite{Hinderer:2007mb}.
To compute the Love numbers, it is therefore sufficient to evaluate the metric perturbations at the radius of the object after a numerical integration from the interior. The same procedure can be straightforwardly applied to any of the tidal Love numbers defined in Eqs.~\eqref{Love0}.

The case of the rotational Love numbers defined in Eq.~\eqref{Lovem} is similar in spirit but slightly more involved. In this case the corresponding perturbation equations are inhomogeneous ODEs. To match the interior solution with the one in the exterior, we adopt the following procedure. First, we construct a particular solution of the inhomogeneous equations which is regular at the origin. Then, we also construct the solution to the associate homogeneous system which is regular at the origin. 
The amplitude of this solution is a free parameter that we adjust to match the \emph{full} interior solution (i.e. a linear combination of the inhomogeneous and of the homogeneous solutions) to the exterior. It is easy to show that imposing continuity at $r=R$ fixes uniquely the constants $\gamma$'s and  also the amplitude of the homogeneous solutions for different values of $\ell$. In order to increase accuracy, we compute higher-order series expansions near the center of the star and use such analytical solutions to start the numerical integration outwards.

As a representative example, we present the schematic form of the ${\cal O}(\chi)$ correction to the mass quadrupole moment:
%%%
\begin{equation}
 \delta \tilde\lambda_E^{(23)} = %\sqrt{\frac{7}{\pi}}
 \chi\frac{A_1+A_2\log(1-2c)}{A_3+A_4\log(1-2c)+A_5\log(1-2c)^2}\,, \label{L23R}
\end{equation}
%%%
where $A_i$ are cumbersome polynomial terms that depend on $C$, $\beta_A^{(3)}=R {h_0^{(3)}}'/h_0^{(3)}$, $\beta_{Ph}^{(2)}=R{\delta H_{0h}^{(2)}}'/\delta H_{0h}^{(2)}$, with $\delta H_{0h}^{(2)}$ being the homogeneous solution associated with the third equation of the system~\eqref{axial_led2}, and on $s=R {\delta H_{0}^{(2)}}/(\chi h_0^{(3)})$. This last term represents the relative amplitude of the polar perturbation ${\delta H_{0}^{(2)}}$ relative to its axial source $h_0^{(3)}$ to first order in the spin. All these perturbations are evaluated at the radius of the star. Note that the value of $s$ might also depend on the compactness so that, to explore the Newtonian regime, one cannot just take the $C\to0$ limit of Eq.~\eqref{L23R}. In Sec.~\ref{sec:results}, we will give numerical evidence that $\tilde\lambda_E^{(23)}\sim 1/C^5$ in this limit, so that the ratio $\tilde\lambda_E^{(23)}/\tilde\lambda_E^{(2)}$ is finite as $C\to0$.
The expressions for the other rotational Love numbers defined in Eq.~\eqref{Lovem} are qualitatively similar to
Eq.~\eqref{L23R} and we avoid presenting them here. We show the full explicit expression for all rotational Love numbers in a {\scshape Mathematica}\textsuperscript{\textregistered} notebook presented as Supplemental Material to this paper.

%%%%%%%%%%%%%%%%%%%%%%%%%
\subsection{Deformed three-hair relation for tidally distorted NSs}
%%%%%%%%%%%%%%%%%%%%%%%%%
As anticipated in Paper~I, the corrections to the multipole moments of a tidally distorted spinning NS will modify the approximate three-hair relations that exist for isolated NSs~\cite{Pappas:2013naa,Stein:2014wpa,Yagi:2014bxa,Pappas:2015mba}. For such stars these relations read~\cite{Stein:2014wpa}
%%%
\begin{equation}
 M_\ell+i\frac{q}{a} S_\ell = \bar B_{n,\left[\frac{\ell-1}{2}\right]}M(iq)^\ell\,, \label{3hair}
\end{equation}
%%%
where $a=S_1/M$, $iq=\sqrt{M_2/M}$, $\bar B_{n\ell}$ is a coefficient that depends on the EoS only mildly, and $[x]$ denotes the largest integer not exceeding $x$. Because of the mild dependence on the EoS, all moments with $\ell>2$ can be approximately computed from $M_0\equiv M$, $S_1=J$ and $M_2$ through relation~\eqref{3hair}~\cite{Pappas:2013naa,Stein:2014wpa,Yagi:2014bxa,Pappas:2015mba}.

As a result of an external tidal field, the relation~\eqref{3hair} is deformed in two ways.
First of all, we note that such relation is valid for axisymmetric and reflection symmetric solutions. An electric (respectively, magnetic) tidal field with odd (respectively, even) values of $\ell$ would break the equatorial symmetry, introducing moments such as $S_2$, $M_3$, $S_4$, $M_5$, etc, which are not included in Eq.~\eqref{3hair}. 
As we discuss in Appendix~\ref{app:noequatorial}, a quadrupolar magnetic tidal field would induce a nonvanishing mass moment $M_3$ at linear order in the spin, whereas an octupolar electric tidal field would induce a current moment $S_2$.

Furthermore, also the standard moments associated with the equatorial symmetry would acquire tidal contributions proportional to the spin. For example, the mass quadrupole moment $M_2$ would acquire ${\cal O}(\chi)$ corrections proportional to the $\ell=3$ magnetic component of the external tidal field (and would also acquire ${\cal O}(\chi^2)$ corrections proportional to the $\ell=4$ electric component of the external tidal field, cf. Paper~I). 

An interesting question is whether such deformation are still ``universal'', in the sense that the relation among different multipole moments is still only mildly dependent on the EoS. This problem is discussed in Sec.~\ref{sec:universality} below.

%%%%%%%%%%%%%%%%%%%%%%%%%
\section{Numerical Results}\label{sec:results}
%%%%%%%%%%%%%%%%%%%%%%%%%
In this section, we solve the perturbation equations explicitly for a self-gravitating perfect-fluid configuration. We use some realistic tabulated EoS listed in Table~\ref{tab:EoS} which cover a wide range of NS deformability. We also consider a polytropic equation of state, $P(\rho)\sim \rho^{1+1/n}$ with index $n=1$, which will be denoted by ``POLYn1'' in the following. 

\begin{table}
 \begin{tabular}{cc}
 EoS & Reference\\
 \hline\hline
 APR	&  \cite{Alford:2004pf} \\
 MS1	&  \cite{Mueller:1996pm} \\
 FPS	&  \cite{Friedman:1981qw} \\
 SLy4	&  \cite{Douchin:2001sv} \\
 BCSK20	&  \cite{PhysRevC.82.035804} \\
 \hline
 POLYn1	& $P(\rho)\sim \rho^2$\\
 \hline
 \hline
 \end{tabular}
 \caption{List of tabulated EoS used in this work.}\label{tab:EoS}
\end{table}
%%%%%%

\begin{figure*}[ht]
\begin{center}
\includegraphics[width=7.5cm]{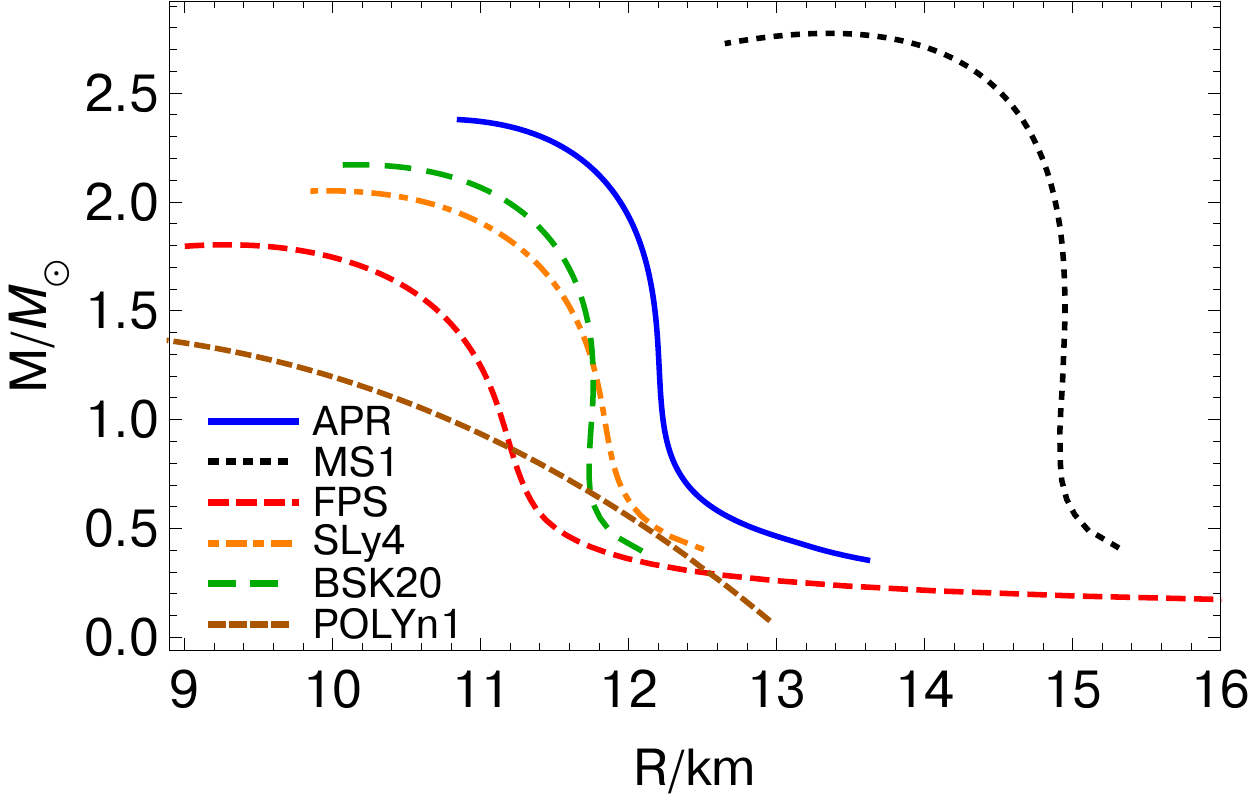}
\includegraphics[width=7.5cm]{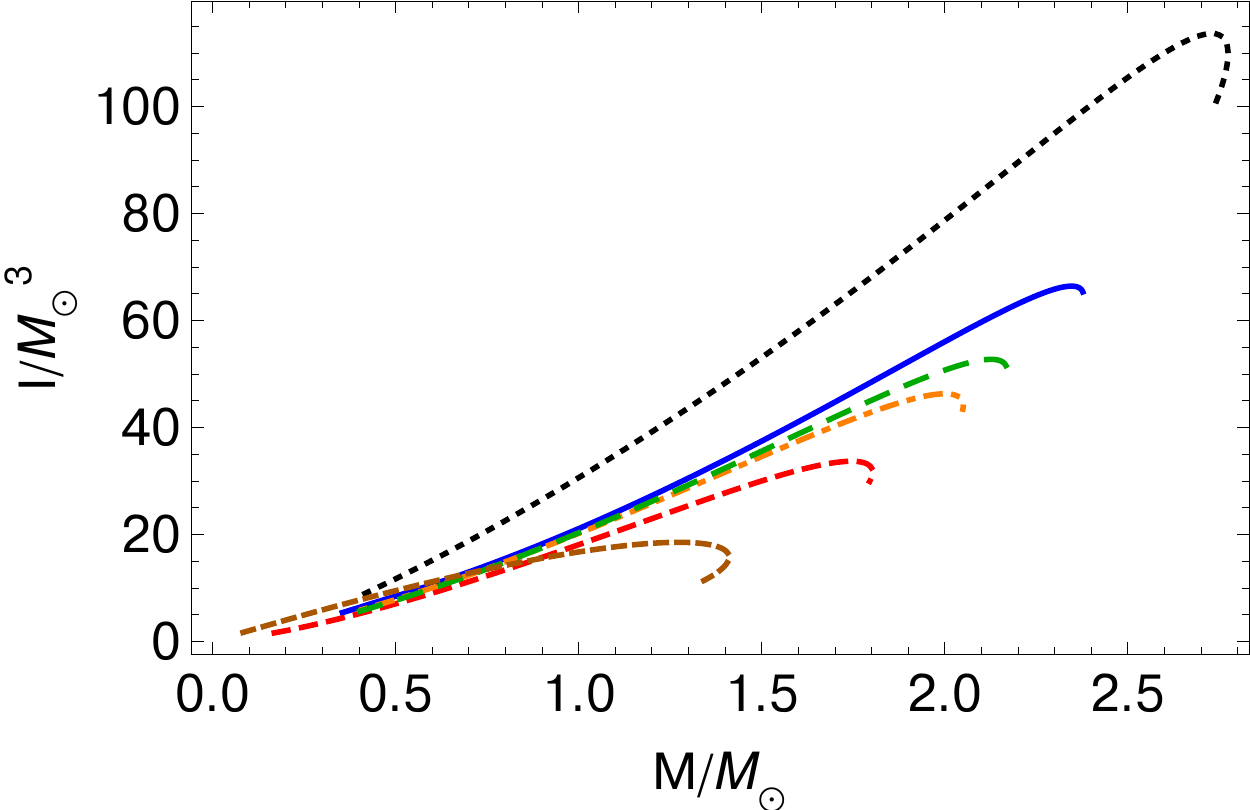}
\caption{(color online). Left panel: mass-radius relation for the slowly-spinning, perfect-fluid stellar models analyzed in this work. We consider various tabulated EoS covering a wide range of NS deformability (cf. Table~\ref{tab:EoS}). Right panel: moment of inertia as a function of the stellar mass.}
\label{fig:background}
\end{center}
\end{figure*}

Figure~\ref{fig:background} shows the mass-radius relation and the moment of inertia as a function of the NS mass for the unperturbed models. As usual, the stellar mass is defined by the value of the radial function ${\cal M}$ at the radius, i.e. $M\equiv {\cal M}(R)$, where the radius $R$ is defined by the relation $P(R)=0$. The moment of inertia reads $I\equiv J/\Omega$, where we recall that $\Omega$ is the fluid angular velocity and $J$ is the angular momentum. To linear order in the spin, the moment of inertia is a ${\cal O}(\chi^0)$ quantity which depends only on the compactness and on the EoS.

%%%%%%%%%%%%%%%%%%%%%%%%%%%%%%%%%%%%%%%%
\subsection{Tidally-deformed, nonrotating stars}
%%%%%%%%%%%%%%%%%%%%%%%%%%%%%%%%%%%%%%%%
We now consider the tidal deformations of the background solution just presented. 
Let us start by discussing nonrotating configurations, which have been studied in the past by several authors (e.g. Refs~\cite{Hinderer:2007mb,Binnington:2009bb,Damour:2009vw}) and can therefore be used to test our numerical code.

\begin{figure*}[ht]
\begin{center}
\includegraphics[width=8.7cm]{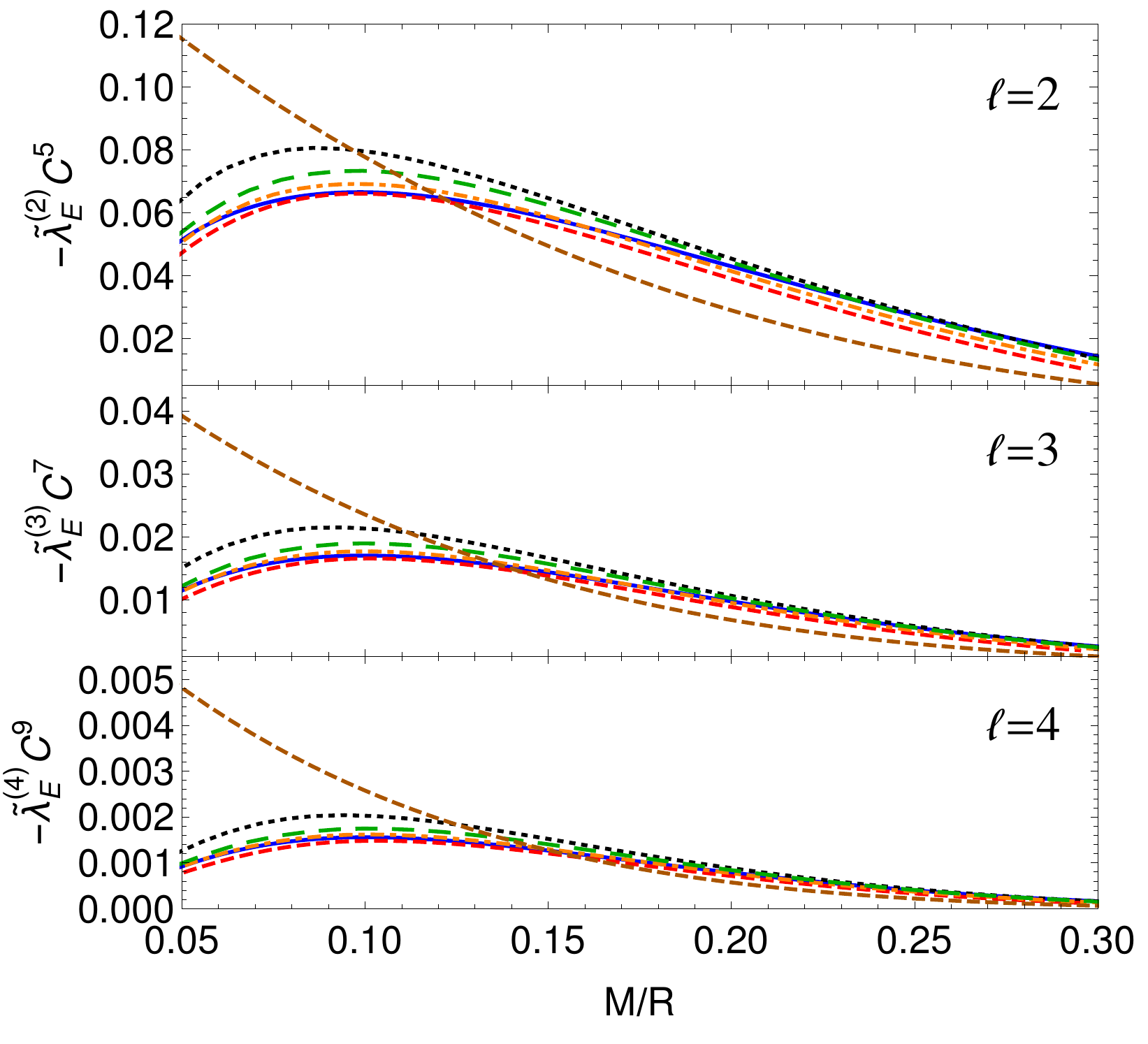}
\includegraphics[width=8.7cm]{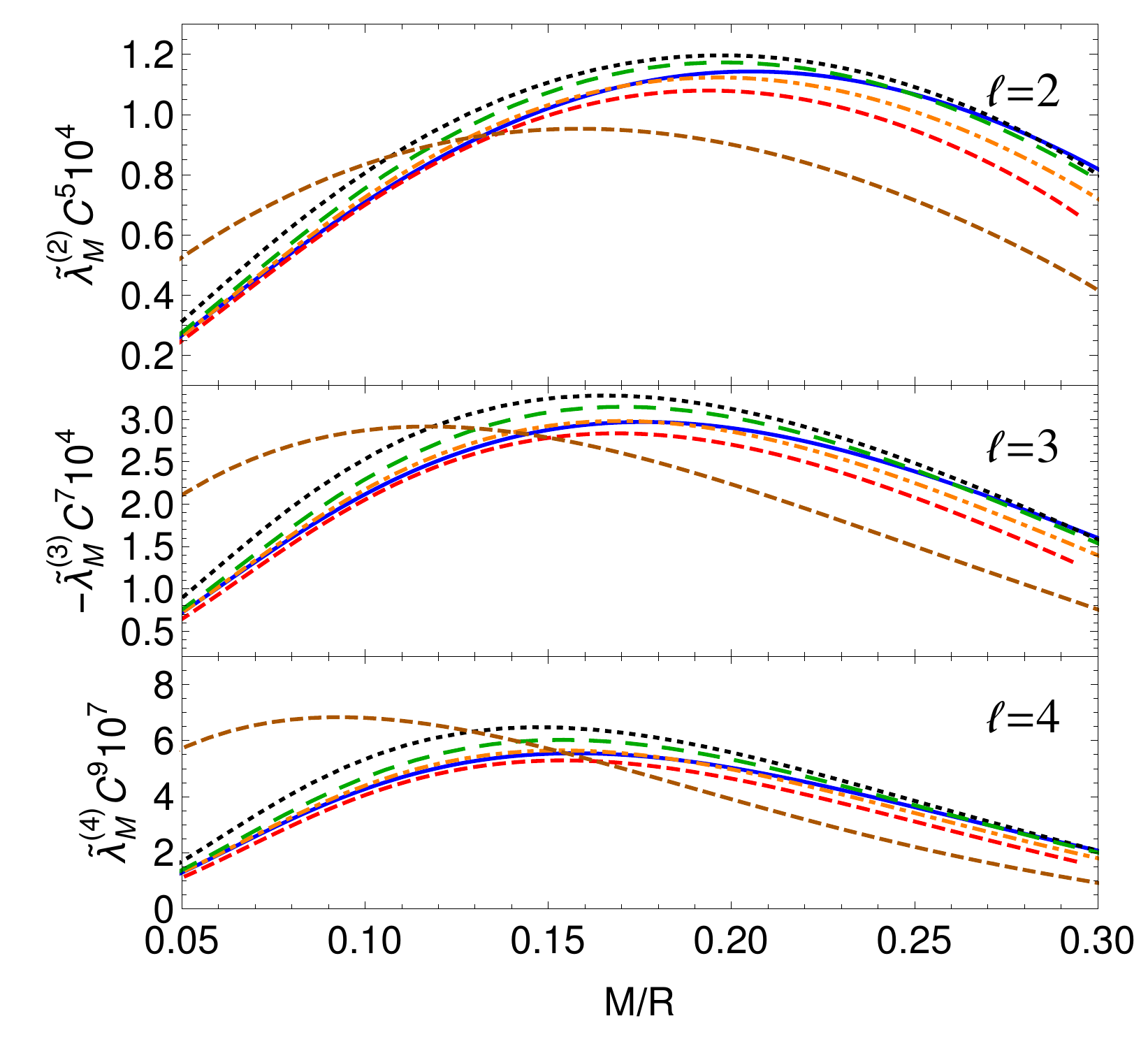}
\caption{(color online). Dimensionless electric (left panels) and magnetic (right panels) Love numbers for $\ell=2,3,4$ (from top to bottom) for various nonrotating NS models with different EoS and as a function of the stellar compactness $C=M/R$. Note that, according to our definitions, the electric Love numbers $\lambda_E^{(\ell)}$ and the magnetic Love number $\lambda_M^{(3)}$ are negative. Nonetheless, modulo a negative factor, our definitions are fully equivalent to the usual ones (cf. e.g. Ref.~\cite{Binnington:2009bb}). Note that $\tilde{\lambda}_E^{(2)}$, $\tilde{\lambda}_M^{(3)}$, $\tilde{\lambda}_E^{(4)}$ are associated with equatorial-symmetric tidal perturbations, whereas $\tilde{\lambda}_M^{(2)}$, $\tilde{\lambda}_E^{(3)}$, $\tilde{\lambda}_M^{(4)}$ break this symmetry.
}
\label{fig:lambda_nonrot}
\end{center}
\end{figure*}
% \clearpage
% \newpage

We solved the perturbation equations for $\ell=2,3,4$ so that we can evaluate the tidal Love numbers in Eq.~\eqref{Love0} for $\ell=2,3,4$. These are shown in Fig.~\ref{fig:lambda_nonrot}, where the left and right panels refer to electric and magnetic tidal Love numbers, respectively. 
For completeness, we show both the tidal Love numbers in Eq.~\eqref{Love0b} (namely $\tilde{\lambda}_E^{(2)}$, $\tilde{\lambda}_M^{(3)}$, $\tilde{\lambda}_E^{(4)}$) which are associated with equatorial-symmetric tidal perturbations and those defined in Eq.~\eqref{Love0c} (namely, $\tilde{\lambda}_M^{(2)}$, $\tilde{\lambda}_E^{(3)}$, $\tilde{\lambda}_M^{(4)}$) which instead break this symmetry.
As previously mentioned, these numbers have been normalized by suitable powers of $M$ to make them dimensionless. Furthermore, in Fig.~\ref{fig:lambda_nonrot} we multiply the dimensionless Love numbers by some suitable power of the compactness $C=M/R$ so that, in the Newtonian limit, these quantities tend to a constant value. The small-compactness limit can be better appreciated for the POLYn1 curves, because the realistic tabulated EoS are meant to describe only configurations with large compactness. This is the reason why the axes of Fig.~\ref{fig:lambda_nonrot} have been truncated at $C\sim0.05$. Were these plots extended to $C\to0$, the electric Love numbers for POLYn1 would approach their Newtonian constant value, whereas the magnetic Love numbers all go to zero in the Newtonian limit, since axial perturbations are not defined in that case. 
In the static case, our results for the electric and magnetic tidal Love numbers are in perfect agreement with those computed in Ref.~\cite{Binnington:2009bb}, cf. Appendix~\ref{app:comment} for a comment about the computation of the magnetic Love numbers in the static case.

%%%%%%%%%%%%%%%%%%%%%%%%%%%%%%%%%%%%%%%%%%%%%%%%%%%%%%%%%%%%%%%%%%%%
\subsection{Rotational tidal Love numbers}
%%%%%%%%%%%%%%%%%%%%%%%%%%%%%%%%%%%%%%%%%%%%%%%%%%%%%%%%%%%%%%%%%%%%

\begin{figure*}[ht]
\begin{center}
\includegraphics[width=8.cm]{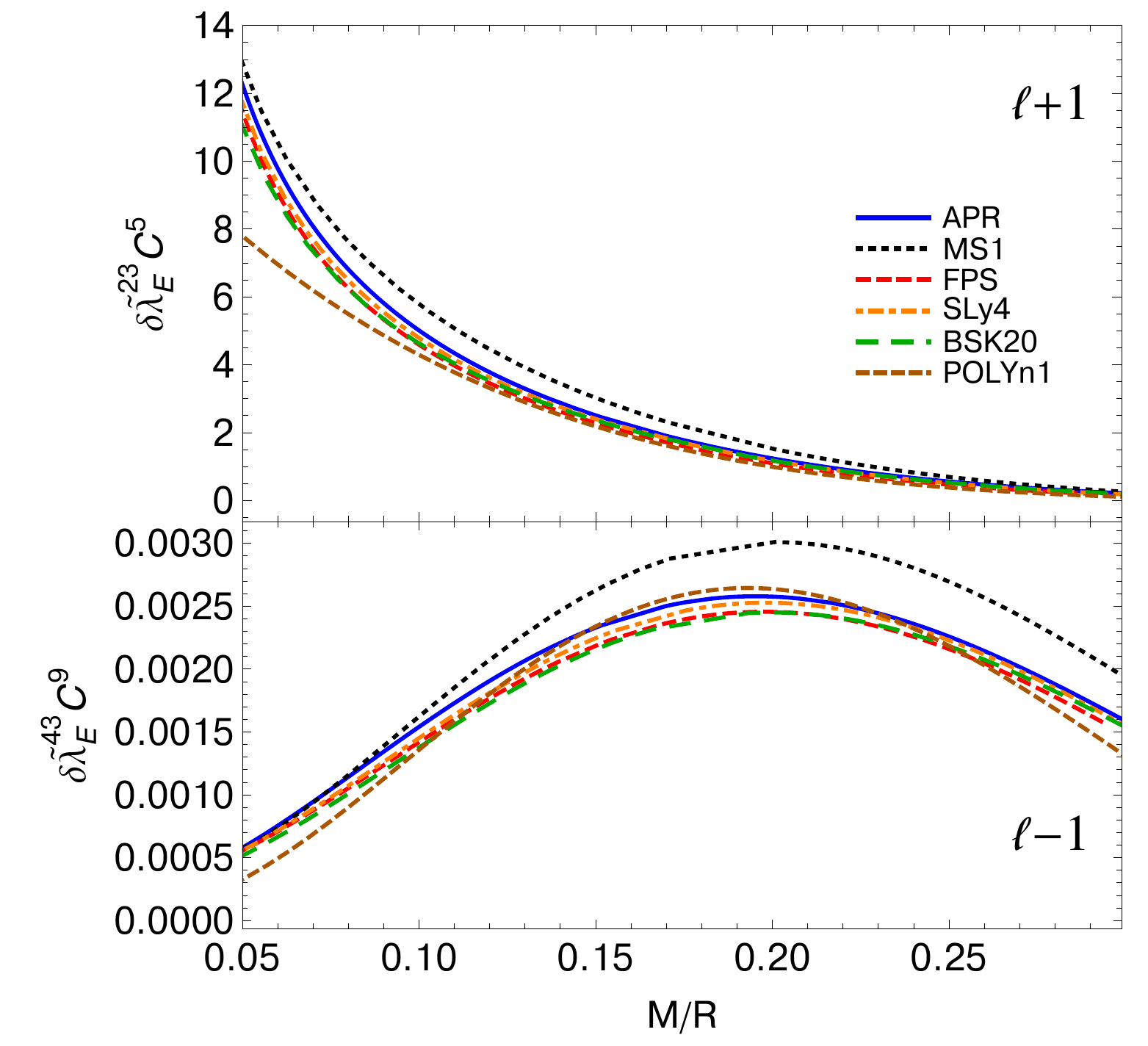}
\includegraphics[width=8.cm]{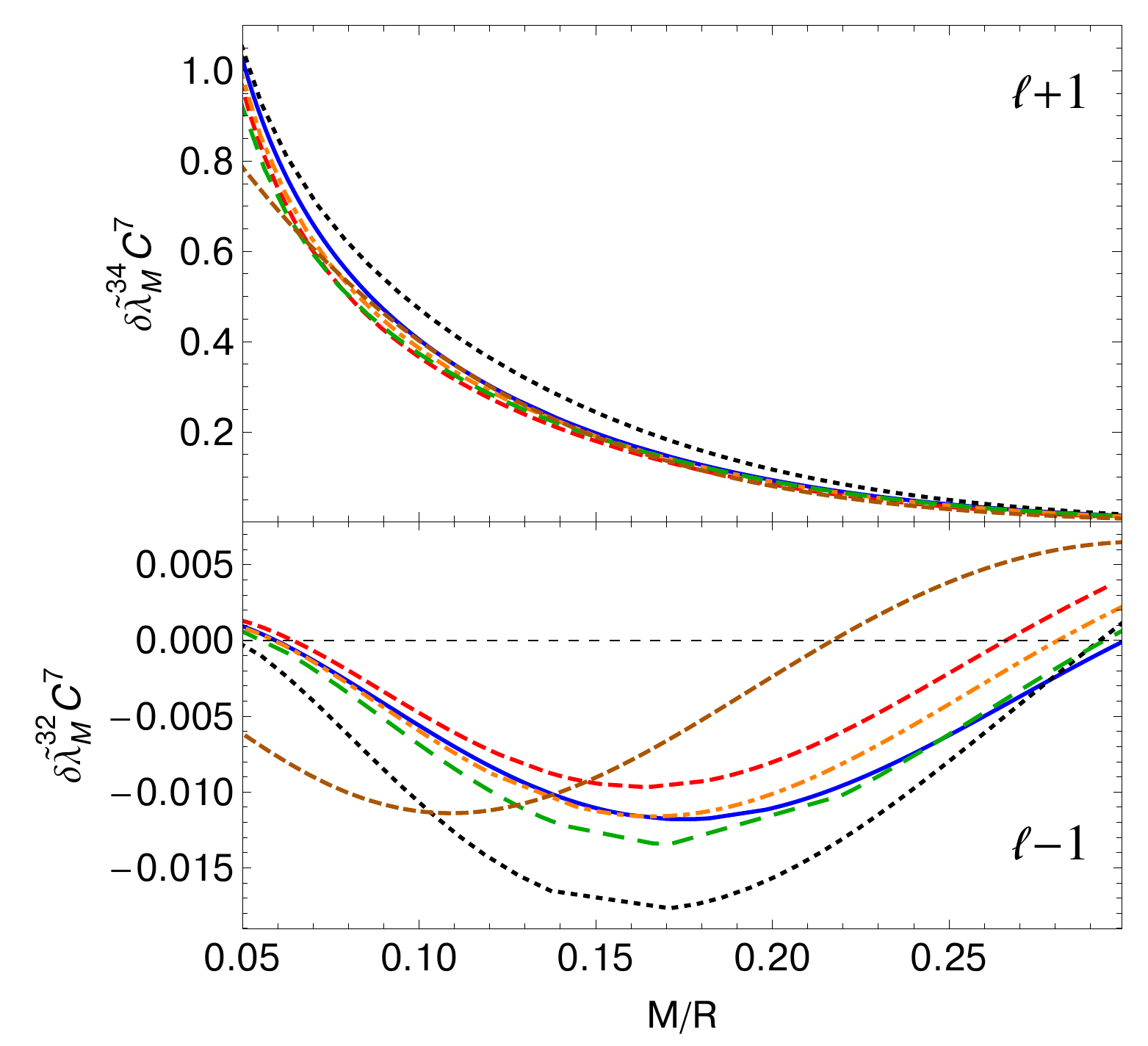}
\caption{(color online). Dimensionless rotational electric (left panels) and magnetic (right panels) tidal Love numbers for a spinning NS as a function of the compactness $M/R$ and for the various tabulated EoS adopted in this work.
In our notation,  $\delta\tilde\lambda_E^{(23)}$ denotes the correction to $\tilde\lambda_E^{(2)}$ arising through the coupling to the magnetic octupolar ($\ell=3$) tidal field to first order in the spin. Note that $\delta\tilde\lambda_M^{(12)}$ would represent a tidally induced spin shift that can be reabsorbed in the definition of $\chi$ and therefore $\delta\tilde\lambda_M^{(12)}=0$. 
}
\label{fig:rot_VS_C}
\end{center}
\end{figure*}

The rotational tidal Love numbers in Eq.~\eqref{Lovem} for various families of spinning NSs are shown in
Fig.~\ref{fig:rot_VS_C}, which is one of the main results of this work.
As a reference, in Table~\ref{tab:Love} we present some rotational tidal Love numbers for selected values of the compactness and for two representative EoS.
%%%
Here we only present the rotational Love numbers associated with equatorial-symmetric perturbations; the case in which the equatorial symmetry is broken in discussed in Appendix~\ref{app:noequatorial} (see Fig.~\ref{fig:rot_VS_C2}).

\begin{table}
 \begin{tabular}{c|ccc|ccc|}
  &  \multicolumn{3}{c}{APS EoS} & \multicolumn{3}{c}{MS1 EoS} \\
 $C$ &  $0.1$ & $0.15$ & $0.2$ & $0.1$ & $0.15$ & $0.2$  \\
  \hline
$\delta\tilde\lambda_E^{(23)}C^5$ & 5.1508  & 2.5522	& 1.2525	& 5.9280 & 3.0727 & 1.5413\\  
$\delta\tilde\lambda_E^{(43)}C^9$ & 0.0016  & 0.0024	& 0.0026	& 0.0017 & 0.0027 & 0.0031\\  
$\delta\tilde\lambda_M^{(34)}C^7$ & 0.4121  & 0.1982	& 0.0930	& 0.4789 & 0.2444 & 0.1163\\  
$-\delta\tilde\lambda_M^{(32)}C^7$& 0.0046  & 0.0098	& 0.0099	& 0.0095 & 0.0160 & 0.0143\\  
 \end{tabular}
\caption{Selected (normalized) rotational Love numbers shown in Fig.~\ref{fig:rot_VS_C} for two EoS (APS and MS1) and selected values of the compactness.}\label{tab:Love}
\end{table}

In Fig.~\ref{fig:rot_VS_C}, we have normalized the rotational Love numbers by suitable powers of the compactness $C$ in order for the Newtonian limit to be independent of $C$. The scaling has been achieved by investigating the polytropic EoS POLYn1 as $C\to0$, since the other tabulated EoS are not realistic in the Newtonian limit. 

It is important to note that $\delta\tilde{\lambda}_E^{(23)}\sim1/C^5$ in the Newtonian limit, precisely as the static tidal Love number $\tilde{\lambda}_E^{(2)}$. For this reason, as already shown in Fig.~\ref{fig:ratio}, the ratio $\delta\tilde{\lambda}_E^{(23)}/\tilde{\lambda}_E^{(2)}$ does not depend on $C$ for small compactness. 

The left (respectively, right) panels of Fig.~\ref{fig:rot_VS_C} (and Fig.~\ref{fig:rot_VS_C2} in Appendix~\ref{app:noequatorial}) show the ``magnetic-led'' (respectively, ``electric-led'') rotational tidal Love numbers corresponding to a tidal field with harmonic index $\ell+1$ (top panels) and $\ell-1$ (bottom panels) that sources a $\ell$-pole moment with opposite parity, in agreement with the spin-selection rules previously discussed.

Finally, it is interesting to note that not only $\delta\tilde{\lambda}_M^{(32)}$ is nonmonotonic, but also it does not have a definite sign as a function of the compactness. This peculiar behavior is observed also for the other rotational Love numbers associated with the broken equatorial symmetry discussed in Appendix~\ref{app:noequatorial} [cf. Fig.~\ref{fig:rot_VS_C2}].

%%%%%%%%%%%%%%%%%%%%%%%%%%%%%%%%%%%%%%%%%%%%%%%%%%%%%%%%%%%%%%%%%%%%
\subsection{Approximate universality} \label{sec:universality} 
%%%%%%%%%%%%%%%%%%%%%%%%%%%%%%%%%%%%%%%%%%%%%%%%%%%%%%%%%%%%%%%%%%%%
Some years ago, Yagi and Yunes~\cite{Yagi:2013bca,Yagi:2013awa} discovered that certain suitably normalized combinations
of the tidal Love numbers of a nonspinning NS and of the multipole moments of a spinning NS are related to each other by
approximately universal relations which depend only mildly on the EoS. In their simplest version, these relations are
known as $I$-Love-$Q$ and connect the moment of inertia $I$, the (rotation-induced) mass
quadrupole moment $Q$, and the electric quadrupolar tidal Love number (all of them suitably normalized) through relations
which are independent of the NS EoS at the percent level. Similar relations also exist among other (electric and
magnetic) tidal Love numbers of a nonspinning NS~\cite{Yagi:2013sva}, although the universality deteriorates for
increasing values of the multipole $\ell$ and in the magnetic sector relative to the electric one with the same $\ell$.

The original I-Love-Q relations were found for slowly, rigidly rotating, barotropic, isotropic, unmagnetized and isolated stars~\cite{Yagi:2013bca,Yagi:2013awa}. Recently, they have been extended to include rapid rotation~\cite{Doneva:2013rha,Pappas:2013naa,Chakrabarti:2013tca,Yagi:2014bxa}, nonbarotropic~\cite{Martinon:2014uua}, anisotropic~\cite{Yagi:2015cda}, and differentially rotating~\cite{Bretz:2015rna} fluids, strong magnetic fields~\cite{Haskell:2013vha}, dynamical configurations~\cite{Maselli:2013mva}, exotic compact objects~\cite{Pani:2015tga}, and even deviations from general relativity~\cite{Yagi:2013bca,Yagi:2013awa,Sham:2013cya,Pani:2014jra,Doneva:2014faa,Doneva:2015hsa}. The outcome of these studies is that the approximate universality is remarkably robust in realistic configurations. 

It is important to stress that the tidal Love numbers entering the $I$-Love-$Q$ relations are those of \emph{nonspinning} NSs, whereas the moment of inertia and the mass quadrupole moment refer to rotating stellar configurations. It is therefore important to understand how spin corrections to the tidal Love numbers impact such universality. The spin corrections to the tidal Love numbers of a NS are computed in this work for the first time.

\begin{figure*}[ht]
\begin{center}
\includegraphics[width=5.5cm]{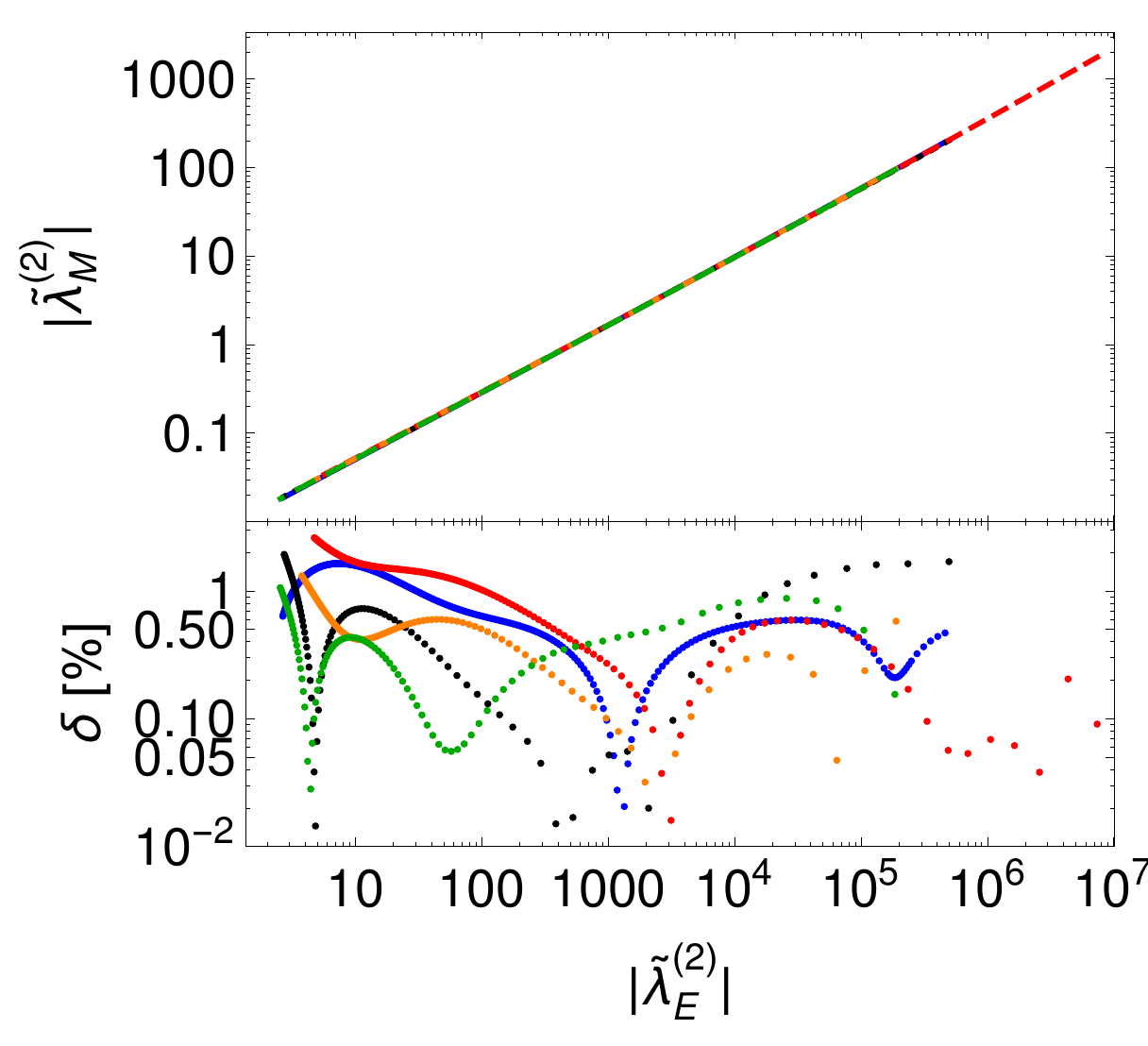}
\includegraphics[width=5.5cm]{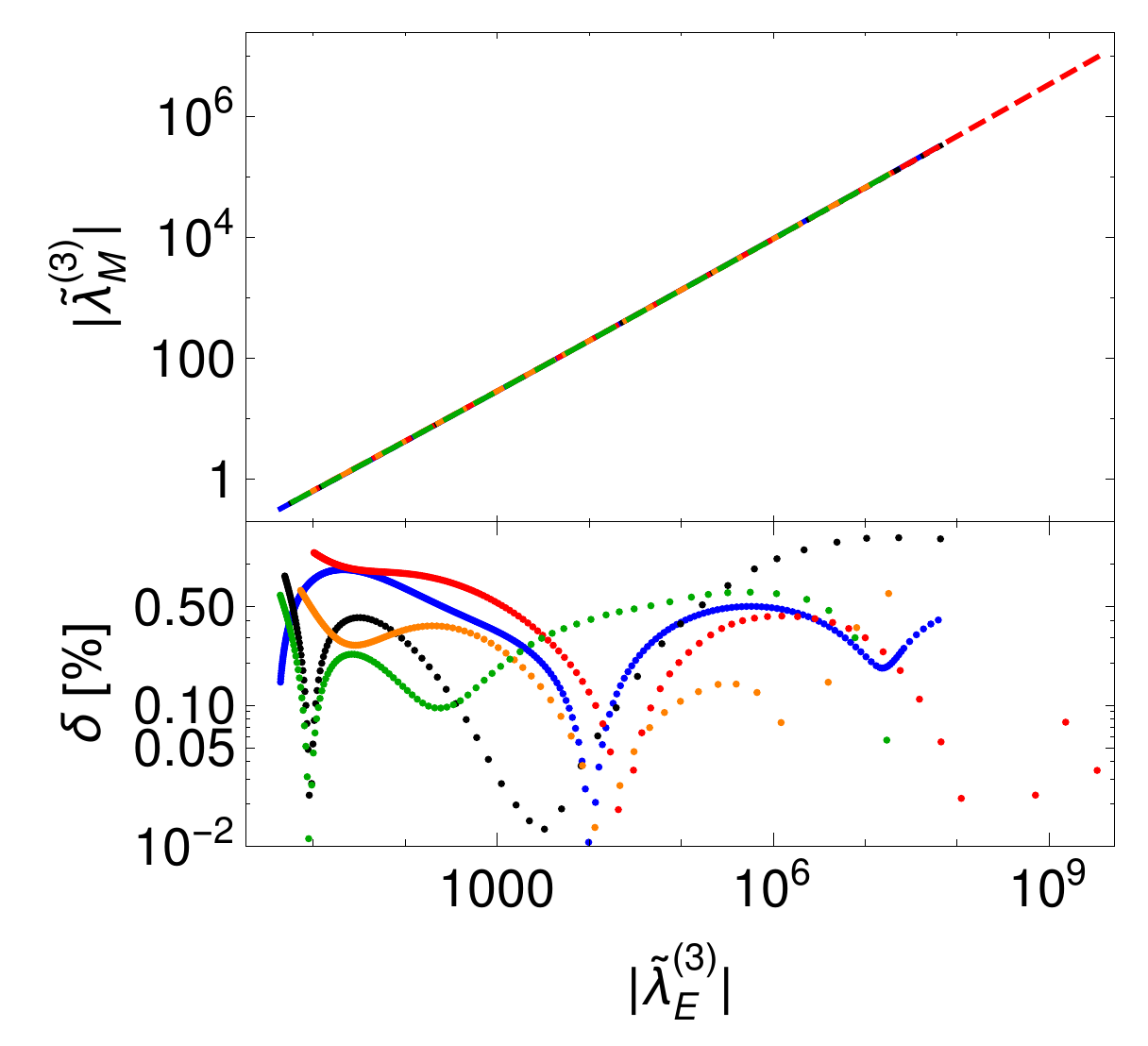}
\includegraphics[width=5.5cm]{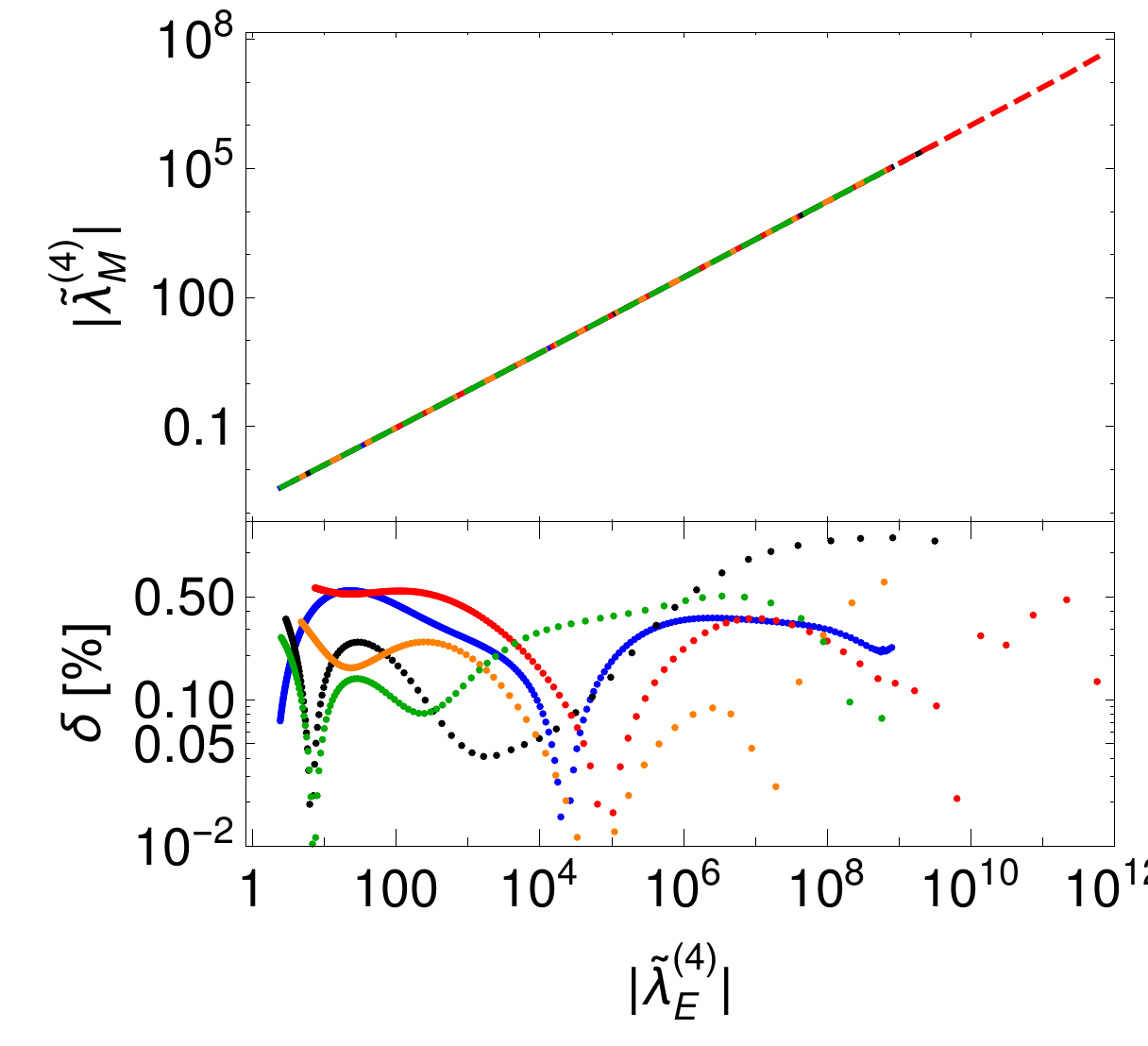}
%%%%
\caption{(color online). Universal relations for electric and magnetic Love numbers of a nonspinning NS with $\ell=2,3,4$ (cf. also Ref.~\cite{Yagi:2013sva}). For each panel, we show the percentual deviation from a EoS-independent fit. The deviations from universality are roughly less than $1\%$ between electric and magnetic Love numbers with same multipolar index $\ell$.}
\label{fig:lambdEBa}
\end{center}
\end{figure*}

\begin{figure*}[ht]
\begin{center}
%%%%
\includegraphics[width=5.5cm]{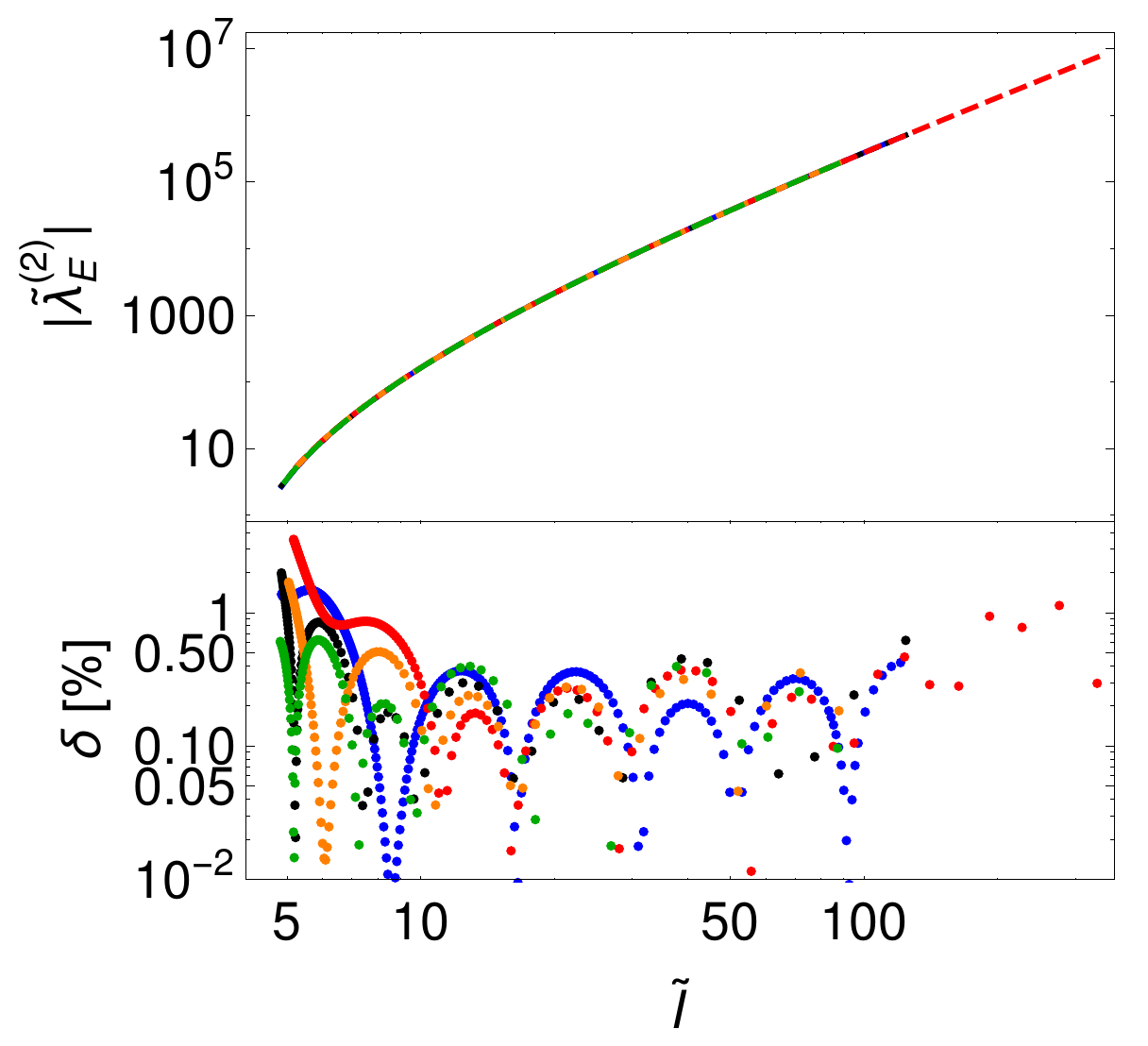}
\includegraphics[width=5.5cm]{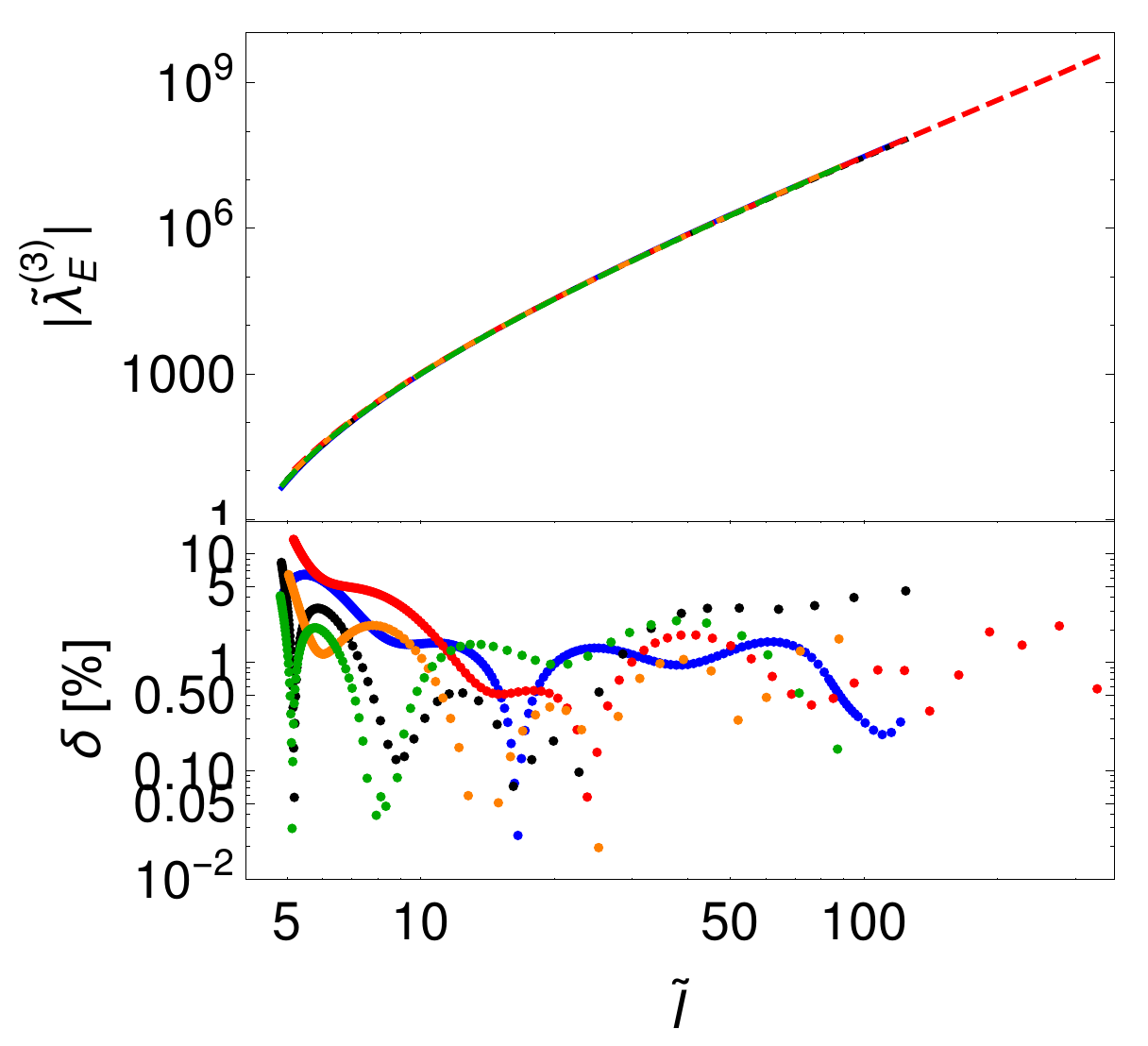}
\includegraphics[width=5.5cm]{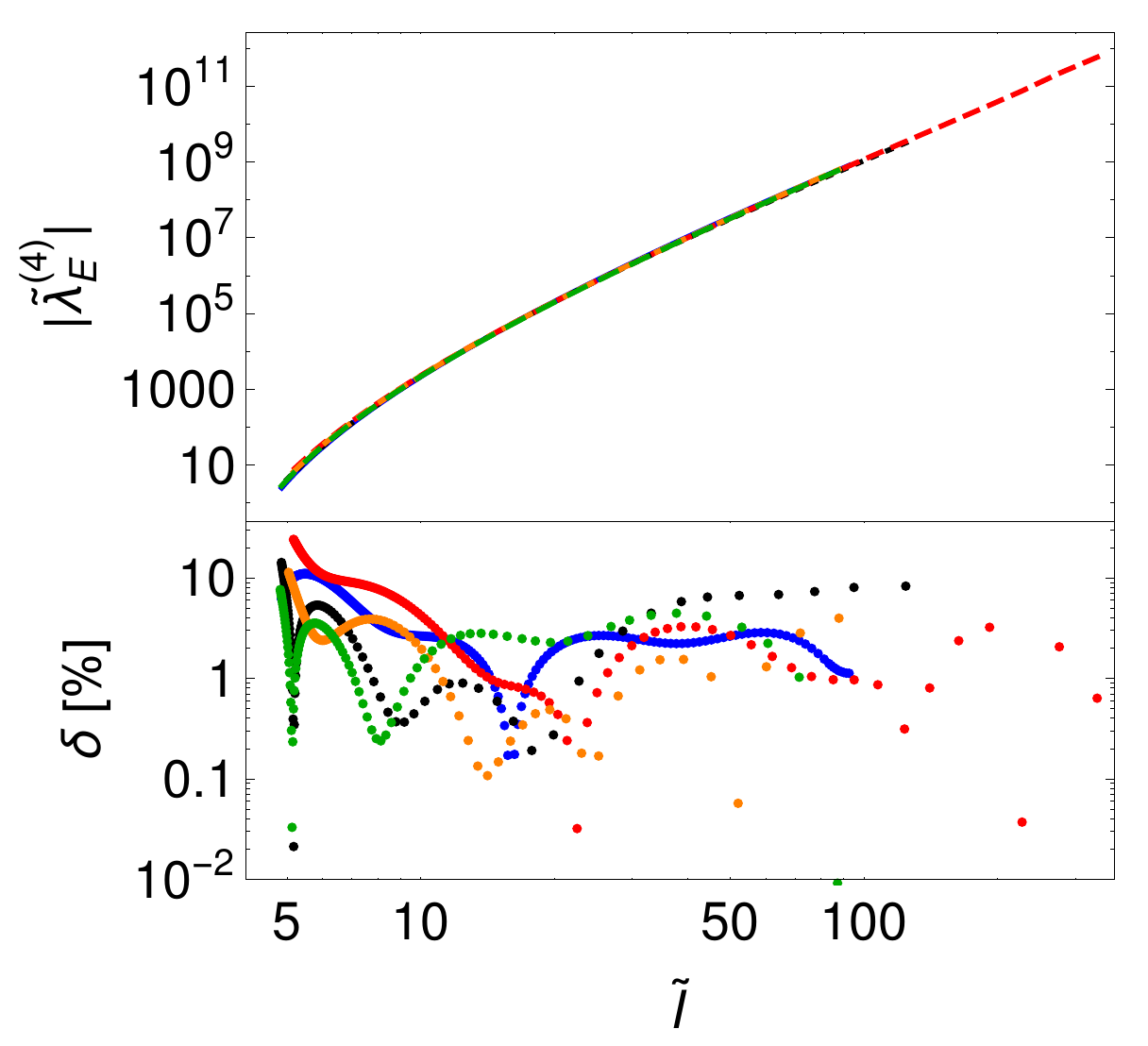}
%%%%
\includegraphics[width=5.5cm]{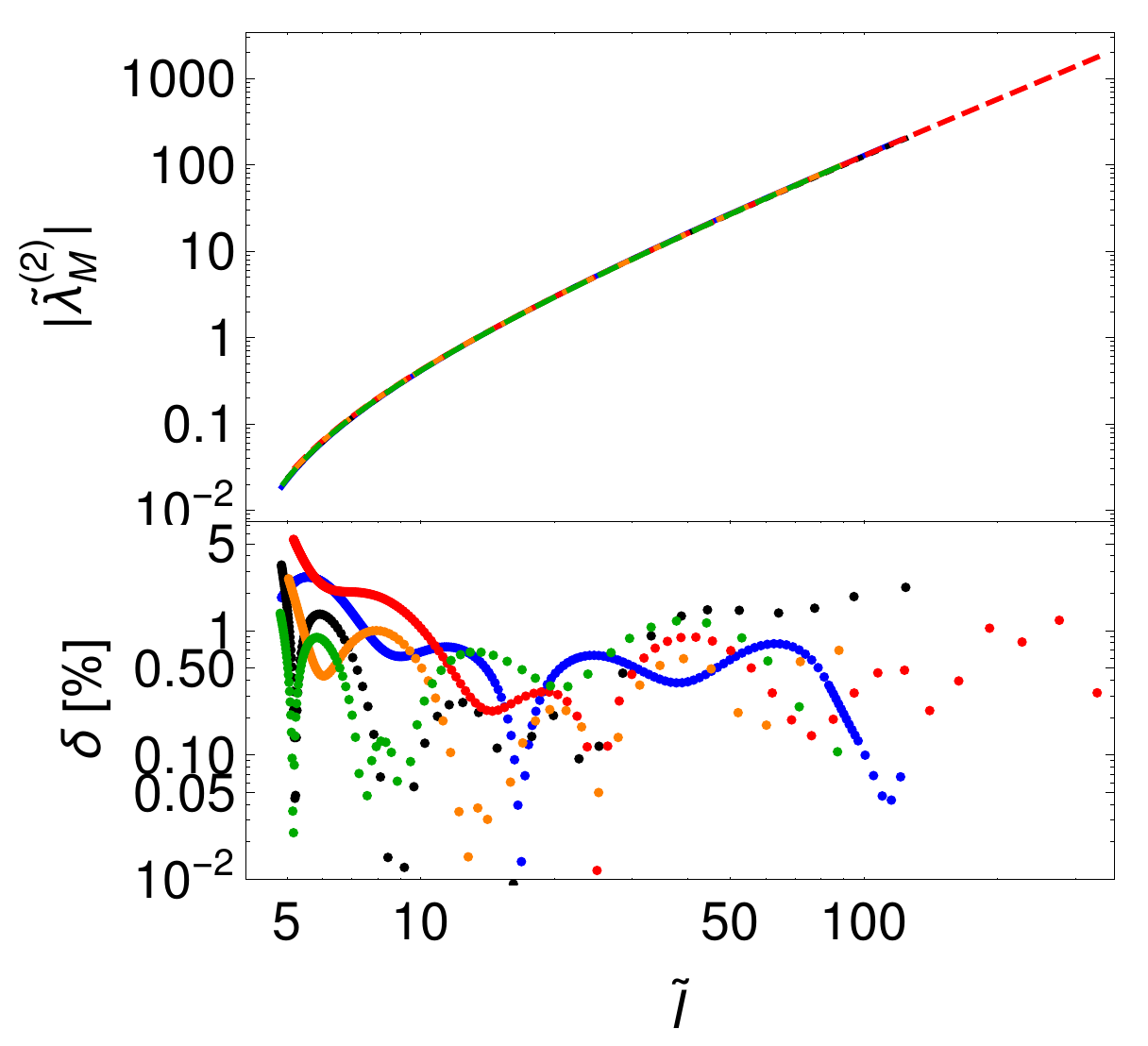}
\includegraphics[width=5.5cm]{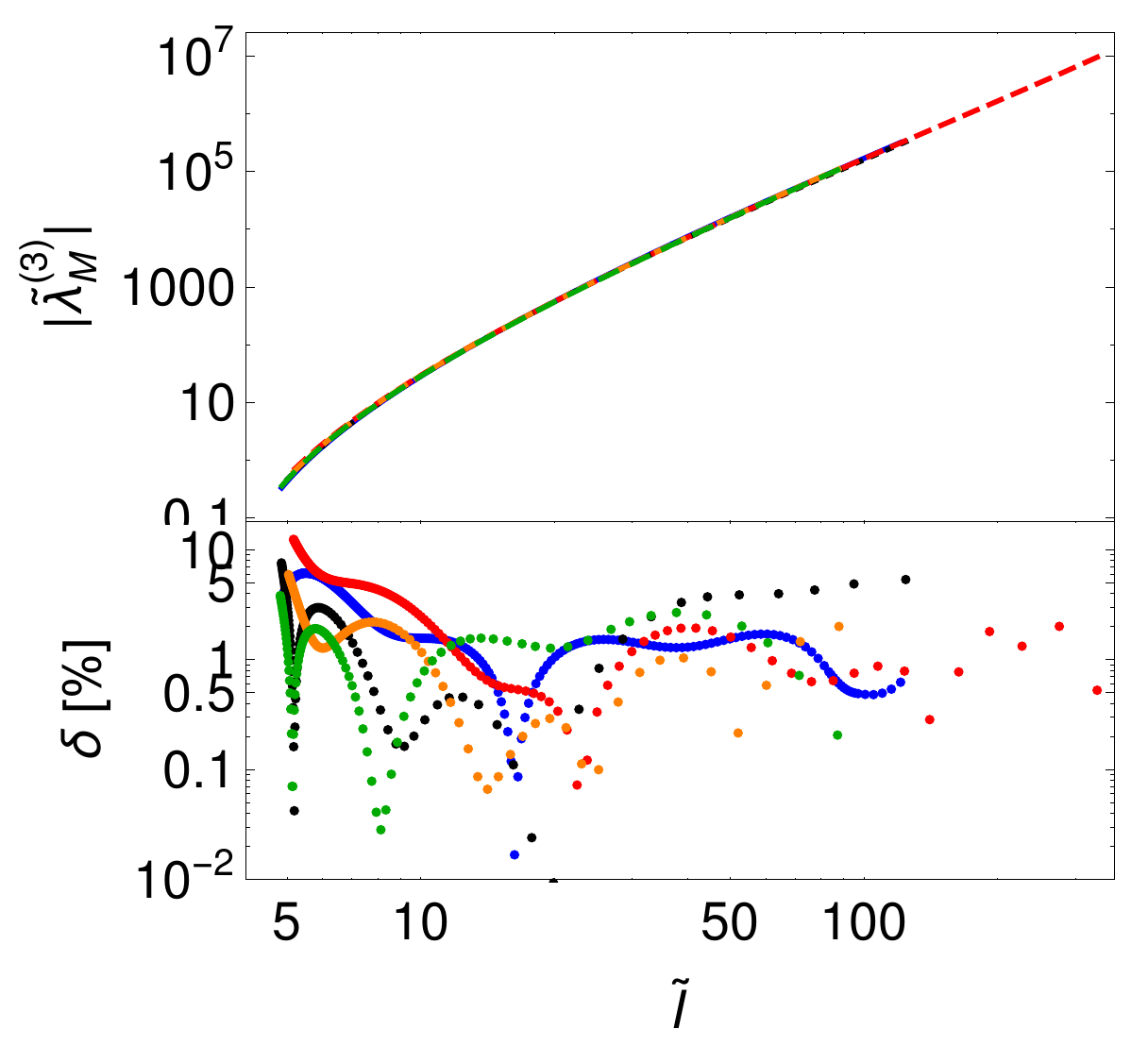}
\includegraphics[width=5.5cm]{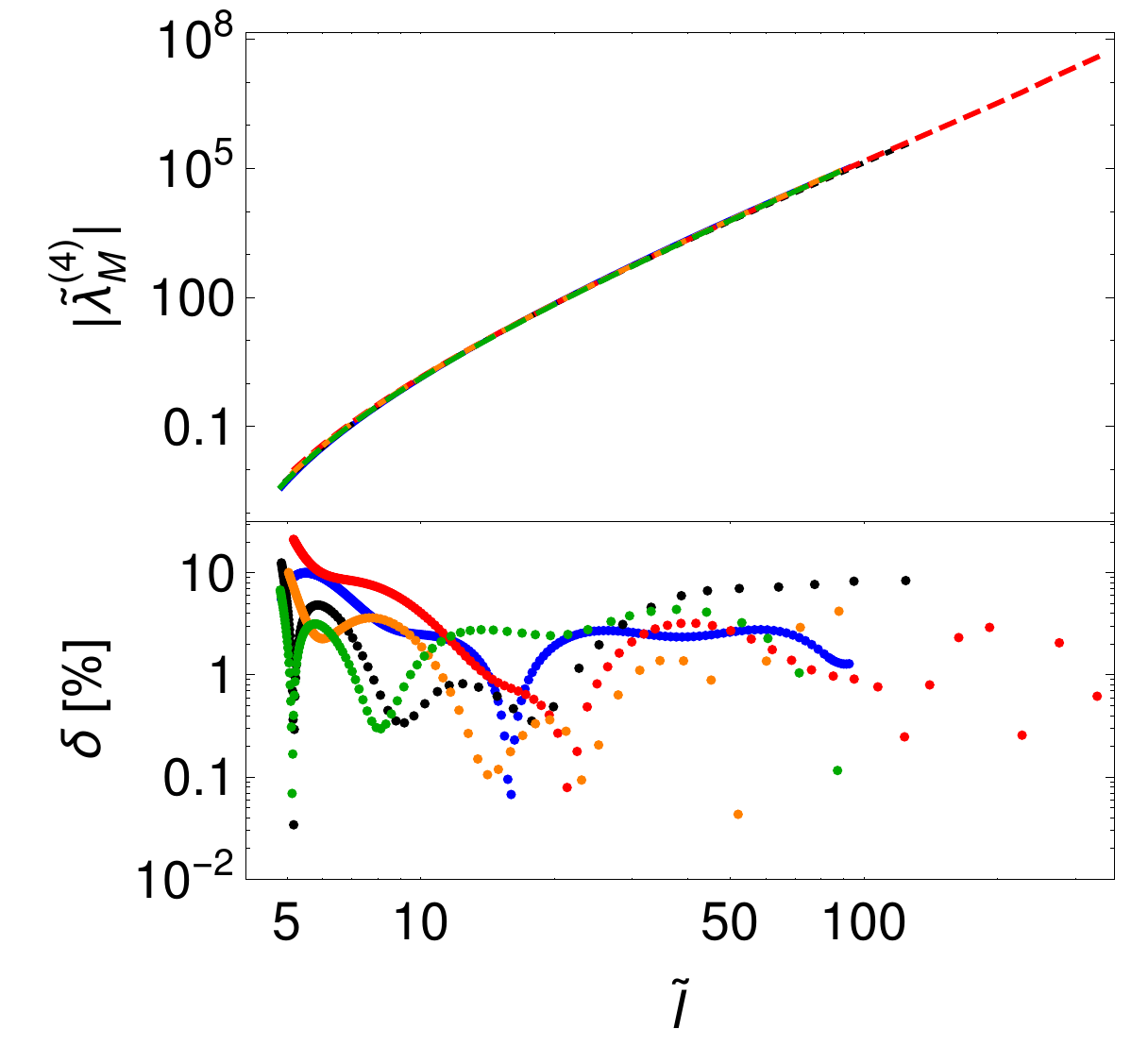}
\caption{(color online). Same as in Fig.~\ref{fig:lambdEBa} but for the universal relations between the dimensionless moment of inertia $\tilde{I}\equiv I/M^3$ and the tidal Love number $\tilde\lambda_E^{(\ell)}$ (top panels) or $\tilde\lambda_M^{(\ell)}$ (bottom panels). With respect to Fig.~\ref{fig:lambdEBa}, the deviations from universality increase with $\ell$.}
\label{fig:lambdEBb}
\end{center}
\end{figure*}

%%%%%%%%%%%%%%%%%%%%%%%%%%%%%%%%%%%
\subsubsection{Nonrotating case}
%%%%%%%%%%%%%%%%%%%%%%%%%%%%%%%%%%%
As a by-product of our analysis, we can first study the universality of the relations among the (dimensionless) tidal Love numbers to zeroth order in the spin. The results shown in Figs.~\ref{fig:lambdEBa} and~\ref{fig:lambdEBb} confirm the analysis of Ref.~\cite{Yagi:2013sva} (but see Appendix~\ref{app:comment} below for a comment on the computation of the magnetic Love numbers performed in Ref.~\cite{Yagi:2013sva}) and extend it by computing $\tilde\lambda_E^{(4)}$, $\tilde\lambda_M^{(3)}$ and $\tilde\lambda_M^{(4)}$.

In the three panels of Fig.~\ref{fig:lambdEBa} we show the relations among $\tilde\lambda_M^{(\ell)}$ and $\tilde\lambda_E^{(\ell)}$, i.e. between magnetic and electric tidal Love numbers with the same multipole $\ell$. In this case the degree of universality is roughly the same for different values of $\ell$, typically below the percent level. 
However, as noted in Ref.~\cite{Yagi:2013sva}, when written more usefully in terms of the lowest 
multipole moments (for example in terms of the dimensionless moment of inertia $\tilde{I}\equiv I/M^3$) the degree of universality deteriorates with increasing $\ell$, as shown in  Fig.~\ref{fig:lambdEBb} for the electric (top panels) and magnetic (bottom panels) tidal Love numbers with $\ell=2,3,4$.
For $\ell=3$ and $\ell=4$, the universality of the $\tilde\lambda_E^{(\ell)}$-$\tilde{I}$ and $\tilde\lambda_M^{(\ell)}$-$\tilde{I}$ relations holds approximately within $5\%$ and $10\%$, respectively.

%%%%%%%%%%%%%%%%%%%%%%%%%%%%%%%%%%%
\subsubsection{Universality-breaking for rotational tidal Love numbers}
%%%%%%%%%%%%%%%%%%%%%%%%%%%%%%%%%%%
We are now in a position to investigate the universality of the rotational Love numbers computed in this paper.

\begin{figure*}[ht]
\begin{center}
\includegraphics[width=7.5cm]{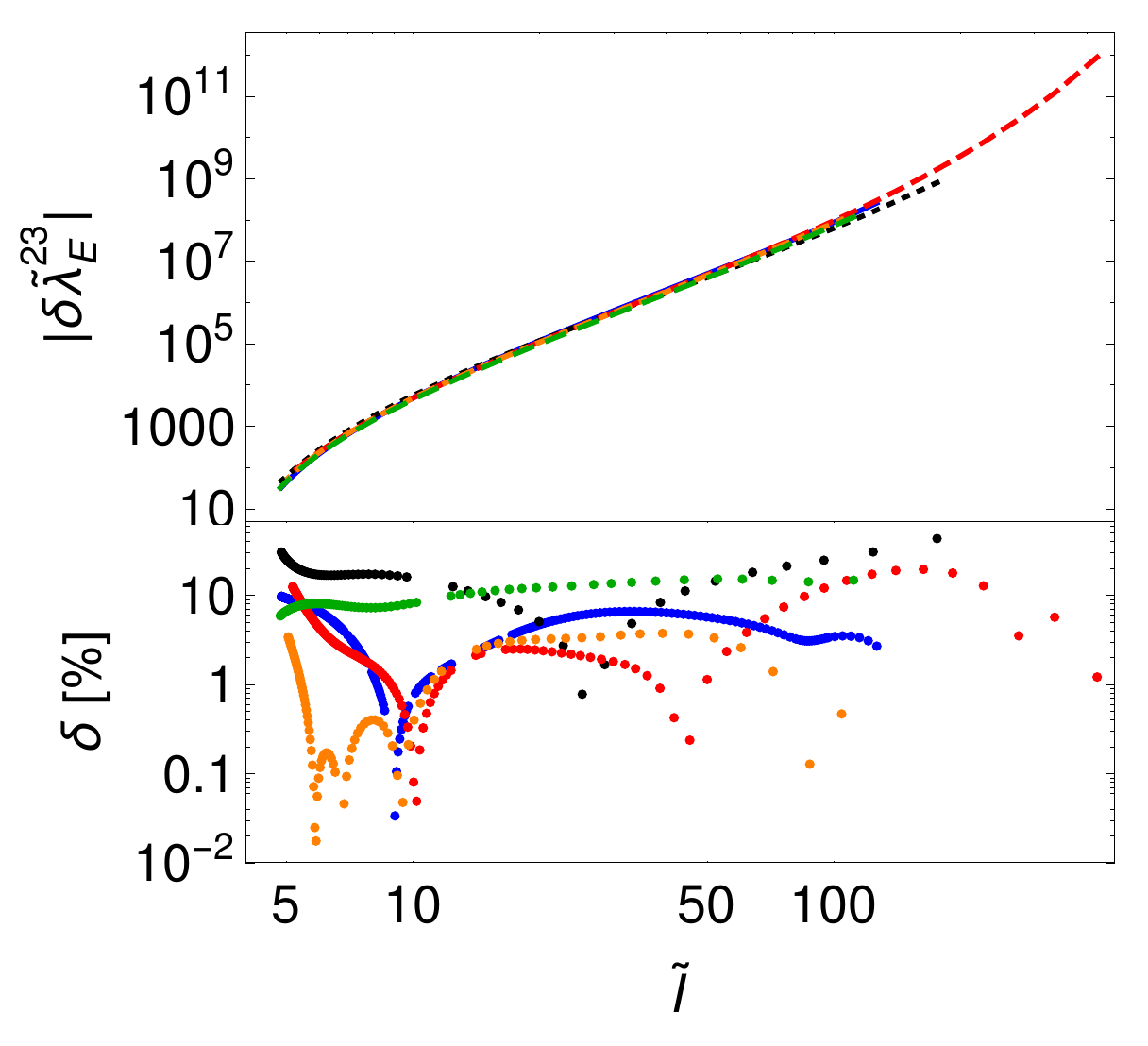}
\includegraphics[width=7.5cm]{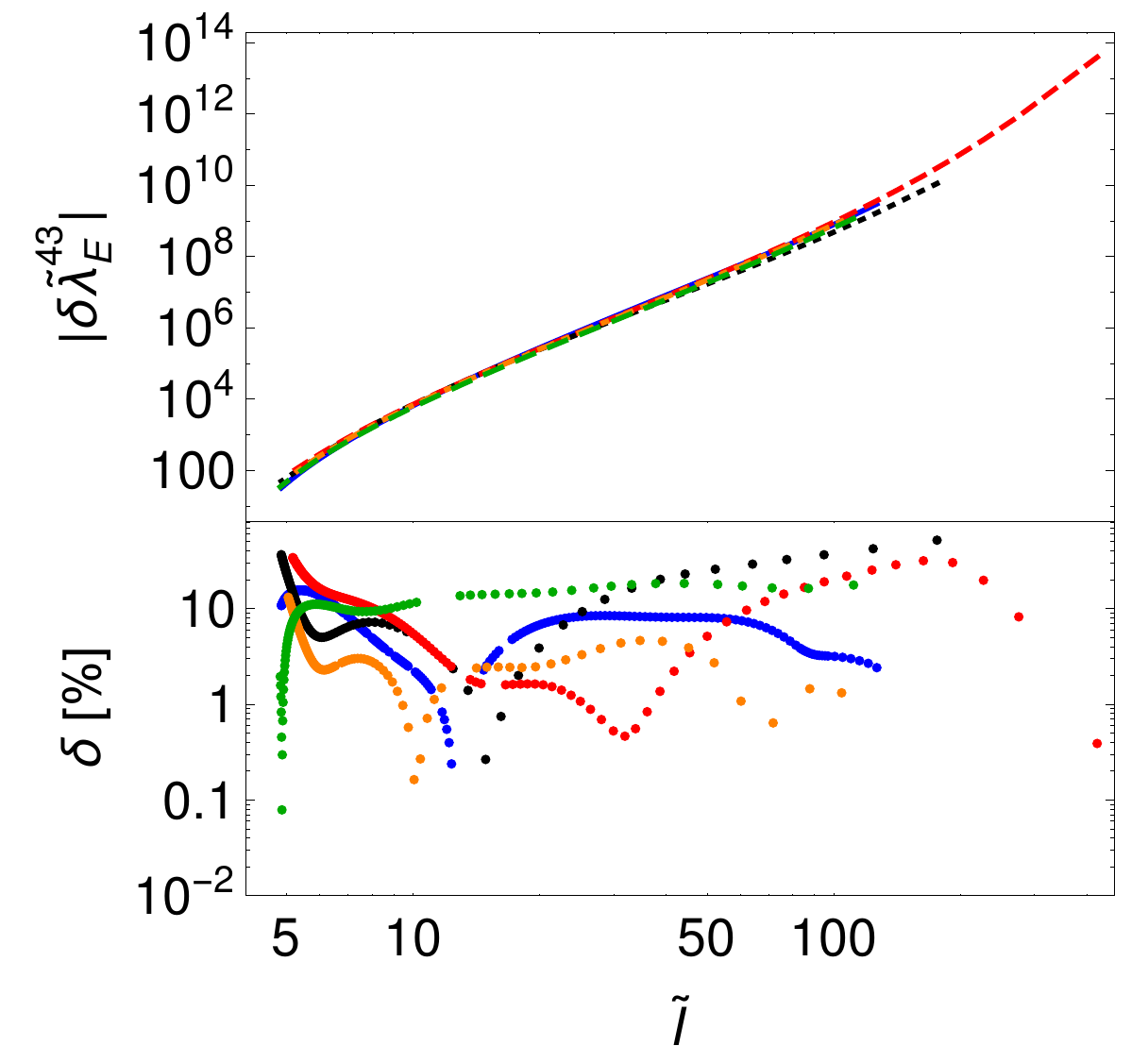}\\
%%%%%
\includegraphics[width=7.5cm]{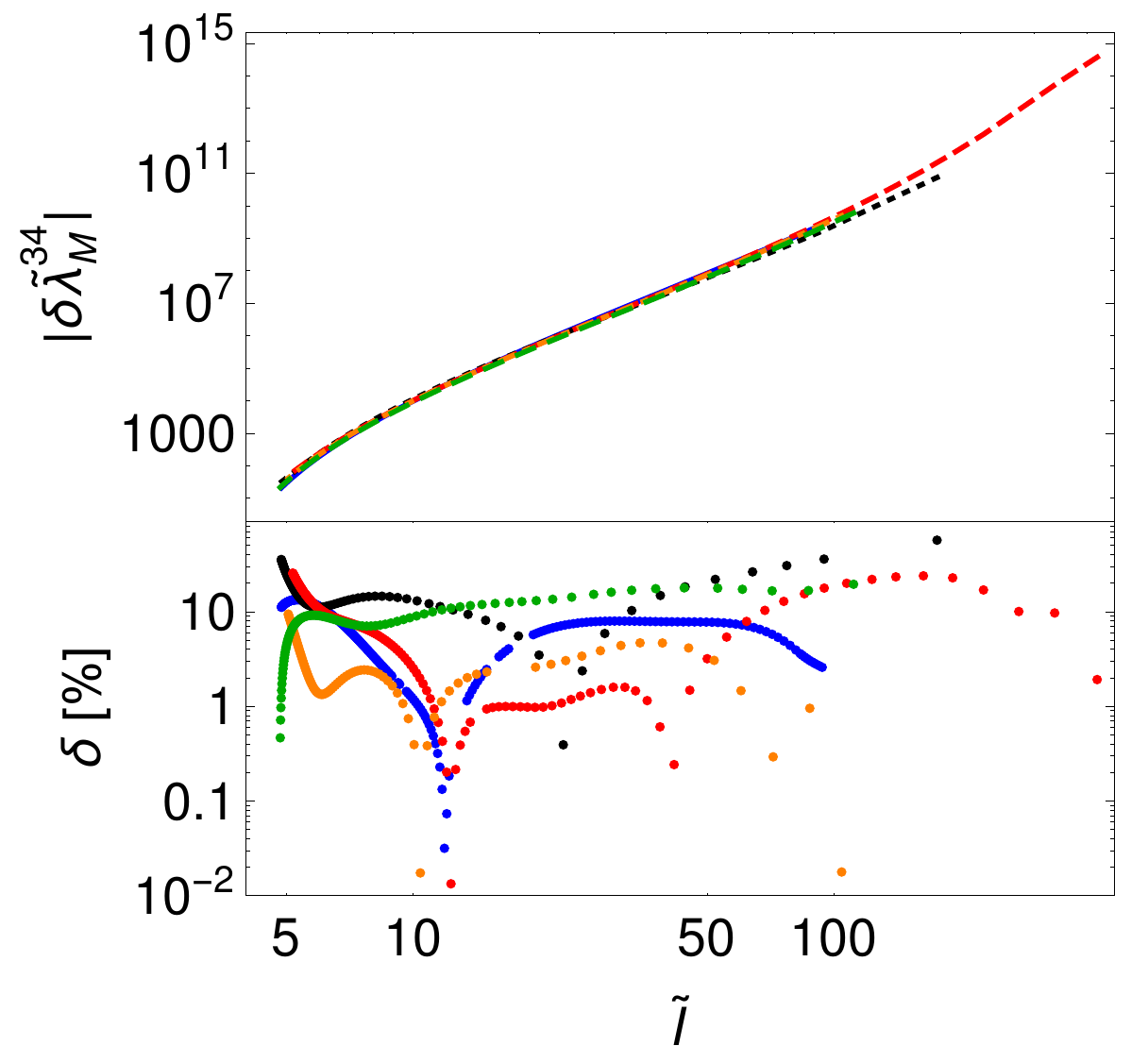}
\includegraphics[width=7.5cm]{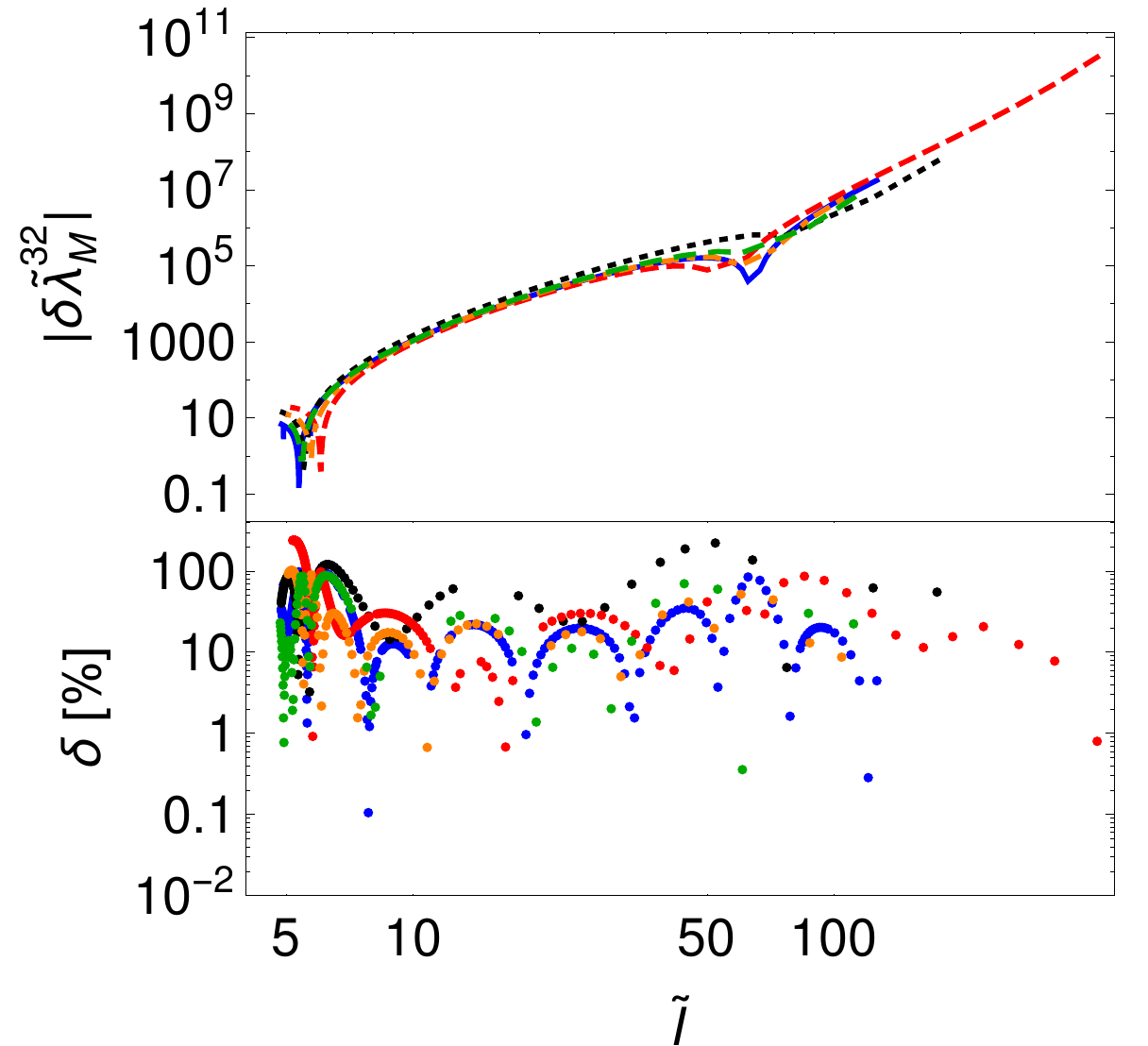}
%%%%%%%%%%%%%%%%%%
\caption{(color online). Spin-induced electric and magnetic Love numbers for spinning NSs as functions of the dimensionless moment of inertia $\tilde{I}$, to be compared to the static case shown in Fig.~\ref{fig:lambdEBb}. Note that the deviations from universality are much larger than in the static case.}
\label{fig:rot_universal}
\end{center}
\end{figure*}

In Fig.~\ref{fig:rot_universal} we show some rotational Love numbers already presented in Fig.~\ref{fig:rot_VS_C} but normalized as explained below Eq.~\eqref{dlambdaE43} as functions of the dimensionless moment of inertia $\tilde{I}$. The top and bottom panels show the magnetic-led and the electric-led rotational Love numbers, respectively. Although the normalized rotational Love numbers are less sensitive to the EoS, the deviations from universality are much larger than in the nonspinning case. For example, the rotational Love numbers $\delta\tilde\lambda_{E}^{(23)}$ (which is related to the deformation of the mass quadrupole moment) shown in Fig.~\ref{fig:rot_universal} displays deviations at the level of $20\%$ and as large as $50\%$, to be compared with the deviations of $\tilde\lambda_{E}^{(2)}$ shown in Fig.~\ref{fig:lambdEBb} which are at the level of $1\%$ or better.

The electric-led case (bottom panels of Fig.~\ref{fig:rot_universal}) is even more dramatic, with $\delta\tilde\lambda_{M}^{(32)}$ displaying deviations larger than $200\%$. Furthermore, as shown in Fig.~\ref{fig:rot_VS_C}, in this case the Love number does not have a definite sign and the zero crossing is also EoS dependent. This poses a serious limitation to the construction of a fitting formula, since at the zero crossing the relative amplitude of $\delta\tilde\lambda_{M}^{(32)}$ for one EoS would be vanishingly small relative to the amplitude of the same quantity for a different EoS. This is the reason for the ``cusps'' shown in the right bottom panel of Fig.~\ref{fig:rot_universal}, which signal a zero crossing in a logarithmic scale.

For completeness, in Fig.~\ref{fig:rot_VS_C2} of Appendix~\ref{app:noequatorial} we present the same analysis but for the rotational Love numbers associated with the breaking of the equatorial symmetry. The results are qualitatively similar to those presented in this section. 

For simplicity, we only present here the fitting formula for the most relevant quantity, namely $\delta\tilde\lambda_{E}^{(23)}$; the fits for other rotational Love numbers are given in the {\scshape Mathematica}\textsuperscript{\textregistered} notebook in the Supplemental Material. The deviations shown in the right top panel of Fig.~\ref{fig:rot_universal} are relative to the following fit:
%%%%
\begin{equation}
 \log \delta\tilde\lambda_{E}^{(23)} = \sum_{i=0}^{10} a_i [\log\tilde{I}]^i\,, \label{fit}
\end{equation}
%%%%
where the constants $a_i$ are given in Table~\ref{tab:coefficients}. As shown in Fig.~\ref{fig:rot_universal}, for the EoS considered in this work (which cover a wide range of NS deformability) the deviations from the fit~\eqref{fit} are typically below $30\%$.

\begin{table}
 \begin{tabular}{rcrrcr}
  $a_0$	&=&	$-471.133$ &$\quad a_6$	&=&	$-49.0000$\\
  $a_1$	&=&	$+1358.55$ &$\quad a_7$	&=&	$+7.50706$\\
  $a_2$	&=&	$-1750.22$ &$\quad a_8$	&=&	$-0.74413$\\
  $a_3$	&=&	$+1327.18$ &$\quad a_9$	&=&	$+0.04314$\\
  $a_4$	&=&	$-651.555$ &$\quad a_{10}$&=&   $-0.00111$\\
  $a_5$	&=&	$+216.063$ &\\
 \end{tabular}
 \caption{Coefficients of the fit~\eqref{fit}.}\label{tab:coefficients}
\end{table}

Although the relation between the rotational Love numbers and the moment of inertia depends more strongly on the EoS than in the case of the static Love numbers, the quantities presented in Fig.~\ref{fig:rot_universal} are first-order corrections in the spin, i.e., they are subdominant relative to the static case. It is therefore interesting to quantify how this variability affects the approximate universality of the multipole moments of a NS. To this end, let us consider Eq.~\eqref{M2final}, which expresses the deformed mass quadrupole moment in the presence of an electric $\ell=2$ component and a magnetic $\ell=3$ component of the tidal field. By defining $\langle A \rangle$ as the relative spread of a quantity $A$ given by different EoS, from Eq.~\eqref{M2final} we obtain 
%%%%
\begin{eqnarray}
 \langle \frac{M_2}{M^3}\rangle &=& \langle \tilde\lambda_E^{(2)}\rangle +\chi \langle \delta\tilde\lambda_E^{(23)}\rangle \left(\frac{\delta\tilde\lambda_E^{(23)}}{\tilde\lambda_E^{(2)}}\right)\frac{{\cal B}_0^{(3)} M}{{\cal E}_0^{(2)}} \nn\\
 %%%%
 &\sim&1\% +5\%\left(\frac{\chi}{0.05}\right)\left(\frac{\delta\tilde\lambda_E^{(23)}}{30 \tilde\lambda_E^{(2)}}\right) \left(\frac{C}{0.2}\right)^{3/2}\left(\frac{5R}{r_0}\right)^{3/2}\,,\nn\\
\end{eqnarray}
%%%%
where in the last step we used Eqs.~\eqref{E2B2} and~\eqref{E3B3}, the estimates $\langle \tilde\lambda_E^{(2)}\rangle\sim 1\%$, $\langle \delta\tilde\lambda_E^{(23)}\rangle \sim 40\%$ and $\delta\tilde\lambda_E^{(23)}\sim 30 \tilde\lambda_E^{(2)}$ [cf. Fig.~\ref{fig:ratio}]. The values for $\chi$, $r_0$ and $C$ adopted in the equation above refer to a typical NS-NS merger.
The contribution of $\delta\tilde\lambda_E^{(23)}$ is larger as the binary approaches the merger and the approximate universality of the induced mass quadrupole moment deteriorates from $1\%$ to roughly $6\%$ due to spin corrections. Such corrections scale as $\chi (R/r_0)^{3/2}$, so they would be more relevant for fastly spinning NSs and negligible at a large orbital distance.

% 
%%%%%%%%%%%%%%%%%%%%%%%%%%%%%%%%%%%%%%%%%%%%%%%%%%%%%%%%%%%%%%%%%%%%%%
\section{Conclusions and prospects} \label{sec:conclusions}
%%%%%%%%%%%%%%%%%%%%%%%%%%%%%%%%%%%%%%%%%%%%%%%%%%%%%%%%%%%%%%%%%%%%%%
By extending the formalism developed in Paper~I we have computed, for the first time to our knowledge, the tidal Love numbers of a spinning NS to first order in the angular moment. The spin of the object introduces couplings between electric and magnetic distortion and new classes of rotational tidal Love numbers emerge.
We have derived the equations governing electric-led and magnetic-led tidal perturbations for generic multipoles and have explicitly solved them for various tabulated EoS [cf. Table~\ref{tab:EoS}] for a tidal field with both an electric and a magnetic component with $\ell=2,3,4$.

We found various interesting features; for instance, some rotational Love numbers are nonmonotonic functions of the compactness and do not have a definite sign. For a binary system close to the merger, various components of the tidal field become relevant. In this case we find that an octupolar magnetic tidal field can significantly modify the mass quadrupole moment of a neutron star. Preliminary estimates, assuming a spin parameter $\chi\approx0.05$ and an orbital distance $r_0\approx 5 R$, show modifications $\gtrsim 10\%$ relative to the static case.  Our results suggest that spin-tidal couplings can introduce important corrections to the gravitational waveforms of a spinning NS binary system.

Furthermore, some of these new rotational Love numbers deviate from a nearly universal relation~\cite{Yagi:2013bca,Yagi:2013awa,Yagi:2013sva} by an amount as large as $200\%$ in the electric-led case and as large as $50\%$ in the magnetic-led case. 
By using these results, we estimated that --~for a NS binary approaching the merger~-- the approximate universality of the induced mass quadrupole moment deteriorates from $1\%$ in the static case to roughly $6\%$ when $\chi\approx0.05$ and $r_0\approx 5R$. Clearly, the effect of the spin would be more relevant for NSs with $\chi\gtrsim 0.05$ and it can only be relevant in the final phase of the inspiral.

The phenomenology of the tidal deformations of a spinning object is very rich and much work remains to be done.
From the technical side, it is natural to extend our study to the nonaxisymmetric case~\cite{Landry:2015zfa}, which would introduce a further class of rotational Love numbers emerging from novel couplings between nonaxisymmetric perturbations [cf. diagram in Fig.~\ref{fig:scheme}]. Likewise, it would be interesting (albeit technically challenging) to include quadratic corrections in the NS angular momentum, thus also extending our analysis in Paper~I to the NS case.

From the phenomenological side, it is important to understand spin-tidal effects for the gravitational waveforms. This is particularly urgent given that NS binaries are the main target of advanced gravitational-wave detectors. Our results suggest that spin-tidal effects might be important to improve gravitational-wave templates~\cite{Bernuzzi:2012ci,Read:2013zra,Buonanno:2014aza} and to estimate precisely the tidal Love numbers through gravitational-wave detections~\cite{Flanagan:2007ix,Hinderer:2007mb,Hinderer:2009ca,Baiotti:2010xh,Baiotti:2011am,Vines:2011ud,Pannarale:2011pk,Lackey:2011vz,Damour:2012yf,Vines:2010ca,Lackey:2013axa,Favata:2013rwa,Yagi:2013baa}.

We have shown that the main spin-tidal contribution to first order in the spin and in the axisymmetric case comes from a deformation of the mass quadrupole moment of the NS induced by an external octupolar tidal field. For a binary system, such an effect is negligible at a large orbital distance but becomes more relevant close to the merger. Therefore, a solid qualitative outcome of our analysis is that spin-tidal effects are more important for high-frequency signals, possibly those in which the merger phase happens within the detector band.

A promising approach to incorporate such effects in the relativistic two-body dynamics is by extending effective point-particle techniques (cf. e.g. Refs.~\cite{Damour:1991yw,Porto:2005ac,Porto:2008jj}) to include spin-tidal couplings. Work in this direction is underway. 

Finally, in Paper~I we discussed some subtleties in the definition of the tidal Love number for a relativistic compact object (see also Refs.~\cite{Damour:2009va,Kol:2011vg}). In brief, in a nonlinear theory the very notion of the multipole moments of a \emph{single} deformed compact object is problematic. We believe that a more rigorous analysis is needed to clarify this important issue, even in the static case. This would likely require a (at least partial) fifth order post-Newtonian expansion of the field equations for a binary system, which is currently not available for generic mass ratios, in order to determine the quantities which actually appear in the post-Newtonian gravitational waveforms.

%%%%%%%%%%%%%%%%%%%%%%%%%%%%%%%%%%%%%%%%%%%%%%%%%%%%%%%%%%%%%%%%%%%%%%
\begin{acknowledgments}
%%%%%%%%%%%%%%%%%%%%%%%%%%%%%%%%%%%%%%%%%%%%%%%%%%%%%%%%%%%%%%%%%%%%%%
We are indebted to Nathan Johnson-McDaniel for pointing out to us the results of  Ref.~\cite{2009PhRvD..80l4039J}, which have been useful to improve the accuracy of our estimates.
We thank Emanuele Berti, T\'erence Delsate, Tanja Hinderer, Davide Gerosa, Phil Landry, Giorgos Pappas, Eric Poisson, Luciano Rezzolla, Jan Steinhoff and Kent Yagi for useful discussions and correspondence.
P.P. was supported by the European Community through
the Intra-European Marie Curie Contract No.~AstroGRAphy-2013-623439 and by FCT-Portugal through the project IF/00293/2013.
This work was partially supported by ``NewCompstar'' (COST action MP1304) and by the NRHEP 295189 FP7-PEOPLE-2011-IRSES
Grant. 
\end{acknowledgments}
%%%%%%%%%%%%%%%%%%%%%%%%%%%%%%%%%%%%%%%%%%%%%%%%%%%%%%%%%%%%%%%%%%%%%%%%%%%%%%
% 
% 
\appendix

%%%%%%%%%%%%%%%%%%%%%%%%%%%%%%%%%%%%%%%%%%%%%%%%%%%%%%%%%%%%%%%%%%%%%%%%%%%%%%%%
\section{Stationary perturbations of a slowly rotating relativistic object}\label{app:Kojima}
%%%%%%%%%%%%%%%%%%%%%%%%%%%%%%%%%%%%%%%%%%%%%%%%%%%%%%%%%%%%%%%%%%%%%%%%%%%%%%%%
As a background, we consider the spinning geometry~\eqref{metric} to first order in the rotation rate, with fluid velocity
$u^\mu=e^{-\nu/2}(1,0,0,\Omega)$ and stress-energy tensor $T^{\mu\nu}$ given in Eq.~\eqref{Tmunu}~(see Sec.~\ref{sec:spback}).
We perform a harmonic decomposition of the (stationary) metric perturbations of this background in the Regge-Wheeler gauge~\cite{Regge:1957td} as
%%%
\begin{equation}
 \delta g_{\mu\nu}(t,r,\vartheta,\varphi)=\delta g_{\mu\nu}^{\rm odd}(r,\vartheta,\varphi)+\delta g_{\mu\nu}^{\rm even}(r,\vartheta,\varphi)
\end{equation}
%%%
with
\begin{equation}\label{oddpart}
\delta g_{\mu\nu}^{\rm odd} =
 \begin{pmatrix}
  0 & 0 & h_0^{(\ell)} S_\vartheta^{\ell} & h_0^{(\ell)} S_\varphi^{\ell} \\
  * & 0 & h_1^{(\ell)} S_\vartheta^{\ell} & h_1^{(\ell)} S_\varphi^{\ell} \\
  *  & *  & 0 & 0  \\
  * & * & * & 0
 \end{pmatrix}\,,
\end{equation}
%%%%%%%%%%%%%%%%%%%%%%
\begin{equation}\label{evenpart}
\delta g_{\mu\nu}^{\rm even}=
\begin{pmatrix}
g_{tt}^{(0)} H_0^{(\ell)}  & H_1^{(\ell)} & 0 & 0\\
  * & g_{rr}^{(0)} H_2^{(\ell)}  &  0 & 0\\
  *  & *  & r^2 K^{(\ell)}  & 0  \\
  * & * & * & r^2\sin^2\vartheta K^{(\ell)} 
 \end{pmatrix}Y^\ell\,,
\end{equation}
%%%%
%
where asterisks represent symmetric components, $g_{tt}^{(0)}=e^\nu$,
$1/g_{rr}^{(0)}-1-2{\cal M}/r$, $Y^{\ell}=Y^{\ell}(\vartheta,\varphi)$ are the scalar spherical harmonics and we have
defined the vector spherical harmonics as
\begin{eqnarray}
(S_\vartheta^{\ell},S_\varphi^{\ell})&\equiv&\left(-\frac{Y^{\ell}_{,\varphi}}{\sin\vartheta}
,\sin\vartheta Y^{\ell}_{,\vartheta}\right)\,. \label{vecspheharm}
\end{eqnarray}
Here and in the following, a sum over the harmonic indices $\ell\ge1$
and $m$ (such that $|m|\leq\ell$) is implicit.
%%%%%%%%%%
The perturbation of the stress-energy tensor reads~\cite{Kojima:1992ie}
%%%
\begin{eqnarray}
  \delta T^{\mu\nu}&=&(P+\rho)(u^\mu\delta u^\nu+u^\nu\delta u^\mu)+(\delta P^{(\ell)}+\delta\rho^{(\ell)})Y^\ell u^\mu u^\nu\nonumber\\
  &&+\delta g^{\mu\nu}P+\delta P^{(\ell)} Y^\ell g^{\mu\nu}\,,
% T^{\mu\nu}=(P+\delta p^{(\ell)}Y^\ell+\rho+\delta\rho^{(\ell)} Y^\ell)u^\mu u^\nu+g^{\mu\nu}(P+ \delta p^{(\ell)} Y^\ell)\,,
\end{eqnarray}
%%%
where $\delta P^{(\ell)}$ and $\delta\rho^{(\ell)}$ are radial functions. The perturbation of the fluid velocity is
%%%%%%%%%%
\begin{eqnarray}
 \delta u^t &=& e^{-\frac{\nu }{2}}\frac{ {\bar \omega} (U^{(\ell)} S_\varphi+V^{(\ell)} Y^\ell_{,\varphi})}{\kappa  (P+\rho )}\,,\\
 \delta u^r &=&   e^{\frac{\nu }{2}-\lambda}\frac{R^{(\ell)} Y^\ell}{\kappa  (P+\rho)}\,,\label{ut}\\
\delta u^\vartheta &=&  e^{\frac{\nu }{2}}\frac{U^{(\ell)} S_\vartheta+V^{(\ell)} Y^\ell_{,\vartheta}}{\kappa  r^2 (P+\rho )}\,,\\
\delta u^\varphi &=&e^{\frac{\nu }{2}}\frac{U^{(\ell)} S_\varphi+V^{(\ell)} Y^\ell_{,\varphi}}{\kappa  r^2 \sin^2\vartheta (P+\rho )}\,,\label{uphi}
\end{eqnarray}
%%%%%
where $U^{(\ell)}$, $V^{(\ell)}$, $R^{(\ell)}$ are purely radial functions. The function $\delta u^t$ has been fixed to ensure $u^2=-1$ to first
order in the perturbations and in the spin. As discussed in the main text, we restrict to a static fluid and therefore
impose $\delta u^\mu=0$ in the above decomposition.
%%%%%%%%%%
%%%%%%%%%%%%%%%%%%%%%%%%%%%%%%%%%%%%%%%%
\section{Coefficients and sources for the inhomogeneous equations governing tidal deformations}\label{app:sources}
%%%%%%%%%%%%%%%%%%%%%%%%%%%%%%%%%%%%%%%%

The coefficients of the ($\ell\ge2$) differential operators appearing in the systems~\eqref{polar_led2} and \eqref{axial_led2} read
%%%%
\begin{eqnarray}
 C_1^{(\ell)}&=& \frac{r \left(\kappa  r^2 (P-\rho )+2\right)-2 {\cal M}}{r (r-2 {\cal M})}\,,\\
 %%%%
 C_0^{(\ell)}&=& \left[{r^2 {c_s^2} (r-2
   {\cal M})^2}\right]^{-1}\left[-\ell(\ell+1) r {c_s^2} (r-2 {\cal M})\right.\nn\\
  && \left.-{c_s^2} \left(2 \kappa  r^3 {\cal M} (13
   P+5 \rho )+4 {\cal M}^2\right.\right.\nn\\
   &&\left.\left.+\kappa  r^4 \left(4 \kappa  r^2 P^2-9 P-5 \rho
   \right)\right)\right.\nn\\
   &&\left.+\kappa  r^3 (r-2 {\cal M}) (P+\rho)\right]\,,\nn \\ \\
 %%%%  
 {C_1^\ast}^{(\ell)}&=& -\frac{\kappa  r^2 (P+\rho )}{r-2 {\cal M}}\,,\\
 %%%%   
 {C_0^\ast}^{(\ell)}&=& -\frac{1}{r(r-2{\cal M}}\left[\ell(\ell+1)-\frac{4{\cal M}}{r}\right.\nn\\
   &&\left.+2\kappa r^2(P+\rho)\right] \,,
\end{eqnarray}
%%%%
where we recall that ${c_s}=\sqrt{\partial P/\partial\rho}$ is the speed of sound in the fluid. The source terms read
\begin{widetext}
\begin{eqnarray}
 {S^\ast_+}^{(\ell)} &=& \frac{1}{(\ell(\ell+1)-2) \sqrt{4 \ell (\ell+2)+3}
   {c_s^2} (r-2 {\cal M})^2}  \left[\kappa\left(2-\ell(\ell+1)\right) r^3 (r-2 {\cal M}) (P+\rho )
   ({\omega_1}+\Omega ) {H_0^{(\ell)}}\right.\nn\\
   &&\left.-{c_s^2}
   \left({H_0^{(\ell)}} \left(2 r {\cal M} \left(-\kappa  r^2 \rho  \left(\left(5
   {\ell(\ell+1)}-22\right) \Omega +(\ell (5 \ell+9)-6) {\omega_1}\right)-2 \left(\ell
   \left({\ell(\ell+1)}-4\right)-2\right) (\Omega -{\omega_1})\right.\right.\right.\right.\nn\\
   &&\left.\left.\left.\left.+\kappa  r^2 P
   \left(\Omega  \left(3 \ell (\ell+5)+32 \kappa  r^2 \rho +14\right)+(14-\ell (13 \ell+25))
   {\omega_1}\right)+24 \kappa ^2 r^4 \Omega  P^2+8 \kappa ^2 r^4 \Omega 
   \rho ^2\right)\right.\right.\right.\nn\\
   &&\left.\left.\left.+r^2 \left(\kappa  r^2 \Omega  \left(\rho  \left(5 {\ell(\ell+1)}+16
   \kappa  r^2 P \left(\kappa  r^2 P-1\right)-14\right)+P \left(4 \kappa  r^2
   P \left(\ell (\ell+3)+4 \kappa  r^2 P-2\right)+(\ell-3) \ell-6\right)\right.\right.\right.\right.\right.\nn\\
   &&\left.\left.\left.\left.\left.-8 \kappa  r^2 \rho
   ^2\right)+{\omega_1} \left(\kappa  r^2 \left(P \left(-4 \kappa  \ell
   (\ell+3) r^2 P+\ell (9 \ell+13)-14\right)+(\ell (5 \ell+9)-6) \rho \right)-2 \ell
   (\ell(\ell+1)-2)\right)\right.\right.\right.\right.\nn\\
   &&\left.\left.\left.\left. +2 \ell (\ell(\ell+1)-2) \Omega \right)+4 {\cal M}^2
   \left(((\ell-1) \ell-4) (\Omega -{\omega_1})-4 \kappa  r^2 \Omega  (P+\rho
   )\right)\right)\right.\right.\nn\\
   &&\left.\left.+2 r (r-2 {\cal M}) {H_0^{(\ell)}}' \left((\ell(\ell+1)-2) r
   ({\omega_1}-\Omega )+{\cal M} \left((\ell (3 \ell+5)-4) (\Omega -{\omega_1})+4 \kappa  r^2 \Omega  (P+\rho )\right)\right.\right.\right.\nn\\
   &&\left.\left.\left.+\kappa  r^3 P \left(\ell (\ell+3)
   (\Omega -{\omega_1})+4 \kappa  r^2 \Omega  \rho \right)+4 \kappa ^2
   r^5 \Omega  P^2\right)\right)\right]\,,
   \end{eqnarray}
   \begin{eqnarray}
   %%%%%%%%%%%%%%%%%%%%%%%%%%%%%%
 {S^\ast_-}^{(\ell)} &=& \frac{1}{(\ell(\ell+1)-2) \sqrt{4 \ell^2-1} {c_s^2} (r-2 {\cal M})^2}\left[ {c_s^2}\left(2 r (r-2 {\cal M}) {H_0^{(\ell)}}' \left({\cal M} \left(\left(3
   {\ell(\ell+1)}-6\right) (\Omega -{\omega_1})+4 \kappa  r^2 \Omega (P+\rho)\right)\right.\right.\right.\nn\\
   &&\left.\left.\left.+r \left(\kappa  r^2 P \left(\left({\ell(\ell-1)}-2\right)
   (\Omega -{\omega_1})+4 \kappa  r^2 \Omega  \rho
   \right)-(\ell(\ell+1)-2) (\Omega -{\omega_1})+4 \kappa ^2 r^4
   \Omega  P^2\right)\right) \right.\right.\nn\\
   &&\left.\left.+{H_0^{(\ell)}} \left(4 {\cal M}^2 \left(\left(\ell^2+3
   \ell-2\right) (\Omega -{\omega_1})-4 \kappa  r^2 \Omega  P-4 \kappa  r^2
   \Omega  \rho \right)+2 r {\cal M} \left(\kappa  r^2 P \left(-\left(13
   {\ell(\ell+1)}-26\right) {\omega_1} \right.\right.\right.\right.\right.\nn\\
   &&\left.\left.\left.\left.\left. +\left(3 \ell^2-9 \ell+2\right) \Omega +32 \kappa 
   r^2 \Omega  \rho \right)-\kappa  r^2 \rho  \left(\left(5 {\ell(\ell+1)}-10\right)
   {\omega_1}+\left(5 \ell^2+9 \ell-18\right) \Omega \right) \right.\right.\right.\right.\nn\\
   &&\left.\left.\left.\left. +2 \left(\ell^3+2 \ell^2-3
   \ell-2\right) (\Omega -{\omega_1})+24 \kappa ^2 r^4 \Omega  P^2+8 \kappa
   ^2 r^4 \Omega  \rho ^2\right)+r^2 \left(\kappa  r^2 P \left(\left(9 \ell^2+5
   \ell-18\right) {\omega_1}\right.\right.\right.\right.\right.\nn\\
   &&\left.\left.\left.\left.\left. +\left(\ell^2+5 \ell-2\right) \Omega -16 \kappa  r^2
   \Omega  \rho \right)+4 \kappa ^2 r^4 P^2 \left(\left(-{\ell(\ell+1)}+2\right)
   {\omega_1}+\left({\ell(\ell-1)}-4\right) \Omega +4 \kappa  r^2 \Omega  \rho
   \right)\right.\right.\right.\right. \nn\\
   &&\left.\left.\left.\left. +\kappa  r^2 \rho  \left(\left(5 {\ell(\ell+1)}-10\right) {\omega_1}+\left(5 \ell^2+9 \ell-10\right) \Omega \right)-2 \left(\ell^3+2 {\ell(\ell-1)}-2\right)
   (\Omega -{\omega_1})+16 \kappa ^3 r^6 \Omega  P^3-8 \kappa ^2 r^4
   \Omega  \rho ^2\right)\right)\right)\right.\nn\\
   &&\left.+\kappa  (\ell(\ell+1)-2) r^3 (r-2 {\cal M})    (P+\rho ) ({\omega_1}+\Omega )
   {H_0^{(\ell)}}\right]\,,
      \end{eqnarray}
   \begin{eqnarray}
   %%%%%%%%%%%%%%%%%%%%%%%%%%%%%%
 {S}^{(\ell)}_+ &=&  \frac{2 e^{-\nu }}{\sqrt{4 \ell^2+8 \ell+3} r^2 {c_s^2} (r-2{\cal M})^2} \left[{c_s^2} \left(r (r-2 {\cal M}) {h_0^{(\ell)}}'
   \left(-2 {\cal M} \left(r {\omega_1}' \left({\ell(\ell+1)}+2 \kappa  r^2 P+2 \kappa
    r^2 \rho +3\right)\right.\right.\right.\right.\nn\\
    &&\left.\left.\left.\left.+{\omega_1} \left(3 \ell(\ell+1)-4 \kappa  r^2 P-4
   \kappa  r^2 \rho \right)-3 \ell(\ell+1) \Omega \right)+r \left(2 \kappa  r^2 P
   \left(\ell(\ell+1) \Omega -{\omega_1} \left({\ell(\ell+1)}-4 \kappa  r^2 \rho
   \right)\right)+\ell^2 r {\omega_1}'\right.\right.\right.\right.\nn\\
   &&\left.\left.\left.\left.-2 \ell^2 \Omega +\ell r {\omega_1}'+2 \ell(\ell+1) {\omega_1}-2 \ell \Omega -2 \kappa ^2 r^4 P^2 \left(r
   {\omega_1}'-4 {\omega_1}\right)+2 \kappa  r^3 \rho 
   {\omega_1}'\right)+10 {\cal M}^2 {\omega_1}'\right)\right.\right.\nn\\
   &&\left.\left.+{h_0^{(\ell)}} \left(r^2 \left(\kappa  r^2 P \left(2
   {\omega_1} \left({\ell(\ell+1)}+12 \kappa  r^2 \rho \right)-2 {\omega_1}' \left(\ell r-4 \kappa  r^3 \rho +r\right)-5 \ell(\ell+1) \Omega \right)+\kappa 
   r^2 \rho  \left(2 \ell(\ell+1) {\omega_1}\right.\right.\right.\right.\right.\nn\\
   &&\left.\left.\left.\left.\left.-5 \ell(\ell+1) \Omega -4 r
   {\omega_1}'\right)+2 \ell(\ell+1)^2 (\Omega -{\omega_1})-20 \kappa
   ^3 r^6 P^3 {\omega_1}-4 \kappa ^2 r^4 P^2 \left({\omega_1} \left(5 \kappa  r^2 \rho -4\right)-3 r {\omega_1}'\right)+8
   \kappa ^2 r^4 \rho ^2 {\omega_1}\right)\right.\right.\right.\nn\\
   &&\left.\left.\left.-2 r {\cal M} \left(-2 \ell^3
   {\omega_1}+2 \ell^3 \Omega +\kappa  r^2 P \left(2 {\omega_1}
   \left({\ell(\ell+1)}+22 \kappa  r^2 \rho +2\right)+2 r {\omega_1}' \left(4 \kappa  r^2 \rho -5-\ell\right)-5 \ell(\ell+1) \Omega \right)\right.\right.\right.\right.\nn\\
   &&\left.\left.\left.\left.+\kappa  r^2 \rho  \left(2
   \left({\ell(\ell+1)}+2\right) {\omega_1}-5 \ell(\ell+1) \Omega -12 r {\omega_1}'\right)-2 \ell^2 {\omega_1}+2 \ell^2 \Omega +\ell r {\omega_1}'+12 \kappa ^2 r^4 P^2 \left(r {\omega_1}'+3 {\omega_1}\right)\right.\right.\right.\right.\nn\\
   &&\left.\left.\left.\left.+8 \kappa ^2 r^4 \rho ^2 {\omega_1}-5 r {\omega_1}'\right)-4 {\cal M}^2 \left(r {\omega_1}' \left(-\ell+8 \kappa  r^2 P+8
   \kappa  r^2 \rho +10\right)+{\omega_1} \left(-2 \ell(\ell+1)+\kappa  r^2
   P+\kappa  r^2 \rho \right)+2 \ell(\ell+1) \Omega \right)\right.\right.\right.\nn\\
   &&\left.\left.\left.+40 {\cal M}^3 {\omega_1}'\right)\right)-\kappa  r^2 (P+\rho ) {h_0^{(\ell)}} \left(-2 r {\cal M}
   \left(\ell(\ell+1) \Omega -4 \kappa  r^2 P {\omega_1}\right)+r^2 \left(\ell
   (\ell+1) \Omega +4 \kappa ^2 r^4 P^2 {\omega_1}\right)+4 {\cal M}^2
   {\omega_1}\right)\right]\,,
      \end{eqnarray}
   \begin{eqnarray}
 %%%%%%%%%%%%%%%%%%%%%%%%%%%%%%
 {S}^{(\ell)}_- &=& \frac{2 e^{-\nu }}{\sqrt{4 \ell^2-1} r^2 {c_s^2} (r-2 {\cal M})^2} \left[{c_s^2} \left({h_0^{(\ell)}} \left(r^2
   \left(-\kappa  r^2 P \left(2 {\omega_1} \left({\ell(\ell+1)}+12 \kappa  r^2
   \rho \right)+2 r {\omega_1}' \left(\ell+4 \kappa  r^2 \rho \right)-5 \ell
   (\ell+1) \Omega \right)\right.\right.\right.\right.\nn\\
   &&\left.\left.\left.\left.+2 \ell^2 (\ell+1) (\Omega -{\omega_1})+\kappa  r^2 \rho
    \left(-2 \ell(\ell+1) {\omega_1}+5 \ell(\ell+1) \Omega +4 r {\omega_1}'\right)+20 \kappa ^3 r^6 P^3 {\omega_1}\right.\right.\right.\right. \nn\\
    &&\left.\left.\left.\left.+4 \kappa ^2 r^4 P^2
   \left({\omega_1} \left(5 \kappa  r^2 \rho -4\right)-3 r {\omega_1}'\right)-8 \kappa ^2 r^4 \rho ^2 {\omega_1}\right)+2 r {\cal M}
   \left(2 \ell^3 {\omega_1}-2 \ell^3 \Omega +\kappa  r^2 P \left(2
   {\omega_1} \left({\ell(\ell+1)}+22 \kappa  r^2 \rho +2\right)\right.\right.\right.\right.\right.\nn\\
   &&\left.\left.\left.\left.\left.+2 r {\omega_1}' \left(\ell+4 \kappa  r^2 \rho -4\right)-5 \ell(\ell+1) \Omega \right)+\kappa 
   r^2 \rho  \left(2 \left({\ell(\ell+1)}+2\right) {\omega_1}-5 \ell(\ell+1) \Omega -12
   r {\omega_1}'\right)+4 \ell^2 (\omega_1-\Omega) \right.\right.\right.\right. \nn\\
   &&\left.\left.\left.\left.-\ell r
   {\omega_1}'+2 \ell {\omega_1}-2 \ell \Omega +12 \kappa ^2 r^4 P^2
   \left(r {\omega_1}'+3 {\omega_1}\right)+8 \kappa ^2 r^4 \rho
   ^2 {\omega_1}-6 r {\omega_1}'\right)+4 {\cal M}^2 \left(r
   {\omega_1}' \left(\ell+8 \kappa  r^2 P+8 \kappa  r^2 \rho
   +11\right)\right.\right.\right.\right.\nn\\
   &&\left.\left.\left.\left.+{\omega_1} \left(-2 \ell(\ell+1)+\kappa  r^2 P+\kappa  r^2
   \rho \right)+2 \ell(\ell+1) \Omega \right)-40 {\cal M}^3 {\omega_1}'\right)-r
   (r-2 {\cal M}) {h_0^{(\ell)}}' \left(-2 {\cal M} \left(r {\omega_1}'
   \left({\ell(\ell+1)}\right.\right.\right.\right.\right.\nn\\
   &&\left.\left.\left.\left.\left.+2 \kappa  r^2 (P+\rho) +3\right)+{\omega_1}
   \left(3 \ell(\ell+1)-4 \kappa  r^2 P-4 \kappa  r^2 \rho \right)-3 \ell(\ell+1) \Omega
   \right)\right.\right.\right.\nn\\
   &&\left.\left.\left.+r \left(2 \kappa  r^2 P \left(\ell(\ell+1) \Omega -{\omega_1}
   \left({\ell(\ell+1)}-4 \kappa  r^2 \rho \right)\right)+\ell^2 r {\omega_1}'-2 \ell^2
   \Omega +\ell r {\omega_1}'+2 \ell(\ell+1) {\omega_1}-2 \ell \Omega\right.\right.\right.\right.\nn\\
   &&\left.\left.\left.\left.-2
   \kappa ^2 r^4 P^2 \left(r {\omega_1}'-4 {\omega_1}\right)+2
   \kappa  r^3 \rho  {\omega_1}'\right)+10 {\cal M}^2 \omega_1'\right)\right)+\kappa  r^2 (P+\rho ) {h_0^{(\ell)}} \left(-2 r {\cal M}
   \left(\ell(\ell+1) \Omega -4 \kappa  r^2 P {\omega_1}\right)\right.\right.\nn\\
   &&\left.\left.+r^2 \left(\ell
   (\ell+1) \Omega +4 \kappa ^2 r^4 P^2 {\omega_1}\right)+4 {\cal M}^2
   {\omega_1}\right)\right]\,.
\end{eqnarray}
\end{widetext}
%%%%%
\subsection{Other metric components}
%%%%%%
In their reduced form, the electric-led and magnetic-led systems only depend on the variables $H_0^{(\ell)}$, $\delta h_0^{(\ell\pm1)}$ and $h_0^{(\ell)}$, $\delta H_0^{(\ell\pm1)}$, respectively. Once such quantities are computed by solving the corresponding system of ODEs, all other nonvanishing metric components can be obtained through algebraic relations. For completeness we give such relations here.

To zero order in the spin, the nonvanishing components of the polar sector can be algebraically written in terms of the function $H_0^{(\ell)}$ and its derivatives:
%%%
\begin{eqnarray}
 K^{(\ell)}&=&\frac{{H_0^{(\ell)}}}{(\ell(\ell+1)-2) r (r-2 {\cal M})}\left[ \left(2 r {\cal M} \left(-{L(\ell-1)}\right.\right.\right.\nn\\
 &&\left.\left.\left.+6 \kappa  r^2 P+2 \kappa  r^2
   \rho +4\right)+r^2 \left({\ell(\ell+1)}+4 \kappa ^2 r^4 P^2\right.\right.\right.\nn\\
   &&\left.\left.\left.-2 \kappa  r^2 P-2
   \kappa  r^2 \rho -2\right)-4 {\cal M}^2\right)\right.\nn\\
   &&\left.+2 r (r-2 {\cal M}) \left({\cal M}+\kappa  r^3
   P\right) {H_0^{(\ell)}}'\right]\,,\\
   %%%%%%%
 H_2^{(\ell)}&=&H_0^{(\ell)}\,, \\
 \delta p^{(\ell)}&=&\frac{1}{2} (P+\rho ) {H_0^{(\ell)}}\,,
\end{eqnarray}
%%%
and we stress that $H_1^{(\ell)}=0$ as a results of the field equations in the stationary case. Likewise, $h_1^{(\ell)}=0$, so that the only axial variable in the stationary case to zeroth order in the spin is $h_0^{(\ell)}$. 

To first order in the spin, the nonvanishing perturbation variables which are algebraically related to the dynamical ones are
\begin{widetext}
%%%
\begin{eqnarray}
 \delta K^{(\ell+1)}&=&  -\frac{\kappa  r^2}{\ell(\ell+3)} \left[\frac{{\cal M} e^{-\nu } }{\kappa  \sqrt{4 \ell^2+8 \ell+3} r^2}\left[2 {h_0^{(\ell)}}' \left(r
   {\omega_1}' \left({\ell(\ell+1)}+4 \kappa  r^2 P\right)+2 \ell(\ell+1) {\omega_1}-2 \ell(\ell+1) \Omega \right) \right.\right.\nn\\
   &&\left.\left. -2 \sqrt{4 \ell^2+8 \ell+3} e^{\nu }
   {{\delta H_0}^{(\ell+1)}}'\right]-\frac{e^{-\nu }}{\kappa  \sqrt{4 \ell^2+8 \ell+3}}
   \left[2 \kappa  r P \left(2 \ell(\ell+1) (\Omega -{\omega_1})
   {h_0^{(\ell)}}'+\sqrt{4 \ell^2+8 \ell+3} e^{\nu } {{\delta H_0}^{(\ell+1)}}'\right) \right.\right.\nn\\
   &&\left.\left.-4
   \kappa ^2 r^4 P^2 {\omega_1}' {h_0^{(\ell)}}'+\ell(\ell+1){\omega_1}' {h_0^{(\ell)}}'\right]+\frac{4 {\cal M}^2
   e^{-\nu } {\omega_1}' {h_0^{(\ell)}}'}{\kappa  \sqrt{4 \ell^2+8 \ell+3}
   r^2}+\frac{4 \ell(\ell+1) \Omega  e^{-\nu } (P+\rho )
   {h_0^{(\ell)}}}{\sqrt{4 \ell^2+8 \ell+3}}\right.\nn\\
   &&\left.-\frac{2 e^{-\nu } {h_0^{(\ell)}}}{\kappa  \sqrt{4 \ell^2+8 \ell+3} r^3 (r-2
   {\cal M})}
   \left[-2 {\cal M}^2 \left(-4 {\omega_1} \left({\ell(\ell+1)}-\kappa  r^2 P-\kappa 
   r^2 \rho \right)+r {\omega_1}' \left(-{\ell(\ell+1)}+8 \kappa  r^2
   P-2\right)\right.\right.\right.\nn\\
   &&\left.\left.\left.+4 \ell(\ell+1) \Omega \right)-r^2 \left({\omega_1} \left(\ell^2
   \left(\ell^2+4 \ell+3\right)+8 \kappa ^3 r^6 P^2 \rho +8 \kappa ^3 r^6
   P^3\right)-\left(\ell^2+4 \ell+3\right) \ell^2 \Omega \right.\right.\right. \nn\\
   &&\left.\left.\left.+{\omega_1}' \left(\kappa
    \ell (\ell+3) r^3 P+2 \ell r-4 \kappa ^2 r^5 P^2\right)\right)+r {\cal M} \left(r
   {\omega_1}' \left(2 \kappa  \left(\ell^2+3 \ell+4\right) r^2 P-(\ell-5) \ell-8
   \kappa ^2 r^4 P^2\right)\right.\right.\right.\nn\\
   &&\left.\left.\left.+2 {\omega_1} \left(\ell \left(\ell^3+4
   {\ell(\ell+1)}-2\right)-8 \kappa ^2 r^4 P \rho -8 \kappa ^2 r^4 P^2\right)-2 \ell
   \left(\ell^3+4 {\ell(\ell+1)}-2\right) \Omega \right)-8 {\cal M}^3 {\omega_1}'\right]\right.\nn\\
   &&\left.-\frac{{\delta H_0}^{(\ell+1)} \left(2 r {\cal M} \left(-\ell^2-3 \ell+4 \kappa  r^2
   P+2\right)+r^2 \left(\ell (\ell+3)+4 \kappa ^2 r^4 P^2\right)-4
   {\cal M}^2\right)}{\kappa  r^3 (r-2 {\cal M})}+2 (P+\rho) {\delta H_0}^{(\ell+1)}\right]  \,,
   \end{eqnarray}
   \begin{eqnarray}
%    %%%%%%%
 \delta K^{(\ell-1)}&=&   -\frac{e^{-\nu }}{\left({\ell(\ell-1)}-2\right) \sqrt{4 \ell^2-1} r (r-2
   {\cal M})} \left[-r (r-2 {\cal M}) \left(2 {\cal M} \left({h_0^{(\ell)}}' \left(r
   {\omega_1}' \left({\ell(\ell+1)}+4 \kappa  r^2 P\right)+2 \ell(\ell+1) (\omega_1-\Omega)\right)\right.\right.\right.\nn\\
   &&\left.\left.\left.+\sqrt{4 \ell^2-1} e^{\nu }
   {\delta H_0^{(\ell-1)}}'\right)+r^2 \left(-2 \kappa  r P \left(2 \ell(\ell+1) (\Omega
   -{\omega_1}) {h_0^{(\ell)}}'-\sqrt{4 \ell^2-1} e^{\nu }
   {\delta H_0^{(\ell-1)}}'\right)+4 \kappa ^2 r^4 P^2 {\omega_1}'
   {h_0^{(\ell)}}'\right.\right.\right.\nn\\
   &&\left.\left.\left.-\ell(\ell+1) {\omega_1}' {h_0^{(\ell)}}'\right)+4 {\cal M}^2
   {\omega_1}' {h_0^{(\ell)}}'\right)+2 {h_0^{(\ell)}} \left(2 {\cal M}^2
   \left(4 {\omega_1} \left({\ell(\ell+1)}-\kappa  r^2 P-\kappa  r^2 \rho
   \right)+r {\omega_1}' \left(\ell^2+3 \ell-8 \kappa  r^2 P+4\right)\right.\right.\right.\nn\\
   &&\left.\left.\left.-4 \ell
   (\ell+1) \Omega \right)-r {\cal M} \left(-2 \kappa  r^2 P \left(\left({\ell(\ell-1)}+2\right) r
   {\omega_1}'+2 \ell(\ell+1) \Omega -8 \kappa  r^2 \rho  \omega_1\right)\right.\right.\right.\nn\\
   &&\left.\left.\left.+(\ell+1) \left(2 \ell^3 \Omega +2 \left(4-{\ell(\ell+1)}\right) \ell {\omega_1}-2 \ell^2 \Omega -4 \kappa  \ell r^2 \Omega  \rho +\ell r {\omega_1}'-8 \ell
   \Omega +6 r {\omega_1}'\right)+8 \kappa ^2 r^4 P^2 \left(r
   {\omega_1}'+2 {\omega_1}\right)\right)\right.\right.\nn\\
   &&\left.\left.-r^2 \left(-(\ell+1)
   \left(\ell^3 \Omega +\left(-{\ell(\ell+1)}+2\right) \ell {\omega_1}-\ell^2 \Omega -2 \kappa
    \ell r^2 \Omega  \rho -2 \ell \Omega +2 r {\omega_1}'\right)\right.\right.\right.\nn\\
    &&\left.\left.\left.+\kappa  (\ell+1)
   r^2 P \left((\ell-2) r {\omega_1}'+2 \ell \Omega \right)+8 \kappa ^3 r^6
   P^3 {\omega_1}+4 \kappa ^2 r^5 P^2 \left(2 \kappa  r \rho 
   {\omega_1}-{\omega_1}'\right)\right)-8 {\cal M}^3 {\omega_1}'\right)\right.\nn\\
   &&\left.-e^{\nu }\sqrt{4 \ell^2-1}  {\delta H_0^{(\ell-1)}} \left(2
   r {\cal M} \left(-{\ell(\ell+1)}+6 \kappa  r^2 P+2 \kappa  r^2 \rho +4\right)+r^2
   \left({\ell(\ell-1)}\right.\right.\right.\nn\\
   &&\left.\left.\left.+4 \kappa ^2 r^4 P^2-2 \kappa  r^2 (P+\rho)
   -2\right)-4 {\cal M}^2\right)\right] \,,
   \end{eqnarray}
   \begin{eqnarray}
   %%%%%%
  \delta H_2^{(\ell+1)}&=&   \frac{e^{-\nu }}{\sqrt{4 \ell (\ell+2)+3}} \left[2 {h_0^{(\ell)}} \left(-{\omega_1} \left(\ell
   (\ell+1)+4 \kappa  r^2 (P+\rho )\right)+\ell(\ell+1) \Omega +2 (r-2 {\cal M})
   {\omega_1}'\right)-2 r (r-2 {\cal M}) {\omega_1}'
   {h_0^{(\ell)}}'\right.\nn\\
   &&\left.+\sqrt{4 \ell (\ell+2)+3} e^{\nu }
   {\delta H_0}^{(\ell+1)}\right] \,,\\
   %%%%%%%
 \delta H_2^{(\ell-1)}&=&   -\frac{e^{-\nu }}{\ell\sqrt{4 \ell^2-1}} \left[2 \ell {h_0^{(\ell)}} \left(-{\omega_1}
   \left(\ell(\ell+1)+4 \kappa  r^2 (P+\rho )\right)+\ell(\ell+1) \Omega +2 (r-2 {\cal M})
   {\omega_1}'\right)-2 \ell r (r-2 {\cal M}) {\omega_1}'
   {h_0^{(\ell)}}'\right]\nn\\
   &&- {\delta H_0^{(\ell-1)}} \,,\\
   %%%%%%
  \delta\delta p^{(\ell+1)}&=&  \frac{e^{-\nu } (P+\rho )}{2 \sqrt{4 \ell (\ell+2)+3}} \left[2 \ell(\ell+1) \Omega  {h_0^{(\ell)}}+\sqrt{4 \ell
   (\ell+2)+3} e^{\nu } {\delta H_0}^{(\ell+1)}\right]  \,,\\
   %%%%%%%
 \delta\delta p^{(\ell-1)}&=&   \frac{1}{2} e^{-\nu } (P+\rho ) \left(e^{\nu } {\delta H_0^{(\ell-1)}}-\frac{2
   \ell(\ell+1) \Omega  {h_0^{(\ell)}}}{\sqrt{4 \ell^2-1}}\right) \,,
\end{eqnarray}
%%%
\end{widetext}
whereas $\delta H_1^{(\ell\pm1)}=0=\delta h_1^{(\ell\pm1)}$, so that the perturbed metric is symmetric under $t\to-t$ and $\varphi\to-\varphi$.
%%%%%%%%%%%%%%%%%%%%%%%%%%%%%%%%%%%%%%%%%%%
\subsection{Perturbation equations with $\ell=1$}
%%%%%%%%%%%%%%%%%%%%%%%%%%%%%%%%%%%%%%%%%%%
To first order in the spin, a quadrupolar tidal field will source dipolar metric perturbations. As discussed in the main text, such perturbations with $\ell=1$ generically satisfy a different set of equations. 

In the electric-led sector, it turns out that the equations for $\ell=1$ can simply be obtained from those presented above after setting $\ell=1$. However, the $\ell=1$ magnetic-led sector is different. In this case the equations for $H_0^{(1)}$ have the same schematic form as~\eqref{axial_led2}, but with
\begin{widetext}
%%%
\begin{eqnarray}
 C_1^{(1)}&=&0 \,, \\
 C_0^{(1)}&=& \frac{1}{{c_s^2} (r-2 {\cal M})^2 \left({\cal M}+\kappa  r^3 P\right)}\left[{c_s^2} \left({\cal M} \left(\kappa  r^2 \left(10 \rho -\kappa  r^2
   (P-\rho ) (19 P+3 \rho )\right)-6\right)-\kappa  r {\cal M}^2 (35 P+11
   \rho )\right.\right.\nn\\
   &&\left.\left.+\kappa  r^3 \left(\kappa  r^2 \left(P \rho  \left(3 \kappa  r^2
   P+2\right)-3 \rho ^2 \left(\kappa  r^2 P+1\right)+P^2 \left(3-2 \kappa 
   r^2 P\right)\right)+2 \rho \right)\right)\right.\nn\\
   &&\left.+\kappa  r (P+\rho )
   \left(+\kappa  r^3 P+r-{\cal M}\right) \left({\cal M}+\kappa  r^3
   P\right)\right] \,, \\
   %%%%%%%%%%%%%%%
 S_-^{(1)}&=& \frac{\sqrt{\frac{3}{5}} e^{-\nu }}{r^2 {c_s^2} (r-2 {\cal M})^2 \left({\cal M}+\kappa
    r^3 P\right)} \left[{c_s^2} \left(-2 r {\cal M}^2 \left(2
   {h_0^{(2)}} \left(\kappa  r^2 P \left(-10
   {\omega_1}+13 \Omega -7 r {\omega_1}'\right)+\kappa  r^2 \rho  \left(-3 r
   {\omega_1}'-10 {\omega_1}+13 \Omega \right)\right.\right.\right.\right.\nn\\
   &&\left.\left.\left.\left.+r {\omega_1}'-12(\omega_1-\Omega) \right)+r {h_0^{(2)}}' \left(\kappa
    r^2 P \left(r {\omega_1}'+28 {\omega_1}-28 \Omega
   \right)+\kappa  r^2 \rho  \left(-3 r {\omega_1}'-4 {\omega_1}+4 \Omega \right)-7 r {\omega_1}'-8 {\omega_1}+8 \Omega
   \right)\right)\right.\right.\nn\\
   &&\left.\left.+r^3 {\cal M} \left(2 \kappa ^2 r^3 P^2 \left(2 {h_0^{(2)}}
   \left(r {\omega_1}'+6 {\omega_1}+\Omega \right)+r
   {h_0^{(2)}}' \left(r {\omega_1}'-12(\omega_1-\Omega)
   \right)\right)-\kappa  r P \left(2 {h_0^{(2)}} \left(2 \kappa  r^2 \rho
    \left(3 r {\omega_1}'+2 {\omega_1}-12 \Omega \right)\right.\right.\right.\right.\right.\nn\\
    &&\left.\left.\left.\left.\left.+7 r
   {\omega_1}'-28 {\omega_1}+24 \Omega \right)+r
   {h_0^{(2)}}' \left(2 \kappa  r^2 \rho  \left(3 r {\omega_1}'-4
   {\omega_1}+4 \Omega \right)-9 r {\omega_1}'-28 {\omega_1}+28 \Omega \right)\right)+12 \kappa ^2 r^3 \Omega  \rho ^2
   {h_0^{(2)}}\right.\right.\right.\nn\\
   &&\left.\left.\left.-\kappa  r \rho  \left({h_0^{(2)}} \left(18 r
   {\omega_1}'+32 {\omega_1}-40 \Omega \right)+r
   {h_0^{(2)}}' \left(9 r {\omega_1}'+4 {\omega_1}-4 \Omega
   \right)\right)+{\omega_1}' \left(r {h_0^{(2)}}'+10
   {h_0^{(2)}}\right)\right)\right.\right.\nn\\
   &&\left.\left.+16 {\cal M}^3 \left(2 {h_0^{(2)}} (\Omega
   -{\omega_1})-r {h_0^{(2)}}' \left(r {\omega_1}'+2
   {\omega_1}-2 \Omega \right)\right)+r^4 \left(16 \kappa ^3 r^5 \Omega 
   P^3 {h_0^{(2)}}\right.\right.\right.\nn\\
   &&\left.\left.\left.+\kappa  r P \left(r^2 {h_0^{(2)}}' \left(\kappa 
   r \rho  \left(3 r {\omega_1}'-4 {\omega_1}+4 \Omega
   \right)-4 {\omega_1}'\right)-2 {h_0^{(2)}} \left(6 \kappa ^2 r^4
   \Omega  \rho ^2+\kappa  r^2 \rho  \left(-3 r {\omega_1}'+4
   {\omega_1}-4 \Omega \right)+8 {\omega_1}-12 \Omega
   \right)\right)\right.\right.\right.\nn\\
   &&\left.\left.\left.+\kappa ^2 r^3 P^2 \left(2 {h_0^{(2)}} \left(2 \kappa  r^2
   \Omega  \rho -r {\omega_1}'+6 \Omega \right)-r {h_0^{(2)}}'
   \left(r {\omega_1}'-12(\omega_1-\Omega)
   \right)\right)\right.\right.\right.\nn\\
   &&\left.\left.\left.-\left(3 \kappa  r^2 \rho -2\right) \left(4 \kappa  r \Omega  \rho
    {h_0^{(2)}}-{\omega_1}' \left(r {h_0^{(2)}}'+2
   {h_0^{(2)}}\right)\right)\right)\right)+4 \kappa  r^3 \Omega 
   {h_0^{(2)}} (P+\rho ) \left(r {\cal M}-{\cal M}^2+\kappa  r^4 P \left(\kappa
    r^2 P+1\right)\right)\right]\,.
\end{eqnarray}
\end{widetext}
%%%
When $\ell=1$, we can use a residual gauge freedom to set $K^{(1)}=0$ in the ansatz for the metric~\cite{1970ApJ...159..847C,1989ApJ...345..925L}, so that the only further nonvanishing polar components read
%%%
\begin{eqnarray}
 \delta H_2^{(1)}&=& \frac{e^{-\nu }}{5 \left({\cal M}+\kappa  r^3
   P\right)} \left[{\cal M} \left(2 \sqrt{15} r^2 {\omega_1}'
   {h_0^{(2)}}'\right.\right.\nn\\
   &&\left.\left.+4 \sqrt{15} {h_0^{(2)}} \left(r {\omega_1}'+3
   {\omega_1}-3 \Omega \right)+15 e^{\nu } {\delta H_0^{(1)}}\right)\right.\nn\\
   &&\left.+r^2
   \left(\kappa  r \rho  \left(4 \sqrt{15} \Omega  {h_0^{(2)}}-5 e^{\nu }
   {\delta H_0^{(1)}}\right)\right.\right.\nn\\
   &&\left.\left.+\sqrt{15} \left(4 \kappa  r P {\omega_1}
   {h_0^{(2)}}-{\omega_1}' \left(r {h_0^{(2)}}'+2
   {h_0^{(2)}}\right)\right)\right)\right]\,,\nn \\ \\
   %%%%%%
 \delta\delta p^{(1)}&=&   \frac{1}{10} e^{-\nu } (P+\rho ) \left(5 e^{\nu } {\delta H_0^{(1)}}-4
   \sqrt{15} \Omega  {h_0^{(2)}}\right)\,, 
\end{eqnarray}

%%%

%%%%%%%%%%%%%%%%%%%%%%%%%%%%%%%%%%%%%%%%
\section{Exterior solution for $\ell=3$ and $\ell=4$}\label{app:exterior}
%%%%%%%%%%%%%%%%%%%%%%%%%%%%%%%%%%%%%%%%
The explicit solution describing the exterior of a slowly spinning object immersed in an axisymmetric $\ell=3$ electric tidal field reads 
\begin{widetext}
%%%
\begin{eqnarray}
 H_0^{(3)}&=& \frac{2 (y-2)^2 (y-1) y^2 \alpha _3+\left(2 \left(-2-10 y+65 y^2-60 y^3+15
   y^4\right)+15 (y-2)^2 (y-1) y^2 \log\left[1-2/y\right]\right)
   \gamma _3}{2 (y-2) y} \,,\\
%%%%%%%%%%%%%%%%%%%%%  
 \delta h_0^{(4)}&=& -\frac{M \chi}{3360 y^2}  \left[-64 \sqrt{7} y \left(2-9 y+5 y^2\right) \alpha _3+10 \sqrt{7}
   \left(2 \left(16+76 y-762 y^2-2610 y^3+19865 y^4-21315 y^5+6090 y^6\right)\right.\right.\nn\\
   &&\left.\left.+3 y
   \left(-32+144 y-80 y^2-5800 y^3+13050 y^4-9135 y^5+2030 y^6\right)
   \log\left[1-2/y\right]\right) \gamma _3\right.\nn\\
   &&\left.+49 y \left(2 \left(210 y^5-4-18 y-90
   y^2+685 y^3-735 y^4\right)+15 y^3 \left(90 y-63 y^2+14 y^3-40\right)
   \log\left[1-2/y\right]\right) \gamma _{43}\right]\,,\\
%%%%%%%%%%%%%%%%%%%%%  
 \delta h_0^{(2)}&=& -\frac{M \chi}{4200 y^2}  \left[-8 \sqrt{35} y \left(16+40 y-240 y^2+105 y^3\right) \alpha _3+5
   \left(6 \sqrt{35} \left(4 y+304 y^2-774 y^3+354 y^4-8+y \left(80 y\right.\right.\right.\right.\nn\\
   &&\left.\left.\left.\left.+480 y^2-564
    y^3+177 y^4-32\right) \log\left[1-2/y\right]\right) \gamma _3
% \right.\right.\nn\\
%    &&\left.\left.
+35 y
   \left(3 (y-2) y^3 \log\left[1-2/y\right]-4-4 y-6 y^2+6 y^3\right)
   \gamma _{23}\right)\right] \nn\\
\end{eqnarray}
%%%%
where, as in Sec.~\ref{sec:exterior}, $y=r/M$. 
Likewise, for an axisymmetric $\ell=3$ magnetic tidal field, we obtain  
%%%
\begin{eqnarray}
 h_0^{(3)}&=& -\frac{M \left(-64 y^3 \left(8-10 y+3 y^2\right) \alpha _3^\ast+3 \left(8+20 y+60
   y^2-210 y^3+90 y^4+15 y^3 \left(8-10 y+3 y^2\right)
   \log\left[1-2/y\right]\right) \gamma _3^\ast\right)}{192 y} \,,\\
%%%%%%%%%%%%%%%%%%%%%  
 \delta H_0^{(4)}&=&\frac{\chi}{2688 (y-2) y^4}  \left[-512 \sqrt{7} y^4 \left(18-25 y+8 y^2\right) \alpha _3^{\ast }+\sqrt{7} \left(-192-288 y+480 y^2+4520 y^3+4680 y^4-219840 y^5\right.\right.\nn\\
 &&\left.\left.+404200 y^6\!-\!246750
   y^7\!+\!49350 y^8\!+\!15 y^4 \left(432+5040 y-18608 y^2+21150 y^3-9870 y^4+1645 y^5\right)
   \log\left[1-2/y\right]\right) \gamma _3^\ast\right.\nn\\
   &&\left.+672 y^3 \left(2
   \left(4+36 y-480 y^2+860 y^3-525 y^4+105 y^5\right)+15 (y-2)^2 y^2 \left(6-14 y+7
   y^2\right) \log\left[1-2/y\right]\right) \gamma _{43}^{\ast
   }\right] \,,\\
%%%%%%%%%%%%%%%%%%%%%  
 \delta H_0^{(2)}&=& \frac{\chi }{2240 (y-2) y^4} \left[-256 \sqrt{35} y^4 \left(38\!-\!31 y\!+\!6 y^2\right) \alpha _3^\ast\!-\!3
   \sqrt{35} \left(2 \left(-16\!-\!24 y\!-\!72 y^2\!-\!90 y^3\!+\!3120 y^4\!-\!4005 y^5\!+\!1215 y^6\right)\right.\right.\nn\\
   &&\left.\left.+15
   y^4 \left(448 y\!-\!348 y^2\!+\!81 y^3\!-\!152\right)
   \log\left[1\!-\!2/y\right]\right) \gamma _3^\ast\!+\!1120 y^3 \left(4\!+\!8
   y\!-\!18 y^2\!+\!6 y^3\!+\!3 (y\!-\!2)^2 y^2 \log\left[1\!-\!2/y\right]\right) \gamma
   _{23}^\ast\right]\,. \nn\\
\end{eqnarray}
%%%%

Finally, the solution describing the exterior of a slowly spinning object immersed in an axisymmetric $\ell=4$ electric tidal field reads
%%%
\begin{eqnarray}
 H_0^{(4)}&=& \frac{1}{28 (y-2) y}\left[4 (y-2)^2 y^2 \left(6-14 y+7 y^2\right) \alpha _4+7 \left(2 \left(4+36 y-480
   y^2+860 y^3-525 y^4+105 y^5\right)\right.\right.\nn\\
   &&\left.\left.+15 (y-2)^2 y^2 \left(6-14 y+7 y^2\right)
   \log\left[1-2/y\right]\right) \gamma _4\right] \,,\\
%%%%%%%%%%%%%%%%%%%%%  
 \delta h_0^{(5)}&=& -\frac{M \chi}{443520 y^2}  \left[-256 \sqrt{11} y \left(-16\!+\!168 y\!-\!240 y^2\!+\!85 y^3\right) \alpha
   _4\!+\!35 \left(16 \sqrt{11} \left(-48\!-\!344 y\!+\!8256 y^2\!+\!31980 y^3\!-\!459270 y^4 \right.\right.\right.\nn\\
   &&\left.\left.\left.+884520
   y^5-586845 y^6+127575 y^7+3 y \left(128-1344 y+1920 y^2+112720 y^3-396900 y^4+476280
   y^5-238140 y^6\right.\right.\right.\right.\nn\\
   &&\left.\left.\left.\left.+42525 y^7\right) \tanh^{-1}\left[\frac{1}{1-y}\right]\right)
   \gamma _4+99 y \left(2 \left(8+56 y+420 y^2-5670 y^3+10920 y^4-7245 y^5+1575
   y^6\right)\right.\right.\right.\nn\\
   &&\left.\left.\left.+105 y^3 \left(40-140 y+168 y^2-84 y^3+15 y^4\right)
   \log\left[1-2/y\right]\right) \gamma _{54}\right)\right]\,,\\
%%%%%%%%%%%%%%%%%%%%%  
 \delta h_0^{(3)}&=& -\frac{M \chi }{28224 y^2} \left[-128 \sqrt{7} y \left(-4-30 y+210 y^2-224 y^3+63 y^4\right) \alpha
   _4+7 \left(4 \sqrt{7} \left(96+40 y-12300 y^2+52140 y^3\right.\right.\right.\nn\\
   &&\left.\left.\left.-62370 y^4+20250 y^5+15 y
   \left(32+240 y-1680 y^2+3592 y^3-2754 y^4+675 y^5\right)
   \log\left[1-2/y\right]\right) \gamma _4+63 y \left(8+20 y
   \right.\right.\right.\nn\\
   &&\left.\left.\left.+60 y^2-210y^3+90 y^4+15 y^3 \left(8-10 y+3 y^2\right)
   \log\left[1-2/y\right]\right) \gamma _{34}\right)\right]\,,
\end{eqnarray}
%%%%
whereas the magnetic counterpart reads
%%%
\begin{eqnarray}
 h_0^{(4)}&=& -\frac{M}{3360 y} \left[-240 y^3 \left(-40+90 y-63 y^2+14 y^3\right) \alpha _4^\ast+49
   \left(2 \left(-4-18 y-90 y^2+685 y^3-735 y^4+210 y^5\right)\right.\right.\nn\\
   &&\left.\left.+15 y^3 \left(-40+90 y-63
   y^2+14 y^3\right) \log\left[1-2/y\right]\right) \gamma _4^{\ast
   }\right]  \,,\\
%%%%%%%%%%%%%%%%%%%%%  
 \delta H_0^{(5)}&=& \frac{\chi}{55440 (y-2) y^4}  \left[-480 \sqrt{11} y^4 \left(-350+939 y-732 y^2+175 y^3\right) \alpha
   _4^\ast+7 \left(7 \sqrt{11} \left(2 \left(24+68 y-60 y^2-1562 y^3+6882
   y^4\right.\right.\right.\right.\nn\\
   &&\left.\left.\left.\left.+128834 y^5-421890 y^6+462735 y^7-213570 y^8+35595 y^9\right)+15 y^4
   \left(-700-4450 y+30176 y^2-55020 y^3+44296 y^4 \right.\right.\right.\right.\nn\\
   &&\left.\left.\left.\left.-16611 y^5+2373 y^6\right)
   \log\left[1-2/y\right]\right) \gamma _4^\ast+1980 y^3
   \left(-8-112 y+2576 y^2-7560 y^3+8190 y^4-3780 y^5+630 y^6\right.\right.\right.\nn\\
   &&\left.\left.\left.+105 (y-2)^2 y^2
   \left(-2+8 y-9 y^2+3 y^3\right) \log\left[1-2/y\right]\right) \gamma
   _{54}^\ast\right)\right]\,,\\
%%%%%%%%%%%%%%%%%%%%%  
 \delta H_0^{(4)}&=& -\frac{\chi}{35280 (y-2) y^4}  \left[480 \sqrt{7} y^4 \left(-730+1401 y-798 y^2+140 y^3\right) \alpha
   _4^\ast+49 \left(\sqrt{7} \left(2 \left(24+68 y+300 y^2+856 y^3-36360
   y^4\right.\right.\right.\right.\nn\\
   &&\left.\left.\left.\left.+102095 y^5-80220 y^6+19005 y^7\right)+15 y^4 \left(1460-7870 y+11732 y^2-6615
   y^3+1267 y^4\right) \log\left[1-2/y\right]\right) \gamma _4^{\ast
   }\right.\right.\nn\\
   &&\left.\left.-360 y^3 \left(2 \left(-2-10 y+65 y^2-60 y^3+15 y^4\right)+15 (y-2)^2 (y-1) y^2
   \log\left[1-2/y\right]\right) \gamma _{34}^{\ast
   }\right)\right]\,.
\end{eqnarray}
%%%%
\end{widetext}
The solutions above depend on various integration constants. As explained in the main text, the constants $\alpha_i$
(respectively, $\alpha_i^\ast$) are proportional to the electric (respectively, magnetic) components of the external tidal field, whereas
the constants $\gamma_{i}$ and $\gamma_{ij}$ (respectively, $\gamma_{i}^\ast$ and $\gamma_{ij}^\ast$) are associated with the
deformation of the mass (respectively, current) $\ell=i$ multipole moments. Terms proportional to $\gamma_{i}$ and
$\gamma_{i}^\ast$ are of zeroth order in the spin, whereas terms proportional to $\gamma_{ij}$ and $\gamma_{ij}^\ast$
are of linear order in the spin.

%%%%%%%%%%%%%%%%%%%%%%%%%%%%%%%%%%%%%%%%
\section{Tidally-deformed, spinning black-hole solution}\label{app:BH}
%%%%%%%%%%%%%%%%%%%%%%%%%%%%%%%%%%%%%%%%
Although our interest in this work is devoted to NSs, for completeness we present here the explicit solution for a tidally-deformed spinning black hole to first order in the spin, in the presence of both an electric and a magnetic tidal field with $\ell=2,3,4$. This solution complements those presented in Ref.~\cite{Poisson:2014gka} and in Paper~I by presenting the axisymmetric $\ell=3,4$ (both electric and magnetic) terms to linear order in the spin for the first time.

As explained in Paper~I, when the central object is a black hole the requirement of regularity at the horizon simplifies
the metric perturbations considerably. Using the same procedure detailed in Paper~I, it is straightforward to show that
regularity at the horizon implies that all $\gamma$'s be exactly zero. Therefore, in agreement with the findings of
Ref.~\cite{Landry:2015zfa} and of Paper~I, all tidal Love numbers of a spinning black hole are vanishing.

The nonvanishing components of the line element~\eqref{metric} in the case of a spinning black hole deformed by a stationary, axisymmetric tidal field with electric and magnetic components with $\ell=2,3,4$ to linear order in the spin read
\begin{widetext}
 %%%%%
\begin{eqnarray}
 g_{tt} &=& \frac{(y-2)}{896 \sqrt{\pi } y} \left[-896 \sqrt{\pi }+112 \sqrt{5} (y-2) y (1+3 {\cos2\vartheta}) \alpha _2+56 \sqrt{7} y \left(2-3 y+y^2\right) (3 {\cos\vartheta}+5 \cos3\vartheta) \alpha _3-324 y \alpha _4\right.\nn\\
 &&\left.+918 y^2 \alpha _4-756 y^3 \alpha _4+189 y^4 \alpha _4-720 y {\cos2\vartheta} \alpha _4+2040 y^2 {\cos2\vartheta} \alpha
   _4-1680 y^3 {\cos2\vartheta} \alpha _4+420 y^4 {\cos2\vartheta} \alpha _4 \right.\nn\\
   &&\left.-1260 y {\cos4\vartheta} \alpha _4+3570 y^2 {\cos4\vartheta} \alpha _4-2940
   y^3 {\cos4\vartheta} \alpha _4+735 y^4 {\cos4\vartheta} \alpha _4\right]\nn\\
   &&-\frac{(y-2) \chi}{896 \sqrt{\pi } y}  \left[672 \sqrt{5} {\cos\vartheta} (3-4 y+3
   {\cos2\vartheta}) \alpha _2^{\ast }+28 \sqrt{7} (-55+24 y+4 (-39+16 y) {\cos2\vartheta}+5 (8 y-9) {\cos4\vartheta}) \alpha _3^{\ast }\right.\nn\\
   &&\left.+6
   {\cos\vartheta} \left(5 \left(-17-62 y+63 y^2\right)+4 \left(675-854 y+175 y^2\right) {\cos2\vartheta}+7 \left(175-382 y+175 y^2\right) \cos4\vartheta\right) \alpha _4^{\ast }\right] \,,
   \end{eqnarray}
   \begin{eqnarray}
 %%%%%
 g_{t\varphi} &=& -\frac{M (y-2) y^2 {\sin\vartheta}^2 \left(336 \sqrt{5} {\cos\vartheta} \alpha _2^{\ast }+28 \sqrt{7} (-4+3 y) (3+5 {\cos2\vartheta}) \alpha _3^{\ast }+15
   \left(20-35 y+14 y^2\right) (9 {\cos\vartheta}+7 {\cos3\vartheta}) \alpha _4^{\ast }\right)}{224 \sqrt{\pi }}\nn\\
   &&-\frac{M \chi  {\sin\vartheta}^2 }{2240 \sqrt{\pi } y}\left[112
   \sqrt{5} \left(-12+15 y+20 y^2+5 (5y-4) {\cos2\vartheta}\right) \alpha _2+56 \sqrt{7} \left(\left(22-35 y-105 y^2+60 y^3\right) {\cos\vartheta}\right.\right.\nn\\
   &&\left.\left.+5 \left(2-9
   y+5 y^2\right) {\cos3\vartheta}\right) \alpha _3+5 \left(896 \sqrt{\pi }+\left(9 \left(-16+40 y+240 y^2-351 y^3+112 y^4\right)+4 \left(-64+192 y+840 y^2\right.\right.\right.\right.\nn\\
   &&\left.\left.\left.\left.-1295
   y^3+420 y^4\right) {\cos2\vartheta}+7 \left(-16+168 y-240 y^2+85 y^3\right) {\cos4\vartheta}\right) \alpha _4\right)\right]\,,
    \end{eqnarray}
   \begin{eqnarray}
 %%%%%
 g_{rr} &=& \frac{y}{896 \sqrt{\pi } (y-2)} \left[896 \sqrt{\pi }+112 \sqrt{5} (y-2) y (1+3 {\cos2\vartheta}) \alpha _2+56 \sqrt{7} y \left(2-3 y+y^2\right) (3 {\cos\vartheta}+5 \cos3\vartheta) \alpha _3-324 y \alpha _4\right.\nn\\
 &&\left.+918 y^2 \alpha _4-756 y^3 \alpha _4+189 y^4 \alpha _4-720 y {\cos2\vartheta} \alpha _4+2040 y^2 {\cos2\vartheta} \alpha _4-1680
   y^3 {\cos2\vartheta} \alpha _4+420 y^4 {\cos2\vartheta} \alpha _4\right.\nn\\
   &&\left.-1260 y {\cos4\vartheta} \alpha _4+3570 y^2 {\cos4\vartheta} \alpha _4-2940 y^3
   {\cos4\vartheta} \alpha _4+735 y^4 {\cos4\vartheta} \alpha _4\right]\nn\\
   &&-\frac{y \chi }{896
   \sqrt{\pi } (y-2)} \left[1344 \sqrt{5} {\cos\vartheta}^3 \alpha
   _2^{\ast }+28 \sqrt{7} (-61+30 y+4 (-45+22 y) {\cos2\vartheta}+5 (-3+2 y) {\cos4\vartheta}) \alpha _3^{\ast }\right.\nn\\
   &&\left.+6 {\cos\vartheta} \left(-5 \left(147-260
   y+91 y^2\right)+4 \left(1065-1820 y+637 y^2\right) {\cos2\vartheta}+21 \left(15-20 y+7 y^2\right) {\cos4\vartheta}\right) \alpha _4^{\ast }\right]\,, 
   \end{eqnarray}
   \begin{eqnarray}
 %%%%%
 g_{\vartheta\vartheta} &=&  \frac{M^2 y^2 \chi}{1344 \sqrt{\pi }}  \left[14 \sqrt{7} (354-261 y+(1096-804 y) {\cos2\vartheta}+5 (-34+21 y) {\cos4\vartheta}) \alpha _3^{\ast }+18 {\cos\vartheta}
   \left(1369-2295 y+868 y^2\right.\right.\nn\\
   &&\left.\left.-4 \left(1175-2079 y+812 y^2\right) {\cos2\vartheta}+21 \left(55-81 y+28 y^2\right) {\cos4\vartheta}\right) \alpha _4^{\ast
   }\right]\nn\\
   &&+\frac{M^2 y^2 }{4480 \sqrt{\pi }}\left[560 \sqrt{5} \left(y^2-2\right) (1+3 {\cos2\vartheta}) \alpha _2+56 \sqrt{7} \left(4-10 y^2+5 y^3\right) (3
   {\cos\vartheta}+5 {\cos3\vartheta}) \alpha _3\right.\nn\\
   &&\left.+5 \left(896 \sqrt{\pi }+\left(-8+60 y^2-70 y^3+21 y^4\right) (9+20 {\cos2\vartheta}+35 {\cos4\vartheta})
   \alpha _4\right)\right] \,, \\
 %%%%%
 g_{\varphi\varphi} &=& g_{\vartheta\vartheta} \sin^2\vartheta  \,,
 %%%%%
\end{eqnarray}
\end{widetext}
where we recall that $\alpha_\ell$ and $\alpha_\ell^\ast$ are proportional to the amplitude of the electric and magnetic tidal field, respectively [cf. Eq.~\eqref{EB}].

%%%%%%%%%%%%%%%%%%%%%%%%%%%%%%%%%%%%%%%%
\section{Tidal Love numbers with broken equatorial symmetry}\label{app:noequatorial}
%%%%%%%%%%%%%%%%%%%%%%%%%%%%%%%%%%%%%%%%
In the main text we have focused on tidal perturbations that preserve the reflection symmetry of the background relative to the equatorial plane at $\vartheta=\pi/2$. A binary system of two nonspinning compact objects or of two spinning compact objects with (anti)aligned spins enjoys such equatorial symmetry relative to the orbital plane. However, in other configurations --~e.g. if the angular momenta of the two objects are not (anti)aligned~-- the equatorial symmetry can be broken. Supernova kicks after the formation of a NS might misalign the spin, although subsequent tidal dissipation contributes to realign them. The details and the efficiency of these processes are largely unknown (cf. Ref.~\cite{Gerosa:2013laa} and references therein).
For completeness, in this appendix we briefly extended the results discussed in the main text also to nonequatorial-symmetric tidal perturbations.

A tidal perturbation is symmetric under reflection around the equatorial plane (equatorial-symmetric for short) if it is invariant under the transformation $\vartheta\to \pi-\vartheta$. Given the symmetry properties of the spherical harmonics, electric (polar) perturbations are equatorial-symmetric when $\ell+m$ is even, whereas magnetic (axial) tidal perturbations are equatorial-symmetric when $\ell+m$ is odd. In the axisymmetric case, electric (respectively, magnetic) perturbations are equatorial-symmetric when $\ell$ is even (respectively, odd). 

The technique adopted in the main text to extract the deformed multipole moments is based on Ryan's
approach~\cite{Ryan:1995wh,Ryan:1997hg,Pappas:2012nt} and assumes equatorial symmetry. However, the Geroch-Hansen
moments are also defined for geometries that break the equatorial symmetry. For such geometries the mass multipole
moments $M_\ell$ with odd $\ell$, and the current multipole moments and $S_\ell$ with even $\ell$, are different from
zero~\cite{Geroch:1970cd,Hansen:1974zz}. Although it should not be difficult to extend Ryan's approach to the
nonequatorial-symmetric case, this would bring us too far. Therefore, here we opt for a more practical definition; a
detailed analysis will appear elsewhere.

For our purposes, it is sufficient to notice that the multipole moments can be extracted (modulo a normalization factor) from the large-distance behavior of the metric in our coordinates, namely,
%%%
\begin{eqnarray}
 g_{tt}&\sim& \sum_{\ell=2} \frac{M_\ell}{r^{\ell+1}} P_\ell(\cos\vartheta) \,,\\
 g_{t\varphi}&\sim& \sum_{\ell=2} \frac{S_\ell}{r^{\ell}} S_\varphi^\ell(\vartheta) \,.
\end{eqnarray}
%%%
In the following we will use the above large-distance behavior of $g_{tt}$ and $g_{t\varphi}$ to define the multipole moments associated with nonequatorial-symmetric tidal perturbations. 
Clearly, such a procedure does not allow us to fix the overall factors of $M_\ell$ and $S_\ell$ and it is not manifestly gauge invariant. Our procedure here should therefore be extended through a more rigorous analysis, but it is nevertheless sufficient to capture the correct scaling of the multipole moments with the compactness.  

In the static case, in addition to the Love numbers defined in Eq.~\eqref{Love0b}, we can also define
%%%
\begin{eqnarray}
 \tilde\lambda_E^{(3)} &=& -\frac{\gamma_3}{\alpha_3} ,\quad \tilde\lambda_M^{(2)} = \frac{\gamma_2^\ast}{480\alpha_2^\ast},\quad  \tilde\lambda_M^{(4)} = \frac{\gamma_4^\ast}{11520\alpha_4^\ast}, \nn \\ \label{Love0c}
\end{eqnarray}
%%%
where the prefactors have been chosen in order to agree with previous definitions of the static electric and magnetic Love numbers~\cite{Binnington:2009bb,Damour:2009vw}. The Love numbers in Eq.~\eqref{Love0c} are shown in Fig.~\ref{fig:lambda_nonrot} in the main text.

To linear order in the spin, similar to the derivation of Eqs.~\eqref{dlambdaE23}--\eqref{dlambdaE43}, we can define the (dimensionless) quantities
%%%%
\begin{eqnarray}
 \tilde\delta\lambda_M^{(23)} &\propto& \chi \frac{74 \sqrt{35} \gamma _3+35 \gamma _{23}}{16800 \alpha _3}\,,\label{dlambdaE23b} \\
 \tilde\delta\lambda_E^{(32)} &\propto& -\chi\frac{\gamma _{32}^{\ast}}{\alpha _2^{\ast}}\,, \\
 \tilde\delta\lambda_E^{(34)} &\propto&  \chi\frac{637 \sqrt{7} \gamma _4^{\ast}-360 \gamma _{34}^{\ast}}{360\alpha _4^{\ast} }\,, \\
 \tilde\delta\lambda_M^{(43)} &\propto& \chi\frac{374 \sqrt{7} \gamma _3+49 \gamma _{43}}{564480 \alpha _3}\,,  \label{dlambdaE43b}
\end{eqnarray}
%%%%
which are proportional to the rotational Love numbers in Eq.~\eqref{Love0b} for the nonequatorial-symmetric case. Note that a quadrupolar magnetic tidal field (which breaks the equatorial symmetry) would source a mass moment $M_1$. However, such dipolar contribution can be eliminated through a shift of the central object along the axis of symmetry and is therefore irrelevant.

\begin{figure*}[ht]
\begin{center}
\includegraphics[width=8.cm]{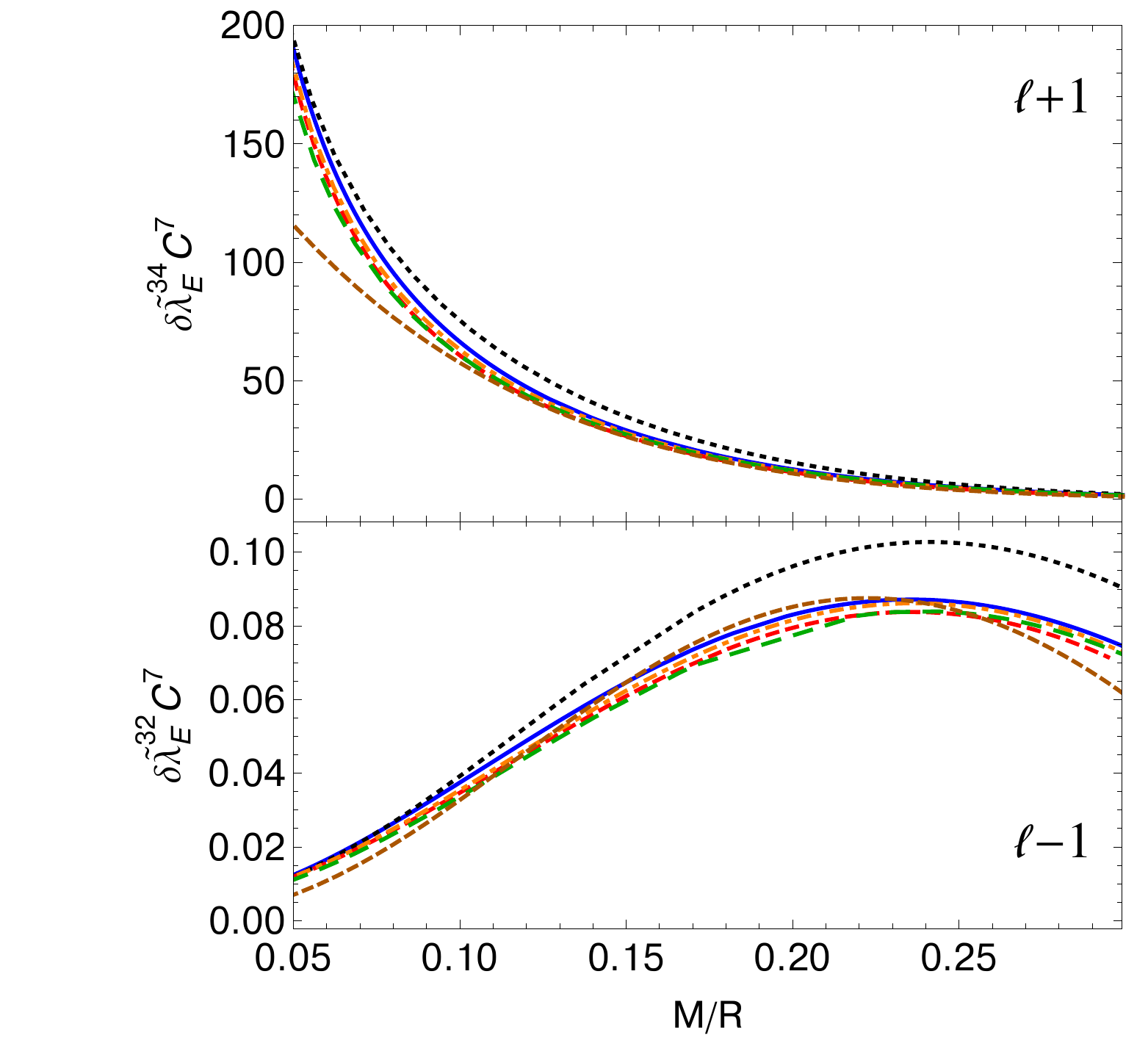}
\includegraphics[width=8.cm]{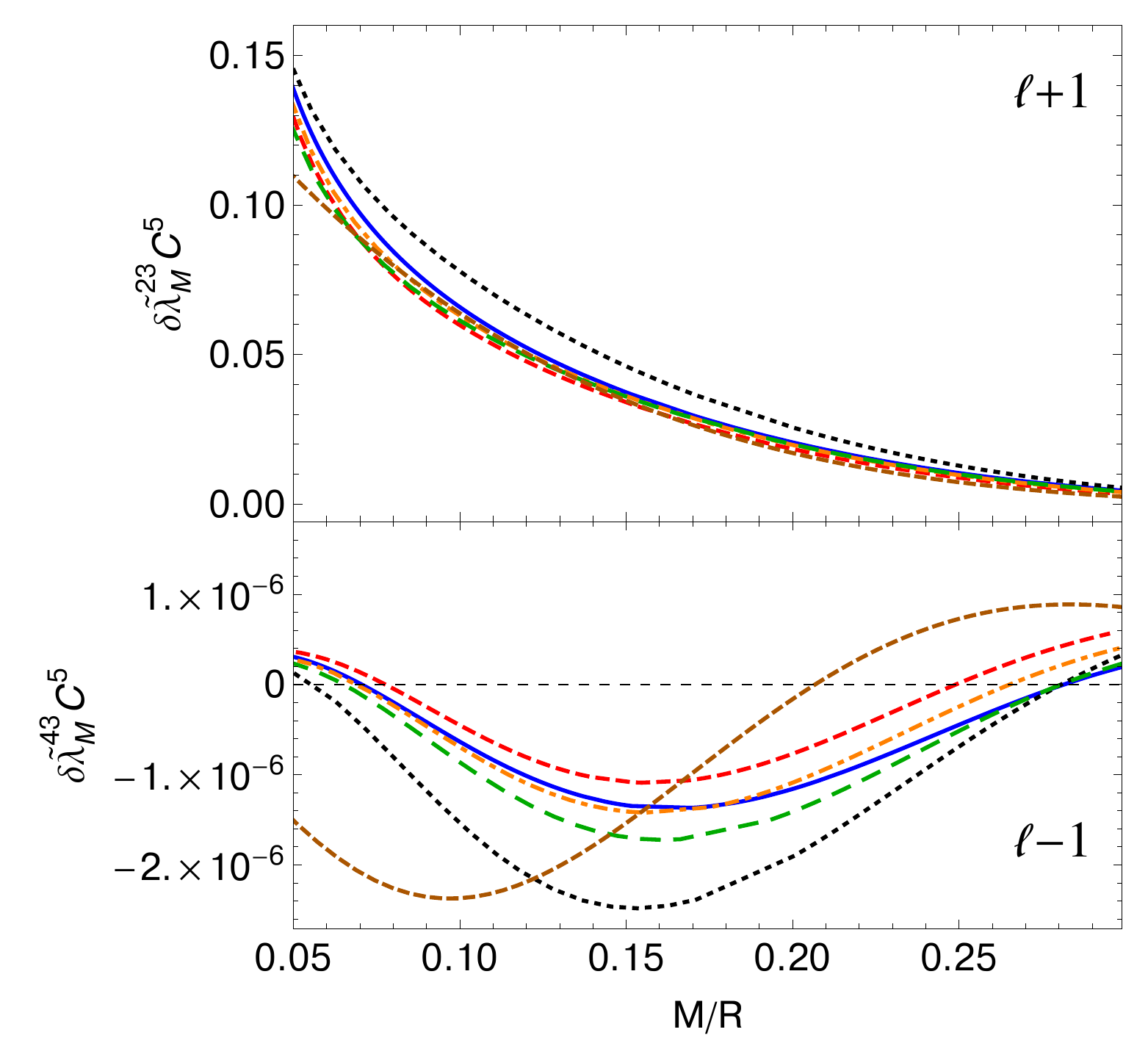}
\caption{(color online). Same as Fig.~\ref{fig:rot_VS_C} but for the case of tidal perturbations that break the equatorial symmetry. Note that in this case the rotational Love numbers are defined modulo a prefactor.
}
\label{fig:rot_VS_C2}
\end{center}
\end{figure*}

\begin{figure*}[ht]
\begin{center}
\includegraphics[width=7.5cm]{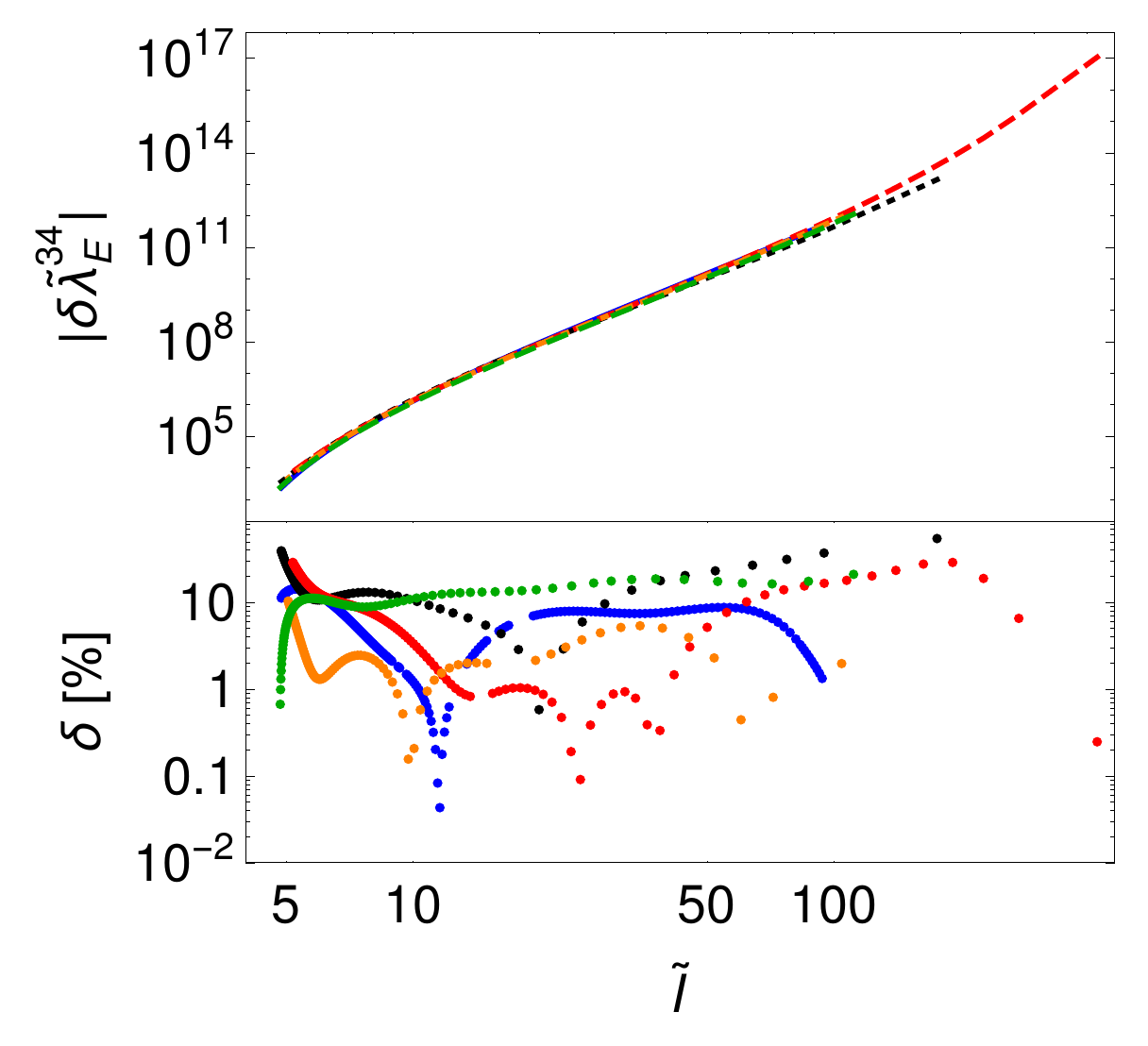}
\includegraphics[width=7.5cm]{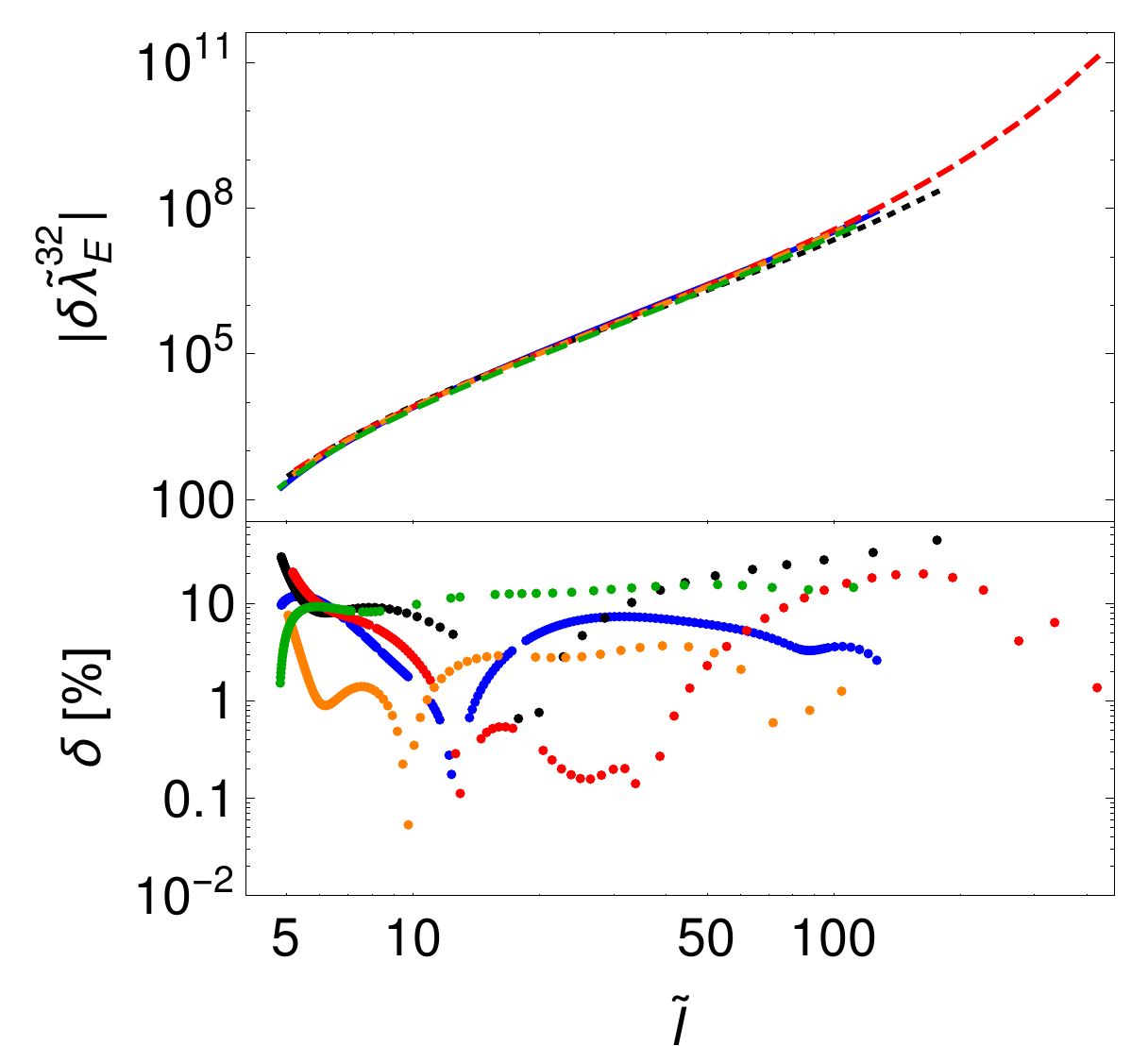}\\
%%%%%
\includegraphics[width=7.5cm]{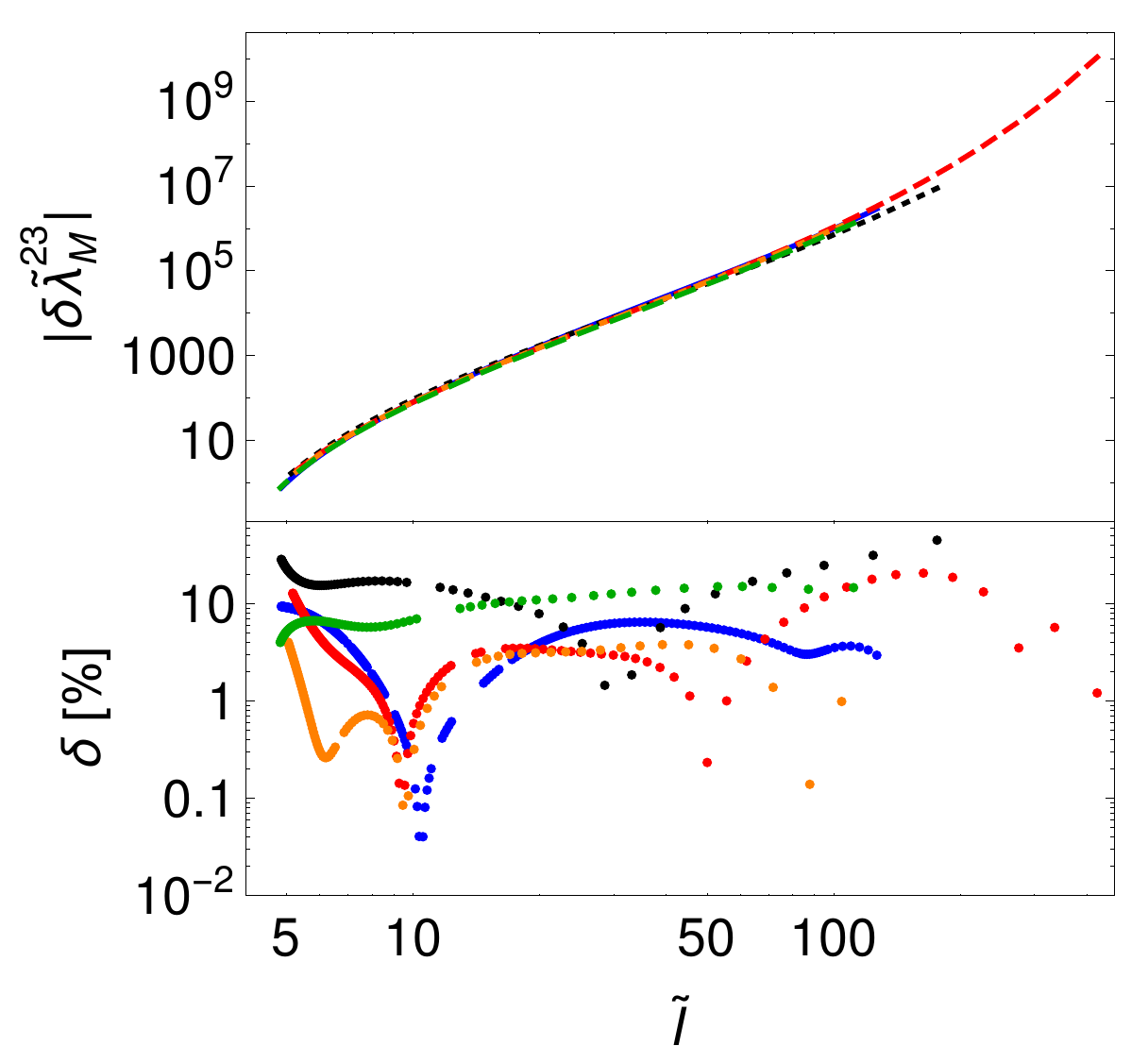}
\includegraphics[width=7.5cm]{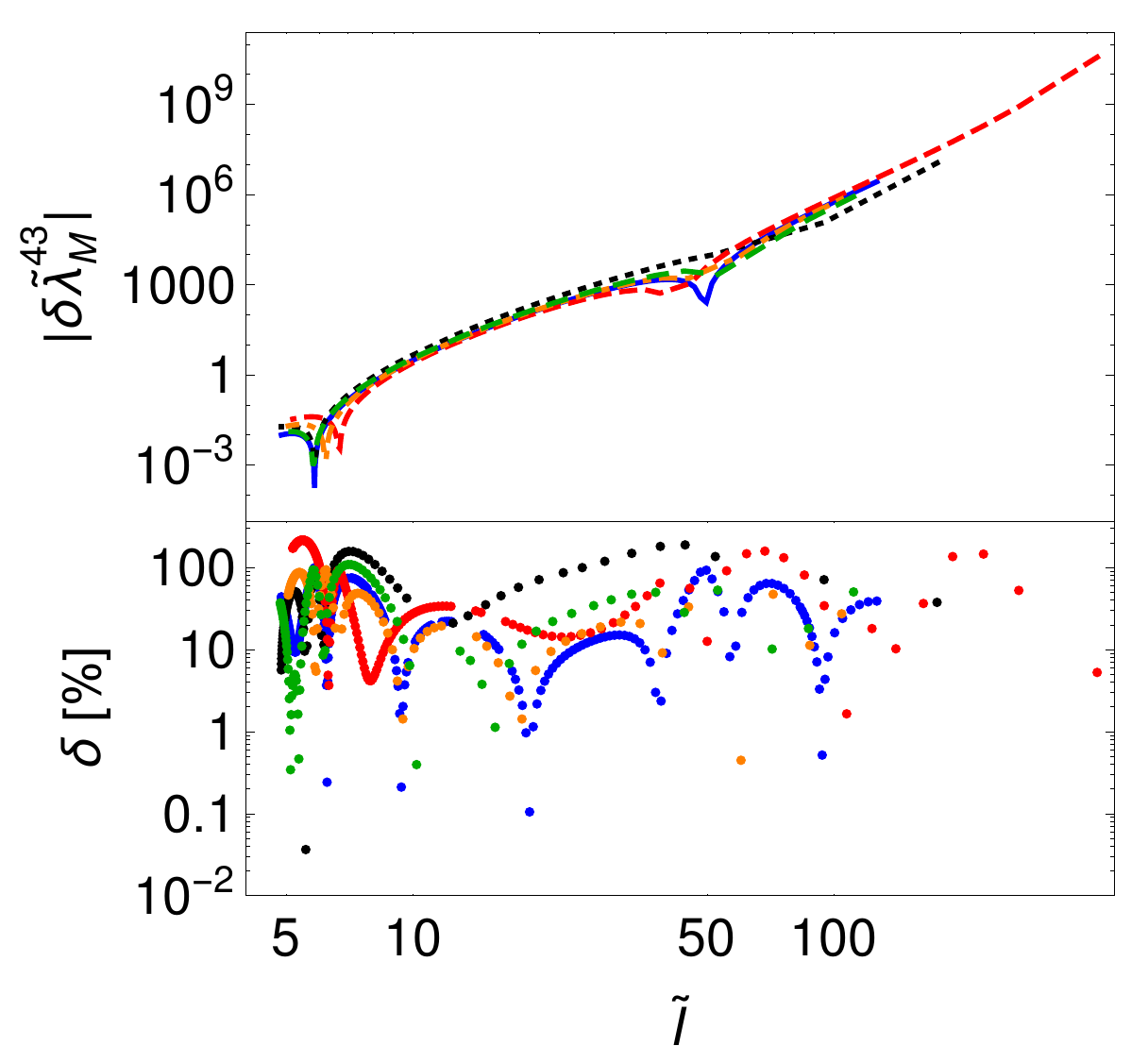}
%%%%%%%%%%%%%%%%%%
\caption{(color online). Same as Fig.~\ref{fig:rot_universal} but for the case of tidal perturbations that break the equatorial symmetry. Note that in this case the rotational Love numbers are defined modulo a prefactor.}
\label{fig:rot_universal2}
\end{center}
\end{figure*}

The rotational Love numbers defined in Eqs.~\eqref{dlambdaE23b}--\eqref{dlambdaE43b} are shown in Figs.~\ref{fig:rot_VS_C2} and \ref{fig:rot_universal2}, which can be compared to their counterparts (Figs.~\ref{fig:rot_VS_C} and \ref{fig:rot_universal}) in the equatorial-symmetric case. We observe that the nonequatorial-symmetric rotational Love numbers present the same qualitative features discussed for the equatorial-symmetric case in the main text.

%%%%%%%%%%%%%%%%%%%%%%%%%%%%%%%%%%%%%%%%%%%%%%%%%%%%%%%%%%%%%%%%%%%%
\section{Note on the magnetic Love numbers for static fluids} \label{app:comment}
%%%%%%%%%%%%%%%%%%%%%%%%%%%%%%%%%%%%%%%%%%%%%%%%%%%%%%%%%%%%%%%%%%%%
In this appendix we point out an inconsistency in the computation of the magnetic Love numbers for static fluids that was not reported in previous work. Although this issue is not directly related to our main goal, here we take the opportunity to correct some results that appeared in the past.

In the static case, magnetic tidal perturbations of a static fluid are governed by the first equation in~\eqref{axial_led2}. Using the differential operator~\eqref{Ds} whose coefficients are given in Appendix~\ref{app:sources}, this equation explicitly reads
%%%%
\begin{eqnarray}
 E_1\equiv&&\frac{d^2 h_0^{(\ell)}}{dr^2}-\frac{\kappa  r^2 (P+\rho )}{r-2 {\cal M}} \frac{d h_0^{(\ell)}}{dr}\nn\\
 &-&\frac{\ell(\ell+1)-{4{\cal M}}/{r}+2\kappa r^2(P+\rho)}{r(r-2{\cal M})}  h_0^{(\ell)} = 0\,.\nn\\ \label{magnetic1}
\end{eqnarray}
%%%%
Such an equation is equivalent to that obtained by Binnington and Poisson (cf. Eq.~(4.29) in Ref.~\cite{Binnington:2009bb}). Although Ref.~\cite{Binnington:2009bb} reports that Eq.~\eqref{magnetic1} is also equivalent to Eq.~(31) in Ref.~\cite{Damour:2009vw}, this is actually not the case.
%%%
To show this, let us first write Eq.~(31) in Ref.~\cite{Damour:2009vw} in our notation:
%%%
\begin{eqnarray}
 E_2&\equiv& \frac{d^2{\psi^{(\ell)}}}{dr^2}+\frac{2 M(r)+\kappa  r^3 [P(r)-\rho (r)]}{r \left(r-2 {\cal M}\right)} \frac{d{\psi^{(\ell)}}}{dr}\nn\\
 &-&\frac{\ell(\ell+1)-{6{\cal M}}/{r}-\kappa r^2(P-\rho)}{r(r-2{\cal M})}  {\psi^{(\ell)}}=0 \,, \label{magnetic2}
\end{eqnarray}
%%%
where ${\psi^{(\ell)}}$ is a gauge-invariant master function~\cite{Cunningham:1978zfa}. As discussed in Ref.~\cite{Damour:2009vw}, for static perturbations ${\psi^{(\ell)}}\equiv r {h_0^{(\ell)}}'-2 h_0^{(\ell)}$. Clearly, by inserting this definition into Eq.~\eqref{magnetic2} one would obtain a third-order ODE for $h_0^{(\ell)}$. By using the background field equations and Eq.~\eqref{magnetic1}, it is straightforward to show that
%%%
\begin{equation}
 E_1-E_2 \equiv \frac{\kappa  r (P+\rho )}{{c_s^2} (r-2 {\cal M})^2}\left( 2 A{h_0^{(\ell)}}+r B {h_0^{(\ell)}}'\right)\,, \label{E1mE2}
\end{equation}
%%%
where 
%%%
\begin{eqnarray}
 A&=& {c_s^2} \left(r \left(\ell(\ell+1)+\kappa  r^2 (P+2 \rho )-1\right)-3 M\right)\nn\\
 &&-M-\kappa  r^3 P \,, \nn\\
 B&=& {c_s^2} \left(r \left(\kappa  r^2 (P+2 \rho )+3\right)-7 M\right)-M-\kappa  r^3 P\,, \nn
\end{eqnarray}
%%%
and we recall that ${c_s}$ is the speed of sound in the fluid. The difference $E_1-E_2$ cannot be further simplified. While the two equations are equivalent in vacuum where $P=0=\rho$, they are generically inequivalent inside matter.
In principle, this is not enough to show that the magnetic Love numbers defined through these two equations are also different, since the matching procedure has to be performed on different exterior solutions. However, as a further proof that Eqs.~\eqref{magnetic1} and \eqref{magnetic2} are inequivalent, we have computed the magnetic Love numbers for $\ell=2$ in both cases, by matching the interior solutions $h_0^{(\ell)}$ and ${\psi^{(\ell)}}$ with their corresponding exterior solution at $r=R$. In the latter case, we define the magnetic Love numbers as in Ref.~\cite{Damour:2009vw} (see also Ref.~\cite{Yagi:2013sva}) modulo an irrelevant prefactor.

%%%%
\begin{figure}[ht]
\begin{center}
\includegraphics[width=7.75cm]{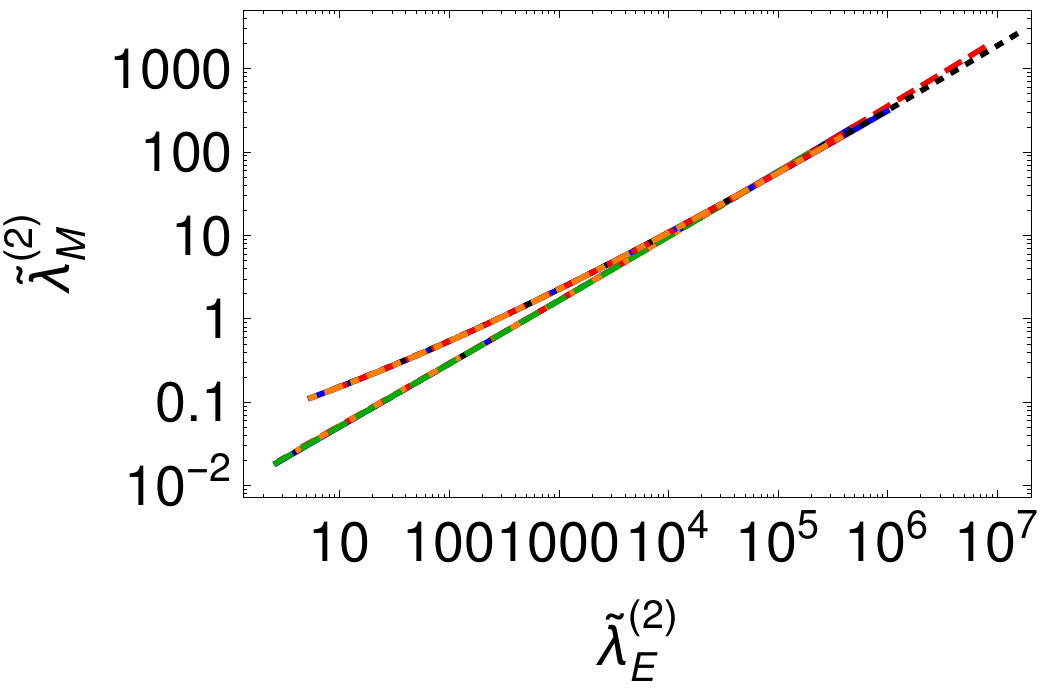}
\caption{(color online). The (dimensionless) magnetic Love numbers computed in the static case by solving Eq.~\eqref{magnetic1} (bottom curves) and compared with those computed by solving Eq.~\eqref{magnetic2} (top curves). The two quantities should be equivalent modulo a constant factor but their slope is instead different, with larger deviations in the relativistic regime. Our derivation independently confirms that performed in Ref.~\cite{Binnington:2009bb}, yielding Eq.~\eqref{magnetic1}.}
\label{fig:check}
\end{center}
\end{figure}
%%%%%

A comparison between the two cases is shown in Fig.~\ref{fig:check}, where the (dimensionless) magnetic Love numbers computed from Eq.~\eqref{magnetic2} have been arbitrarily normalized. 
As this figure clearly shows, the dimensionless magnetic Love numbers are inequivalent, although nearly universal in both cases. We verified that the same result holds also for $\ell=3$ and $\ell=4$.

We believe that Eq.~(31) in Ref.~\cite{Damour:2009vw} is incorrect, probably due to a flawed limiting procedure to the static case. As a consequence, recent results which have used that equation to compute the magnetic Love numbers (e.g. Ref.~\cite{Yagi:2013sva}, whose qualitative results have anyway been confirmed in this work) should be revisited.

Finally, we remark that recent work by Landry and Poisson~\cite{Landry:2015cva} has shown that the axial perturbation equations are much more naturally solved in the case of an irrotational (rather than a static) fluid, as can be seen by taking the static limit of the full perturbation equations. The issue just discussed is not related to the corrections presented in Ref.~\cite{Landry:2015cva}, which we hope to include in our analysis in the future.

%%%%%%%%%%%%%%%%%%%%%%%%%%%%%%%%%%%%%%
% \bibliographystyle{myutphys}
\bibliography{tidalrot}
\end{document}